\documentclass[12pt]{JHEP3}
\usepackage{graphicx}

\usepackage{amsmath}
\usepackage[utf8]{inputenc}

\usepackage{amssymb}
\usepackage{amsfonts}
\usepackage{amsbsy}
\usepackage{url}
\usepackage{enumerate}
\usepackage{caption}

\newcommand{\bea}{\begin{eqnarray}}
\newcommand{\eea}{\end{eqnarray}}
\newcommand{\p}{\partial}

\makeatletter
\@addtoreset{equation}{section}
\makeatother

\newcommand{\refer}[1]{(\ref{#1})}

\newcommand{\abs}[1]{\left|#1\right|}

\title{BPS Boojums in ${\cal N}=2$ supersymmetric gauge theories I}
\author{
Masato Arai$^{1}$,
Filip Blaschke$^{2}$ 
and
Minoru Eto$^{3}$ 
\\
{\it Department of Physics, Yamagata University, Kojirakawa-machi 1-4-12, Yamagata,
Yamagata 990-8560, Japan}\\\\
E-mail: $^1$ \email{meto(at)sci.kj.yamagata-u.ac.jp}\\ 
E-mail: $^2$ \email{arai(at)sci.kj.yamagata-u.ac.jp}\\ 
E-mail: $^3$ \email{fblasch(at)sci.kj.yamagata-u.ac.jp}
}

\abstract{
We study 1/4 Bogomol'nyi-Prasad-Sommerfield (BPS) composite solitons of vortex strings, domain walls and
boojums in ${\cal N}=2$ supersymmetric Abelian gauge theories in four dimensions.  
We obtain solutions to the 1/4 BPS equations with the finite gauge coupling constant.
To obtain numerical solutions for generic coupling constants, we construct globally correct approximate functions which
allow us to easily find fixed points of gradient flow equations. 
We analytically/numerically confirm that the negative mass of a single boojum appearing
at the endpoint of the vortex string on the logarithmically bent domain wall is equal to the half-mass of 
the 't Hooft-Polyakov monopole.
We examine various configurations and clarify how the  shape of the boojum depends on the coupling constants and 
moduli parameters.
We also find analytic solutions to the 1/4 BPS equations for specific values of the coupling constants.
}
\preprint{YGHP-16-01-A}

\begin{document}


\section{Introduction and summary}
Topological solitons have long been appreciated in various fields in modern physics such as string theory, field theory, cosmology, nuclear physics and condensed matter physics. 
They often appear with spontaneously broken symmetries which support the stability of the solutions with conserved topological charges.
Typical examples are Nielsen-Olesen vortex string in the Abelian-Higgs model \cite{Nielsen:1973cs}  and 't Hooft-Polyakov magnetic monopole in $SU(2)$ Yang-Mills theory
\cite{tHooft,Polyakov:1974ek}. The topological charges are associated with the homotopy groups $\pi_1(U(1))$ and $\pi_2(SU(2)/U(1))$, respectively.
Domain walls also often appear in models with degenerate and discrete vacua which are characterized by the homotopy group $\pi_0({\cal M}_{\rm vac})$, where ${\cal M}_{\rm vac}$ is a vacuum manifold.

In addition to these elementary solitons, there are also composite solitons, which include two or more different kinds of elementary solitons.
They are sometimes called D-brane solitons or field theoretical D-branes, because they share many properties with D-branes in superstring theories \cite{Polchinski:1996fm}.
Among various configurations, in this paper, we are especially interested in composite solitons, where vortex strings attach to domain walls. 
Such configurations are natural field theory counterparts to the D-branes on which fundamental strings end in superstring theory. These configurations have been studied for a long time. A simple field theoretical model possessing the composite soliton was provided in Ref.~\cite{Carroll:1997pz}.
Numerical analysis based on the specific scalar field theory was performed in \cite{Bowick:2003au}.
The existence of such configurations in ${\cal N}=1$ supersymmetric (SUSY) QCD was found in \cite{Witten:1997ep}, and  inspired by this result, qualitative explanation of existence of a vortex-string ending on a domain wall was shown in  \cite{Kogan:1997dt, Campos:1998db}. 
It was found that annihilation of domain wall and 
anti-domain
wall produces stable solitons of lower dimensionality \cite{Dvali:2002fi} in a modified model of 
${\cal N}=1$ SUSY QCD \cite{Witten:1997ep}.

Much progress has been done within SUSY gauge theories in the past decade.
In ${\cal N}=2$ SQED, seen as a low energy effective theory of ${\cal N}=2$ SQCD with $SU(2)$  gauge group with $N_F=2$ hypermultiplets,
perturbed by a small mass term for the adjoint scalar field, the 1/4 BPS equations for the Abelian vortex strings ending on the domain walls were first derived 
in \cite{Shifman:2002jm}.  Endpoints of the vortex strings on the domain wall were identified with electric particles in a low energy effective theory of the domain wall. Then the model was extended to  ${\cal N}=2$ 
$SU(2) \times U(1)$ 
SQCD with $N_F=4$ hypermultiplets \cite{Shifman:2003uh}
and the 1/4 BPS equations including the non-Abelian vortex strings \cite{Hanany:2003hp,Auzzi:2003fs} were found. A low energy effective theory on the
composite domain walls was studied and it was found that the endpoints of the non-Abelian vortex strings play the role of a non-Abelian charge  coupled with localized dual non-Abelian gauge fields at linear level of small fluctuations \cite{Shifman:2003uh},
see also related works \cite{Tong2,Eto:2008dm,Arai:2012cx,Arai:2012dm,Arai:2013mwa}.

Almost in the same period, the different kind of the 1/4 BPS composite configurations of the non-Abelian vortex and the magnetic monopoles were found
in ${\cal N}=2$ $U(N_C)$ SQCD with $N_F = N_C$ flavors 
\cite{Auzzi:2003fs, Tong1,Hanany:2004ea,Shifman:2004dr,Gorsky:2004ad}, 
see also recent related works \cite{Cipriani:2011xp,Eto:2011cv,Nitta:2010nd,Cipriani:2012pa,Arai:2014hda}.
Afterward, all these kinds of topological solitons, the vortex strings, domain walls, and the magnetic monopoles were found to coexist as the 1/4 BPS composite
states in ${\cal N}=2$ $U(N_C)$ SQCD with $N_F \ge N_C$ flavors. The most generic 1/4 BPS equations for these were found in \cite{Sakai1}.

The BPS equations are a set of first order differential equations for the gauge fields $A_\mu^a$, the adjoint scalar fields $\Sigma^a$ ($a=1,2,\cdots,N_C^2$),
and $N_C \times N_F$ squark fields $H$ and $\hat H$, which depends on three spatial coordinates 
$x^k~(k=1,2,3)$.
Since, except for several simplest configurations, 
the generic solutions have no spatial symmetries, it is quite hard to solve the BPS equations, even though they are first order differential equations.
To make things worse, no analytic solutions to the 1/4 BPS equations have been found in the model with finite gauge coupling constants.
In order to overcome these difficulties a powerful technique, so-called moduli matrix formalism, was invented for the 1/2 BPS domain walls
\cite{Isozumi:2004jc, Isozumi:2004va}, 1/2 BPS vortex strings \cite{Eto:2005yh} and for the full 1/4 BPS equations \cite{Sakai1}.
It is so powerful that not only the dimension of the moduli space but also all the information on the moduli parameters for all topological sectors are easily exhausted.
Notably, the moduli matrix formalism  provides all exact solutions in the strong gauge coupling limit where the model reduces to the massive
nonlinear sigma model whose target space is a cotangent bundle over Grassmannian manifold 
$Gr_{N_F,N_C} \simeq SU(N_F)/SU(N_C)\times SU(N_F-N_C) \times U(1)$ \cite{Sakai1}.  
There is another advantage of the moduli matrix formalism, though less emphasized in the literature, that the complicated first order differential equations 
for $2N_C^2$ fields in the vector multiplet and $2N_CN_F$ fields
in the hypermultiplet ($\hat H = 0$) are converted into $N_C(N_C+1)/2$ second order equations, collectively called the master equation
\begin{eqnarray}
\frac{1}{g^2}\p_k\left(\Omega^{-1}\p_k\Omega\right) = v^2\left({\bf 1}_{N_C} - \Omega^{-1}\Omega_0\right),
\label{eq:master_NA}
\end{eqnarray}
for real positive $N_C\times N_C$ matrix field $\Omega$ \cite{Sakai1}, where $g$ is the $U(N_C)$ gauge coupling constant, $v^2$ is the FI parameter
and $\Omega_0$ stands for a source term, which can be chosen quite freely up to
certain rules \cite{Sakai1}. Even in the case $N_F=N_C$, for which
the moduli matrix formalism gives a minimal benefit, the degrees of freedom reduces by half.
Details of the moduli matrix formalism are summarized in the review paper \cite{Eto1}.
Also, there are  
good
reviews covering many studies of the topological solitons in ${\cal N}=2$ SQCD \cite{Tong:2005un,Tong:2008qd,Konishi:2007dn,Shifman:2007ce}.

While many properties are shared by BPS topological solitons in field theory and D-branes in string theory,
there is a BPS object, which is inherently field-theoretical and has no analog in string theory. 
It appears at a junction point where two different kinds of topological solitons meet. 
Interestingly, it has a negative BPS mass and can be interpreted as a binding energy.
The first example of this kind of object was found as the domain wall junction \cite{Oda:1999az,Ito:2000zf,Ito:2000mt,Shifman:1999ri,Eto:2005cp,Eto:2005fm}. 
When a vortex string ends on a domain wall, a different type of BPS object with negative mass appears \cite{Sakai1}.
Like
the domain wall junctions, it was shown that the negative BPS mass is nothing but the binding energy of 
the vortex string and the domain wall \cite{Sakai2}. 
In particle physics context it is now called the boojum \cite{Sakai2} because a similar configuration, coined 
Boojum\footnote{The word ``boojum" originates from Lewis Carroll's poem \emph{Hunting of the Snark}.}
by Mermin, is known in the context of ${}^3$He superfluid \cite{Mermin,volovik}.  
Recently, the boojums are getting more popular in various fields. 
For example, it has been studied in the 2 component
Bose-Einstein condensates \cite{Nitta1,Kasamatsu:2013lda}, and also in the dense QCD \cite{Nitta3,Nitta2}.
The negative BPS binding energy was also found for intersections of vortex sheets in 5 dimensions \cite{Hanany:2004ea,Eto:2004rz,Eto:2015obh}.
Typically, these negative BPS masses appear in Abelian-Higgs model and these three different binding energies in 
different dimensions can be reasonably understood through a descent relations given by Kaluza-Klein dimensional 
reductions \cite{Eto2}.

Our main aim in this paper is to study the 1/4 BPS boojums in details in ${\cal N}=2$ SUSY QED with $N_F \ge 2$ flavors in the presence 
of the Fayet-Iliopoulos term. If one desires, one can think of this theory as a low energy effective theory of the mass perturbed
$SU(2)$ SQCD as considered in \cite{Shifman:2002jm}. As mentioned above, fundamental results such as the derivation of the BPS
equations, the BPS masses and the structure of the moduli space were already done \cite{Sakai1,Eto1}, and
the interaction rules of various solitons and some qualitative features were clarified \cite{Sakai2}.
However, we would like to point out that these understandings still remain at the qualitative level.
This is because neither analytic nor numerical solutions to the 1/4 BPS equations have been obtained.
Although all exact solutions to the 1/4 BPS equations were obtained \cite{Sakai1}, it was done only in the strong gauge 
coupling limit where the BPS boojum masses becomes zero and the lump strings in the bulk are singular.
In order to fill this hole, we will provide both numerical solutions for generic case
and analytical solutions for particular cases with the gauge coupling constant kept finite.
To this end, we are greatly helped by the moduli matrix formalism \cite{Sakai1,Eto1}, which reduces
the complicated 1/4 BPS equations to a mere second order equation for a single real scalar field  $u(x^k)$
\begin{eqnarray}
{1 \over 2g^2}\partial_k^2u=v^2\left(1-\Omega_0e^{-u}\right),\quad (k=1,2,3),
\label{deq}
\end{eqnarray}
which is the Abelian version of the master equation given in Eq.~(\ref{eq:master_NA}) with identification $\Omega = e^{u}$.
In this paper, we will, in particular explain how to solve this Abelian master equation.

At this point, it is worth mentioning that the two-dimensional version of Eq.~(\ref{deq}) 
is the 
well-known 
Taubes equation
for the BPS vortices in the Abelian-Higgs theory \cite{Taubes:1979tm}.
Unfortunately, no analytic solutions 
are
known for the Taubes equation in $\mathbb{R}^2$.
This is in contrast to the familiar BPS Yang-Mills instantons, 
for which the well-known Atiyah-Drinfeld-Hitchin-Manin construction is established \cite{Atiyah:1978ri}.
However, Taubes equation on the hyperbolic plane $\mathbb{H}^2$ of curvature $-1/2$ 
is slightly modified from Eq.~(\ref{deq}), such that the first term in the 
parentheses
on the right-hand side vanishes and the equation
becomes integrable \cite{Witten:1976ck}.
This is a consequence of the fact that these vortices are obtained as a dimensional reduction from $SO(3)$-symmetric 
Yang-Mills instantons on $\mathbb{R}^4$ \cite{Witten:1976ck}.
Recent progress for integrable vortex models are found in \cite{Popov:2007ms,Popov:2008gw,Krusch:2009tn,Manton:2009ja,Sutcliffe:2012pu,Manton:2010wu,Eto:2012aa}.
In short, no analytic solutions have been found for the Taubes equation in $\mathbb{R}^2$, much less for the
master equation (\ref{deq}) in $\mathbb{R}^3$.
Thus, it seems almost hopeless to find analytic solutions to the master equations (\ref{deq}). Surprisingly, we will
provide some exact solutions in both $\mathbb{R}^2$ and $\mathbb{R}^3$.

A technical but important result of this paper is finding a {\it global} approximation to solutions of the master equation (\ref{deq}),
which supports almost all other results in this paper. In general, solving the Cauchy boundary problem in $\mathbb{R}^3$ for configuration of identical topological solitons, such as parallel domain walls or parallel vortex strings, 
is not difficult because finding appropriate boundary conditions offers 
no complications. 
However, when two or more topological solitons of a different kind coexist, giving 
an appropriate boundary condition is not always a straightforward task.
A prototype example is a vortex string attached to a domain wall from one side. The vortex string has codimension two
while the domain wall has codimension one. As was mentioned above, the domain wall is pulled and it logarithmically bends
toward spatial infinity along the string axis. Thus, in order to solve the Cauchy boundary problem for this kind of configuration,
we have to have an asymptotic solution for bent domain wall.
We will provide a simple but very generic method of constructing globally correct approximate solutions from the solutions of 
constituent isolated solitons. Such global approximations are very close to the exact solution almost everywhere  except around vicinity of the junction points. Furthermore, they are regular everywhere unlike the solutions in the strong gauge coupling limit, where vortex-strings develop singular cores. 
We will use these global approximations not only for determining the boundary condition of 
the master equation (\ref{deq}) but also for a suitable initial function consistent with the boundary condition to
solve Eq.~(\ref{deq}) by the gradient flow (imaginary time relaxation) method.
The closer initial function is to the true solution, the faster the gradient flow converges. Therefore, the global 
approximate solution is useful for numerical works. 
Whatever the solutions for the constituent solitons are, numerical or analytical, our method works very efficiently.
Namely, with the global approximations, we can solve the Cauchy problem for any kind of configuration in the 
theory with arbitrary coupling constants.

As a direct consequence of solving the BPS equations for the finite gauge coupling, we reveal shape of the boojums,
see Fig.~\ref{fig:simple_config}.
So far, only a schematic picture such as a simple hemisphere for the boojum have been given in the literature. 
We will also show how the shape of the boojums is modified when the coupling constants of the model are changed. It is also interesting
to observe how they deform when multiple boojums coalesce. 
Do they behave similarly to the BPS magnetic monopoles, that it that two separate balls collide and form a donut \cite{Manton:2004tk}? Our numerical computations show
that the boojums behave very differently from the monopoles, see Figs.~\ref{fig:boojum_a}, \ref{fig:boojum_b} and 
\ref{fig:boojum_two_vor}.

Furthermore, the global approximation developed in this paper settles down a problem raised in \cite{Auzzi}:
Based on approximate solutions it was pointed out that there is an ambiguity in the definition of the Boojum mass, 
which stems from the ambiguity of the definition of the geometric parameters such as domain wall area and vortex
string length caused by logarithmic bending of the domain wall.
As mentioned above, the boojum mass was originally computed in \cite{Sakai2} and it was shown that
it is a negative half mass of 't Hooft-Polyakov monopole without any ambiguity. 
To reach this value, they simplified their calculation by taking a mean value of the scalar field in the vector multiplet at a cross section of logarithmically 
bend domain wall \cite{Sakai2}. Although the result -- the negative half of 't Hooft-Polyakov monopole --
seems to be plausible, the validity for taking the average is not very clear because the cross section is exponentially large
far away from the junction point. We will recalculate the boojum mass using the global approximation
and confirm that the calculation in \cite{Sakai2} is correct without any ambiguity.
In \cite{Auzzi}, they also considered the case of two vortex-strings attached from the both sides of the domain wall, ending at different points (see Fig. \ref{fig:vs}). According to \cite{Auzzi}, this setting raises another problem about localization of the binding energy. Two possibilities are considered: One is that the binding energy locates around the junction points only and the other is that the binding energy is also localized half way between the strings endings. Our numerical
computation confirms the former scenario, as is clearly visible in Fig.~\ref{fig:11_A}.

We have another 
progress.
We have stressed that no analytic solutions have been found for the master equation (\ref{deq}) so far.
The situation is the same even for a much simpler case of the Taubes equation, Eq.~(\ref{deq}) in $\mathbb{R}^2$,
for the local vortices. 
Thus, one might think that there is no hope to find analytic solutions to the master equations (\ref{deq})
either in $\mathbb{R}^2$ or $\mathbb{R}^3$.
Contrary to expectations,  we will
obtain several analytic exact solutions to the Taubes equation for specific semi-local vortices in $\mathbb{R}^2$.
Furthermore, combining this with the known exact solutions to the master equation in $\mathbb{R}^1$ for
the domain walls \cite{Sakai3} in a similar way with which we made the global approximate solution, we will, surprisingly, succeed in constructing analytic solutions to the master equation (\ref{deq}) in $\mathbb{R}^3$.
In connection with these results, in Appendix B we also develop accurate approximation to the individual vortex string
and domain wall solutions, which are useful to quickly obtain the global approximation for the 1/4 BPS solutions.

This paper is only a first half of a two-paper miniseries where we present our results on 1/4 BPS solitons. Here, we primarily deal with the technical issues connected with solving 1/4 BPS equations and we investigate elementary properties of the Boojum, such as its mass and shape. In the second paper \cite{Boojum2}, we collect our results about 1/4 BPS solitons which are more in-depth. First, by observing the two-dimensional spreading of magnetic field lines from the boojum inside a thick domain wall and matching it to the Coulomb law for a point magnetic charge at an asymptotic distance, we cement the notion of a Boojum being a confined fractional magnetic monopole. We further pursue this analogy by defining a ``magnetic scalar potential'', which we identify as a solution to the vortex part of the master equation. A novel solution of a semi-local boojum, which arise when a semi-local string with a size moduli attaches to a domain wall, is also investigated. In addition, we study configurations with an infinite number of vortex-strings aligned on one or both sides of the domain wall and quantify the ability of such configurations to store magnetic charge as ``magnetic capacitors''. Dyonic extensions of 1/4 BPS solitons are also pursued. Most notably, we discover that the positive and negative electric charge density is stored on opposite skins of the domain wall, making a thick domain wall an ``electric capacitor''. Finally, we also study 1/4 BPS configuration from the viewpoint of the low energy effective action, the Nambu-Goto action and the DBI action, for the domain wall.

This paper is organized as follows. In Section 2, we review our model, which is ${\cal N}=2$ supersymmetric 
$U(1)$ gauge theory in $(3+1)$-dimensions coupled to $N_F$ hypermultiplets. 
We derive the BPS equations for the boojum configuration, preserving $1/4$ supersymmetry and shows that they reduce to one differential equation (\ref{deq}).  In Section 3, we deal with the axially symmetric configuration, namely one vortex string
ending on the domain wall.  We show all the basic properties of the boojum there. We also calculate the boojum mass there. Then, we demonstrate more complicated
solutions, where multiple vortex strings end on a single domain wall in Section 4. We study similar configuration with two or more domain walls in Appendix A.
We summarize all global approximations for 1/4 BPS solutions in section 5. These are complemented by accurate approximations to the ANO vortex and the domain wall which we develop in Appendix B. Several 
 exact solutions to the 1/2 and 1/4 BPS equations in presented Section 6.
A brief discussion of the future work is given in Section~7. 


\section{The Model}\label{model}

\subsection{Abelian vortex-wall system}

Let us consider ${\cal N}=2$ supersymmetric $U(1)$ gauge theory in (3+1)-dimensions with $2N_F$ 
complex scalar fields in the charged hypermultiplets. We assemble them 
into a row vector $H \equiv (H^1,H^2,\ldots ,H^{N_F})$ and 
$\hat H^\dagger \equiv (\hat H^{1*},\hat H^{2*},\ldots ,\hat H^{N_F*})$. 
The vector multiplet includes the photon $A_\mu$ and 
a real scalar field~$\sigma$. The bosonic Lagrangian is given as
\begin{align}
\label{eq:lag} {\mathcal L} & = -\frac{1}{4 g^2}(F_{\mu\nu})^2+\frac{1}{2g^2}(\partial_{\mu}\sigma)^2+\abs{D_{\mu}H}^2
+\abs{D_{\mu}\hat H^\dagger}^2-V\,,  \\
V & = \frac{g^2}{2}\bigl(v^2-HH^{\dagger} + \hat H^\dagger \hat H\bigr)^2
+ \frac{g^2}{2}\left|H\hat H\right|^2
+\abs{\sigma H-H M}^2
+\abs{\sigma \hat H^\dagger -\hat H^\dagger M}^2\,, 
\end{align}
where $g$ is a gauge coupling constant, $M$ is a real diagonal matrix
\begin{eqnarray}
M = \mbox{diag}(m_1,\ldots, m_{N_F}),
\end{eqnarray}
and $v$ is the Fayet-Illiopoulos D-term.
Without loss of generality we can take $M$ to be traceless, 
namely $\sum_{A=1}^{N_F} m_A = 0$,\footnote{Any overall factor 
$M=m \mathbf{1}_{N_F}+\ldots$ can be absorbed into $\sigma$ by shifting $\sigma\to \sigma-m$}
and align the masses as $m_A > m_{A+1}$. Since $\tilde H$ will play no role,
we will set $\tilde H = 0$ in the rest of this paper.
Throughout this paper, we use the conventions:
\begin{align}
\eta_{\mu\nu} & = \mbox{diag}(+,-,-,-)\,, \\
F_{\mu\nu} & = \partial_{\mu}A_{\nu}-\partial_{\nu}A_{\mu}\,, \\
D_{\mu}H & = \partial_{\mu}H+ iA_{\mu} H\,.
\end{align}
%

Since our model is supersymmetric, we can reduce the full equations of motion
\begin{gather}
\label{eq:eom1} \frac{1}{g^2}\partial_{\mu}F^{\mu\nu} +i \bigl(H D^{\nu}H^{\dagger}-D^{\nu}H H^{\dagger}\bigr) = 0\,, \\
\label{eq:eom2} \frac{1}{g^2}\partial_\mu\partial^{\mu}\sigma + H\bigl(H^{\dagger}\sigma-M H^{\dagger}\bigr)+\bigl(H \sigma - H M\bigr)H^{\dagger} = 0\,, \\
\label{eq:eom3} D_{\mu}D^{\mu}H = g^2\bigl(v^2-\abs{H}^2\bigr)H-\sigma\bigl(\sigma H-H M\bigr)+\bigl(\sigma H-H M\bigr)M\,.
\end{gather} 
into 
BPS equations, which are first order differential equations.
Solutions 
to the BPS equations are called BPS solitons. 
These field configurations possess many special properties, such as 
saturating the energy bound in a given topological sector and preserving 
a fraction of supercharges in 
the supersymmetric theory.
These facts simplify construction and 
discussion of topological solutions significantly.
Of course, one can consider more general values of coupling constants without qualitatively 
changing the results. We will focus on the BPS solutions because the generic case would involve
much-complicated analysis for solving the full equations of motion.

In the absence of the mass matrix $M$, the Lagrangian \refer{eq:lag} is invariant under $SU(N_F)$ 
flavour transformation of Higgs fields $H\to H U$, $U \in SU(N_F)$. 
The non-degenerate masses in $M$ explicitly break this down to  $U(1)^{N_F-1}$, 
which we from now on assume to be the case unless stated otherwise. 

There are $N_F$ discrete vacua with representative field values
\begin{equation}
H_{\langle A\rangle} = (0,\ldots , \underbrace{v}_{A-\mbox{th}}, \ldots , 0)\,, \hspace{5mm} \sigma_{\langle k\rangle} = 
m_A\,, \hspace{5mm} (A=1,\ldots , N_F)\,.
\end{equation} 
We can act on each of these configurations by $N_F$ global $U(1)$ transformations, 
which make  each vacuum topologically a circle. Thus, 
the space of all vacuum configurations -- the manifold $\mathcal V$ -- is isomorphic to 
$N_F$-torus ${\mathcal V} \sim S^{1}\times S^{1}\times \ldots \times S^{1}$. 
A direct consequence of this fact is that the model \refer{eq:lag} has a rich spectrum of 
topological excitations \cite{Tong1, Sakai1, Sakai2}. 

Non-trivial homotopy groups $\pi_0({\mathcal V}) = \mathbb{Z}_{N_F}$ and 
$\pi_1({\mathcal V}) = \mathbb{Z}\times\mathbb{Z}\times \ldots \times \mathbb{Z} $ give rise to 
a large variety of domains walls 
and vortex strings. 
These solitons are $1/2$ BPS solitons, as they preserve half of the supercharges if the model is extended to have supersymmetry. 
In addition, there are also 
lots of composite solitons, where vortex strings attach to the domain walls. 
These configurations preserve 1/4 of supercharges and hence are known as 1/4 BPS solitons, which
we are interested in.

\subsection{1/4 BPS state}
\label{sec:1/4BPS}

Let us now construct the 1/4 BPS solitons. We will arrange vortex strings to be parallel to the $x^3$ axis and the domain walls 
to be perpendicular to the $x^3$ axis. 
We let $A_0=0$ and consider static configurations, so that the Higgs field $H$, the scalar $\sigma$ and the gauge field 
$A_k(k=1,2,3)$ are independent on $x^0$. The Bogomol'nyi bound is then obtained as\footnote{
The following useful identity holds: 
\begin{eqnarray*}
D_a H(D_aH)^\dagger = (D_1 + i \xi D_2)H((D_1 + i \xi D_2)H)^\dagger + \p_a j_a - \xi F_{12}HH^\dagger.
\end{eqnarray*}
}
\begin{align}
{\mathcal E} & = \frac{1}{2g^2}{\Big \{}
                             (\xi F_{12}-\eta \partial_3\sigma-g^2(HH^\dagger-v^2))^2
                             +(\xi F_{23}-\eta \partial_1\sigma)^2+(\xi F_{31}-\eta\partial_2\sigma)^2
                          {\Big \}} \nonumber \\
                    & ~~ + |(D_1+i\xi D_2)H|^2 + |D_3H +\eta (\sigma H-HM)|^2 \nonumber \\
                    & ~~ + \eta v^2\partial_3\sigma-\xi v^2 F_{12}+{\xi\eta \over g^2}\epsilon_{klm}\partial_k(\sigma F_{lm})
                             +\partial_k j_k\,,  \label{ec}     
\end{align}
where $\varepsilon_{klm}$ is a totally antisymmetric symbol in three dimensions ($\varepsilon_{123} = 1$) 
and  $j_k$ are non-topological currents defined by
\begin{eqnarray}
 j_a&=& i\frac{\xi}{2}\epsilon_{ab}(H D_b H^\dagger - D_b H H^\dagger)\,, \quad (a=1,2) \label{ja}\\
 j_3&=&- \eta (\sigma H-HM)H^\dagger\,. \label{jW}
\end{eqnarray}
The energy density (\ref{ec}) saturates the so-called Bogomol'nyi bound
\begin{equation}
{\mathcal E} \geq {\cal T}_W + {\cal T}_S + {\cal T}_B + \partial_k j_k\,,
\end{equation}
with
\begin{eqnarray}
{\mathcal T}_W  = \eta v^2  \partial_3 \sigma\,, \quad
{\mathcal T}_S  = - \xi v^2  F_{12}\,, \quad
{\mathcal T}_B = \frac{\eta\, \xi}{g^2} \epsilon_{klm}\partial_k (F_{lm}\sigma)\,, \label{TB}
\end{eqnarray}
if the following 1/4 BPS equations
\begin{eqnarray}
& \label{eq:bpsws1}D_3 H +\eta \bigl(\sigma H-H M\bigr) = 0\,, & \\
& \label{eq:bpsws2}\bigl(D_1+ i \xi D_2\bigr) H = 0\,,  & \\
& \label{eq:bpsws3} \eta \partial_1 \sigma = \xi F_{23}\,, \hspace{5mm} \eta \partial_2 \sigma = \xi F_{31}\,, & \\ 
& \label{eq:bpsws4} \xi F_{12}-\eta \partial_3 \sigma+g^2\bigl(v^2-\abs{H}^2\bigr) = 0\,. &
\end{eqnarray}
are satisfied. 
It is easy to verify that these equations are compatible with the equations of motion \refer{eq:eom1}-\refer{eq:eom3} 
for all values of parameters $\eta^2 = \xi^2 = 1$, where $\xi = (-1)1$ labels (anti-)vortices and $\eta = (-1)1$ 
denotes (anti-)walls. 

${\mathcal T}_W$ and ${\mathcal T}_S$, the domain wall and the vortex string energy density respectively,
are positive definite. ${\mathcal T}_B$ is the so-called \emph{boojum} energy density, which is interpreted as binding energy 
of vortex string attached to the domain wall, since it is negative irrespective the signs of $\eta$ and $\xi$ \cite{Sakai2, Auzzi}. 
The total energy of $1/4$ BPS soliton is obtained upon the space integration and it consists of three parts
%
\begin{equation}
E_{1/4} =  T_W A +T_S L + T_B\,, \label{evw}
\end{equation}
where we have denoted sum of tensions of the domain walls $T_W = \int dx^3\ {\mathcal T}_W$, 
and that of the vortex strings $T_S = \int dx^1dx^2\ {\mathcal T}_S$, respectively.
$A = \int dx^1dx^2$ and $L=\int dx^3$ stand for
the domain wall's area and length of  the vortex string.
Only masses of the boojums $T_B = \int d^3x\ {\mathcal T}_B$ are finite.
Let $t_W(A,A+1)$ be the tension of elementary domain wall interpolating the vacua $\left<A\right>$
and $\left<A+1\right>$  \cite{Sakai1}
\begin{eqnarray}
t_W(A,A+1) = v^2 |m_{A+1}-m_A|.
\end{eqnarray}
Summing up all the elementary domain walls and vortex strings, we have
\begin{eqnarray}
T_W = \sum t_W = v^2 |\Delta m|,\quad T_S = 2\pi v^2 |k|,
\label{eq:TW}
\end{eqnarray}
where we have denoted $\Delta m = \left[\sigma\right]^{x^3 = +\infty}_{x^3 = -\infty}$ and
$k \in \mathbb{Z}$ stands for the number of vortex strings.
$T_B$ has been also calculated in \cite{Sakai1,Sakai2,Auzzi}. But, there is a discussion
about determining the mass of a single boojum which appears as a junction of the semi-infinite
string ending on a logarithmically bent domain wall. In what follows, we will confirm 
without doubt that the boojum mass even for bent domain walls
is given by the following formula \cite{Sakai2}
\begin{eqnarray}
T_B = - \frac{2\pi}{g^2}\sum  |m_{A+1} - m_{A}|,
\end{eqnarray}
where the sum is taken for all the junctions of domain walls and vortex strings in the solution
under consideration.

\subsection{The moduli matrix formalism}
\label{sec:mmf}

With the use of the moduli matrix approach \cite{Isozumi:2004jc, Sakai1, Eto1} the set of the equations (\ref{eq:bpsws1})--(\ref{eq:bpsws4}) amounts to solving one equation called the master equation. It is easy to see that the following ansatz solves (\ref{eq:bpsws1})--(\ref{eq:bpsws3}) 
\begin{eqnarray}
H = v e^{-\frac{u}{2}}H_0(z) e^{\eta M x^3}\,,\quad
A_1+i\xi A_2 = -i \partial_{\bar z} u\,, \quad
\eta\sigma+i A_3 = \frac{1}{2}\partial_3 u\,, \label{a3}
\end{eqnarray}
where $H_0(z)$ is the so-called the moduli matrix which is holomorphic in a complex coordinate $z\equiv x^1+i\xi x^2$. 
Our notation is $\partial_{\bar{z}}= (\partial_1+i \xi \partial_2)/2$.
By using the $U(1)$
gauge transformation, we fix 
$u=u(z,\bar{z},x^3)$ to be real. Then we have $A_3 = 0$. 
The last equation \refer{eq:bpsws4} turns into the master equation
\begin{equation}\label{eq:quatermaster}
\frac{1}{2g^2v^2}\p_k^2 u = 1-\Omega_0 e^{-u}\,, \hspace{5mm} \Omega_0 = H_{0}(z)e^{2\eta M x^3} H_0^{\dagger}(\bar z)\,.
\end{equation}
Now, all fields can be expressed  in terms of $u$ as follows
\begin{eqnarray}
\sigma = \frac{\eta}{2}\p_3 u,\ 
F_{12} = -2\xi \p_z\p_{\bar z} u,\ 
F_{23} = \frac{\xi}{2}\p_3\p_1 u,\ 
F_{31} = \frac{\xi}{2}\p_2\p_3 u.
\label{eq:F_from_u}
\end{eqnarray}
The energy densities are also written as
\begin{eqnarray}
{\mathcal T}_W &=& \frac{v^2}{2}\p_3^2 u, \label{eq:tw}\\
{\mathcal T}_S &=& 2 v^2 \p_z\p_{\bar z}u = \frac{v^2}{2}\left(\p_1^2 + \p_2^2\right)u, \label{eq:ts}\\ 
{\mathcal T}_B &=& \frac{1}{2g^2}\left\{\left(\p_1\p_3 u\right)^2 + \left(\p_2\p_3u\right)^2 - \left(\p_1^2+\p_2^2\right)\! u\  \p_3^2 u\right\}. \label{eq:tb}
\end{eqnarray}
The non-topological current $j_k$ given in Eqs.~(\ref{ja}) and (\ref{jW}) can be rewritten in
the following expression by using the BPS equations
\begin{eqnarray}
j_k = \frac{1}{2}\p_k(HH^\dagger).
\end{eqnarray}
Thus, we also have
\begin{eqnarray}
{\cal T}_4 = \partial_k j_k  = \frac{1}{2}\p^2 (HH^\dagger) = - \frac{1}{4g^2}\p^2 \p^2 u,
\end{eqnarray}
with $\p^2 \equiv \p_k^2$.
Collecting all pieces, the total energy density is given by
\begin{eqnarray}
{\mathcal E} = \frac{v^2}{2}\p_k^2 u 
+ \frac{1}{2g^2}\left\{\left(\p_1\p_3 u\right)^2 + \left(\p_2\p_3u\right)^2 - \left(\p_1^2+\p_2^2\right) u \p_3^2 u\right\}
-\frac{1}{4g^2}(\p_k^2)^2 u. \label{eq:ted}
\end{eqnarray}
Thus, the scalar function $u$ determines everything.

The moduli matrix formalism is an easy tool for describing the 1/4 BPS solitons.
Indeed, one can construct any kind of configurations by preparing appropriately the moduli matrix $H_0(z) = (h_1(z),h_2(z),\cdots,h_{N_F}(z))$.
As a simple example, let us consider the case that only $h_1$ and $h_2$ are non-zero.
The vacuum $\left<i\right>$ is given by $h_j = \delta_{ij}$ ($j=1,2$).
If they are both non zero constants, 
the corresponding configuration has the domain wall separating  $\left<1\right>$ and $\left<2\right>$ 
vacua. Since we assume $m_A > m_{A+1}$, the vacuum $\left<1\right>$ appears on the left hand side of the domain wall
while the vacuum $\left<2\right>$ on the right hand side. When one wants to put $N_1$ vortex strings in the
vacua $\left<1\right>$, one only needs to prepare $h_1 = P_1(z)$ where $P_1(z)$ stands for
a polynomial with degree $N_1$. This can be straightforwardly extended for generic $H_0(z)$.

Hence, the moduli matrix formalism allows us to handle the complicated 1/4 BPS solitons.
One can even deal with the solutions belonging to different topological sectors simultaneously.
It is nice that all information about the moduli parameters are included in $H_0$.
For these reasons, the moduli matrix formalism has been frequently used in the literature since its inception.
Nevertheless, a finishing touch has been omitted in the sense that no analytic or numerical solutions have been
constructed for the master equation (\ref{eq:quatermaster}) except for the strong gauge coupling limit $g^2 \to \infty$
\cite{Isozumi:2004jc} as was mentioned in the Introduction.
In the literature, only schematic pictures in the weak/strong gauge coupling region has been given.
This lack of solutions also caused some confusion about the definition of the boojum \cite{Auzzi}, which we will mention later.


\section{Axially symmetric solutions}
\label{sec:details_axially_symmetric}

In this section, we explain how we numerically solve the master equation (\ref{eq:quatermaster}).
Throughout this section, we will concentrate on $N_F=2$ case but everything addressed here can be
straightforwardly extended to generic case.
For ease of notation, we will use the following dimensionless coordinates and mass
\begin{eqnarray}
\tilde x^k = \sqrt2 gv x^k,\quad \tilde M = \frac{1}{\sqrt2 gv } M= {\rm diag}\left(\frac{\tilde m}{2},-\frac{\tilde m}{2}\right).
\label{eq:dimless}
\end{eqnarray}
The dimensionless fields are similarly defined by
\begin{eqnarray}
\tilde H  = \frac{H}{v} =  e^{-\frac{u}{2}} H_0 e^{\eta \tilde M \tilde x^3},\quad
\tilde \sigma = \frac{\sigma}{\sqrt{2} gv} = \frac{\eta}{2}\tilde \p_3 u.
\end{eqnarray}
We will also use the dimensionless magnetic fields 
\begin{eqnarray}
\tilde F_{12} &=& \frac{1}{g^2v^2}F_{12} = -\xi \left(\tilde \p_\rho^2 + \frac{1}{\tilde \rho}\tilde \p_\rho\right) u,\ \\
\tilde F_{23} &=& \frac{1}{g^2v^2}F_{23} = \xi \tilde \p_3\tilde \p_\rho u \cos\theta,\ \\
\tilde F_{31} &=& \frac{1}{g^2v^2}F_{31} = \xi \tilde \p_3\tilde \p_\rho u \sin\theta.
\end{eqnarray}
Then, the dimensionless energy density $\tilde {\cal E}$ is defined by
\begin{eqnarray}
{\cal E} = g^2 v^4 \tilde{\cal E} = \tilde {\cal T}_W + \tilde {\cal T}_S + \tilde {\cal T}_B + \tilde {\cal T}_4,
\end{eqnarray}
where
\begin{eqnarray}
\tilde {\cal T}_W  &=& 2 \eta \tilde\p_3 \tilde \sigma = \tilde\p_3^2 u,\\
\tilde {\cal T}_S  &=& - \xi \tilde F_{12} = \left(\tilde\p_\rho^2 + \frac{1}{\tilde\rho}\tilde\p_\rho \right) u,\\
\tilde {\cal T}_B &=& 2\eta\xi \tilde\p_i\left(\epsilon_{ijk} \tilde \sigma \tilde F_{jk}\right) 
=  2 \left[\left(\tilde\p_\rho\tilde\p_3 u\right)^2 - \left(\tilde\p_\rho^2 + \frac{1}{\tilde\rho}\tilde\p_\rho\right)\! u\ \tilde\p_3^2 u\right],\\
\tilde {\cal T}_4 &=& - \left(\tilde\p_\rho^2 + \frac{1}{\tilde\rho}\tilde\p_\rho + \tilde\p_3^2\right)^2 u.
\end{eqnarray}
The relations to the original values are given as
\begin{eqnarray}
T_W 
&=&  \int dx^3\ {\cal T}_W 
= \frac{gv^3}{\sqrt 2}  \int d\tilde x^3\ \tilde{\cal T}_W =  \frac{gv^3}{\sqrt 2}  \tilde T_W \label{eq:tw2}
\\
T_S &=& \int d^2x\ {\cal T}_S = \frac{v^2}{2}\int d^2\tilde x\ \tilde{\cal T}_S = \frac{v^2}{2}\tilde T_S,\\
T_B &=& \int d^3x\ {\cal T}_B = \frac{v}{2\sqrt2 g}  \int d^3\tilde x\ \tilde{\cal T}_B = \frac{v}{2\sqrt2 g} \tilde T_B.
\end{eqnarray}

In what follows, we will not distinguish $x^k$ and $\tilde x^k$, unless stated otherwise. 
An exception is the mass: we will use the notation $\tilde m$ in order not to forget that we are using the 
dimensionless variables.
Then the master equation (\ref{eq:quatermaster}) is expressed as
\begin{equation}
\p_k^2 u - 1 + \Omega_0 e^{-u} = 0,\quad
\Omega_0 = H_0^\dagger e^{2\eta \tilde M x^3} H_0^\dagger\,.
\label{eq:master_dimless}
\end{equation}

If the moduli matrix is constant, we have flat domain walls and the function $u = u_W(x^3)$ depend only on  $x^3$ coordinate. The reduced master equation reads
\begin{equation}
\p_3^2 u_W - 1 + \Omega_0 e^{-u_W} = 0.
\label{eq:master_dimless_wall}
\end{equation}
Similarly, when the moduli matrix has only one non-zero component, say $H_0 = (P_n(z),0,\cdots,0)$,
the configuration has $n$ vortex strings in the vacuum $\left<1\right>$. 
One can eliminate the $x^3$ dependence by defining 
\begin{eqnarray}
u(x^k) = u_S(x^1,x^2) + \eta \tilde m x^3.
\end{eqnarray}
Then the master  equation reduces to
\begin{eqnarray}
(\p_1^2+\p_2^2) u_S - 1 + |P_n|^2 e^{-u_S} = 0.
\label{eq:master_dimless_vortex}
\end{eqnarray}

\subsection{Gradient flow}

Now we are ready to solve the master equation Eq.~(\ref{eq:master_dimless}) numerically. But, 
instead of solving it directly, we will solve the so-called gradient flow equation
\begin{eqnarray}\label{eq:gradflow}
\p_k^2 U - 1 + \Omega_0 e^{-U} = \p_t U,
\label{eq:GF}
\end{eqnarray}
for $U = U(x^k,t)$, using appropriate initial configuration $U(x^k,0)$ and
boundary conditions.  Normally, $U(x^i,t)$ rapidly converges to a static function.
When $\p_t U(x^k,t)$ becomes negligibly small at a sufficiently large $t$, we regard 
$U(x^k,t)$ as a solution $u(x^k)$ of the original master equation,
\begin{eqnarray}
u(x^k) = \lim_{t\to\infty} U(x^k,t).
\end{eqnarray}
The final state obtained in this way depends on the initial configuration and the boundary conditions.
Thus, the  necessary step to successfully solve Eq.~\refer{eq:gradflow} is to provide an initial
configuration compatible with the boundary conditions.

\begin{figure}[t]
\begin{center}
\includegraphics[width=15cm]{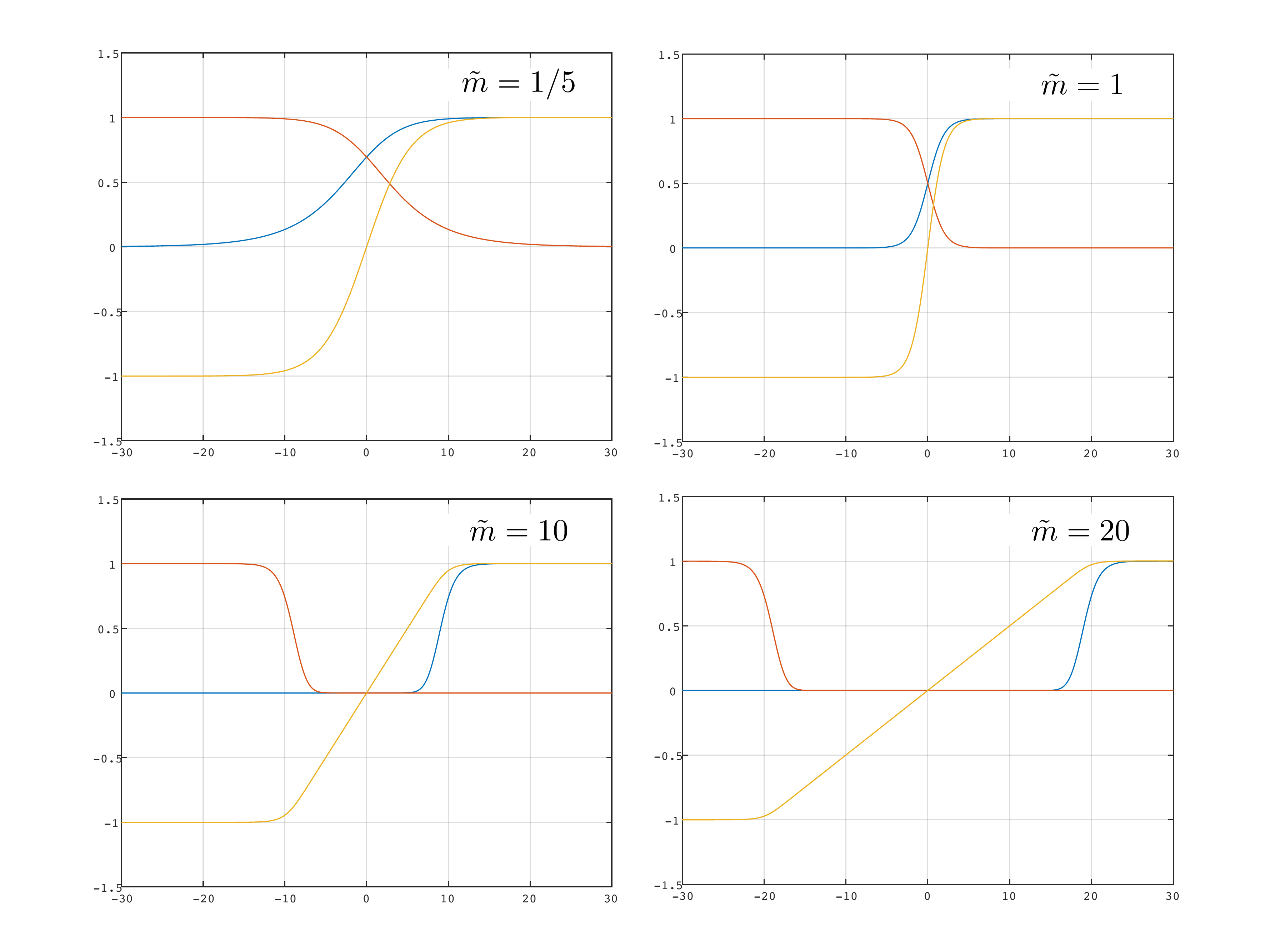}
\caption{The numerical solutions of the domain wall 
in $N_F=2$ case with $\tilde M = (\tilde m/2,-\tilde m/2)$ where $\tilde m$ is taken
to be $1/5$ (strong coupling), $1$, and $10, 20$ (weak coupling), respectively.
The horizontal axis is in dimensionless coordinate $\tilde x^3 = gv x^3$. The blue and red
lines are $\tilde H_1$ and $\tilde H_2$. The yellow line is $2\tilde \sigma /\tilde m$.}
\label{fig:walls}
\end{center}
\end{figure}

\subsubsection{Domain wall}
\label{sec:DW_sol}

Solving the gradient flow equation is not difficult especially for a single type of solitons like the flat domain walls or
the straight vortex strings.
For the domain walls perpendicular to the $x^3$ axis, the master equation reduces to
Eq.~(\ref{eq:master_dimless_wall}). The solution $u_W(x^3)$ can be obtained
by solving the corresponding gradient flow equation for $U = U_W(x^3,t)$
\begin{eqnarray}
\p_3^2 U_W - 1 + \Omega_0(x^3)  e^{-U_W} = \p_t U_W.
\end{eqnarray}
A convenient choice of the initial configuration and the boundary conditions at $x^3 = \pm L_3$ 
$(L_3 \gg \tilde m)$ is
\begin{eqnarray}
U_W(x^3,0) = \log \Omega_0(x^3),\quad
U_W(\pm L_3, t) = \log \Omega_0(\pm L_3).
\end{eqnarray}
The true solution can be obtained as the asymptotic function 
$u_W(x^3) = \lim_{t\to\infty}U_W(x^3,t)$ which satisfies
\begin{eqnarray}
u_W(x^3) \to \left\{
\begin{array}{lcccl}
\tilde mx^3 & \quad&\text{for} & \quad & x^3\gg \tilde m\,,\\
-\tilde mx^3 & &\text{for} & \quad & x^3\to-\tilde m\,.
\end{array}
\right.
\label{eq:asm_wall}
\end{eqnarray}

Several numerical solutions are shown in Fig.~\ref{fig:walls}.
There are two qualitatively different domain walls according to $\tilde m$ \cite{Shifman:2002jm}. 
In the strong gauge coupling limit, $\tilde m \ll 1$, the domain wall has a simple structure
as shown in the left panel of Fig.~\ref{fig:walls}. The traverse size of the domain wall is $\tilde d_W = 2/\tilde m$.
On the other hand, in the weak coupling region, $\tilde m \gg 1$, the domain walls consists of two layers \cite{Shifman:2002jm}:
the fundamental Higgs field decays in the outer thin layer whose transverse size is of order 1 (in the dimensionless 
unit), whereas
the singlet scalar field $\tilde \sigma$ interpolates $\tilde m/2$ and $-\tilde m/2$ through the inner fat layer whose 
transverse size is about $\tilde d_W = 2 \tilde m$.
In the latter case, the inside of the domain wall remains almost the Coulomb phase since the charged fundamental fields
are exponentially small there.

\subsubsection{Vortex string}

The vortex strings can be obtained similarly. As an example,
let us consider $n$ vortex strings in the $\left<1\right>$ vacuum. It is given by
the moduli matrix $H_0 = (P_n(z),0)$ with $P_n(z)$ being the $n$-th order polynomial.
This yields $\Omega_0 = |P_n|^2 e^{\eta \tilde m x^3}$. Then we define 
\begin{eqnarray}
U(x^k,t) = U_S(x^1,x^2)
+ \eta \tilde m x^3. 
\label{eq:U_shift}
\end{eqnarray}
This yields $\tilde \sigma = \eta \tilde m$.
The master equation to $U_S$ for the vortex strings reduces to
\begin{eqnarray}
(\p_1^2+\p_2^2) U_S - 1 + |P_n|^2  e^{-U_S} = \p_t U_S.
\end{eqnarray}
Note that $\tilde m$ disappears due to the shift given in Eq.~(\ref{eq:U_shift}).
There are no dimensionful parameters in this equation so that everything is of order one.
A convenient choice for the initial and boundary conditions is
\begin{eqnarray}
U_S(x^1,x^2,0) &=& \log \left( |P_n|^2 + e^{-\rho^2}\right),
\label{eq:init_S}\\
U_S(\pm L_1,\pm L_2,t) &=& \log \left( |P_n|^2\bigg|_{x^1=\pm L_1,x^2 = \pm L_2}\right),
\end{eqnarray}
where, of course, we assume $|L_{1,2}| \gg 1$ so that $e^{-\rho^2}$ can be neglected.
The exponential term appearing in Eq.~(\ref{eq:init_S}) is there in order to avoid singular behavior 
at the center of the vortex strings in the initial configuration.
It is exponentially small at the boundary $x^1 = \pm L_1$ and $x^2 = \pm L_2$, so that 
the initial configuration does not contradict to the boundary conditions.
As before, the true vortex solution $u_S$ for Eq.~(\ref{eq:master_dimless_vortex}) can be obtained as the asymptotic function
$u_S(x^1,x^2) = \lim_{t\to\infty}U_S(x^1,x^2,t)$.

\begin{figure}[t]
\begin{center}
\includegraphics[width=15cm]{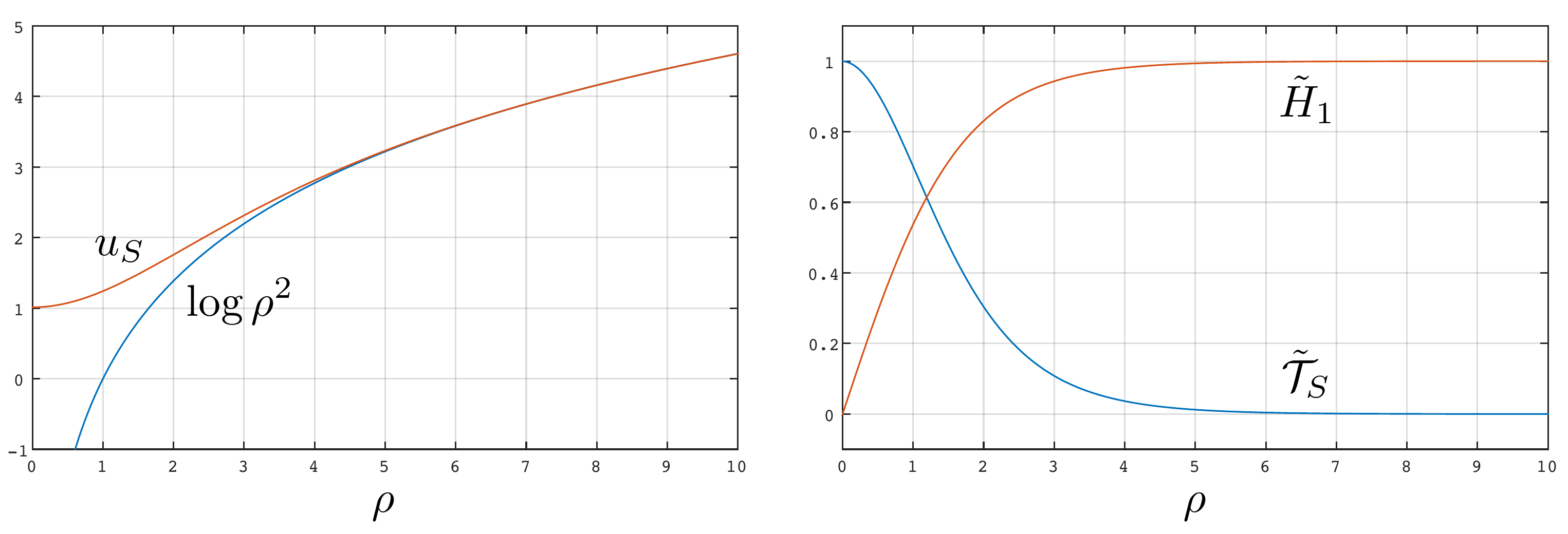}
\caption{Profiles of the vortex string solutions. The red and blue curves correspond to $u_S$ and $\log \rho^2$ in
the left panel. In the right panel, the red line stands for $\tilde H_1$ and  the blue line shows
the magnetic flux density $\tilde {\cal T}_S$.}
\label{fig:vortex}
\end{center}
\end{figure}

The single vortex string ($P_1 = z$) leads to an axially symmetric configuration $U_S(\rho,t)$ satisfying
the master equation
\begin{eqnarray}
\left(\p_\rho^2 + \frac{1}{\rho}\p_\rho\right) U_S - 1 + \rho^2  e^{-U_S} = \p_t U_S.
\end{eqnarray}
This should be solved with the boundary conditions
\begin{eqnarray}
\p_\rho U_S(0,t) = 0,\quad
U_S(L_\rho,t) = \log L_\rho^2.
\end{eqnarray}
The asymptotic behavior of $u_S(\rho) = \lim_{t\to\infty}U_S(\rho,t)$ is
\begin{eqnarray}
u_S \to \log \rho^2 \qquad (\rho \gg 1).
\label{eq:asym_vortex}
\end{eqnarray}
The numerical solution for a single vortex string is given in Fig.~\ref{fig:vortex}.

\subsection{Vortex-string attached to a domain wall}
\label{sec:1/4_BPS_solutions}

\begin{figure}[t] 
\begin{center}
\includegraphics[width=13cm]{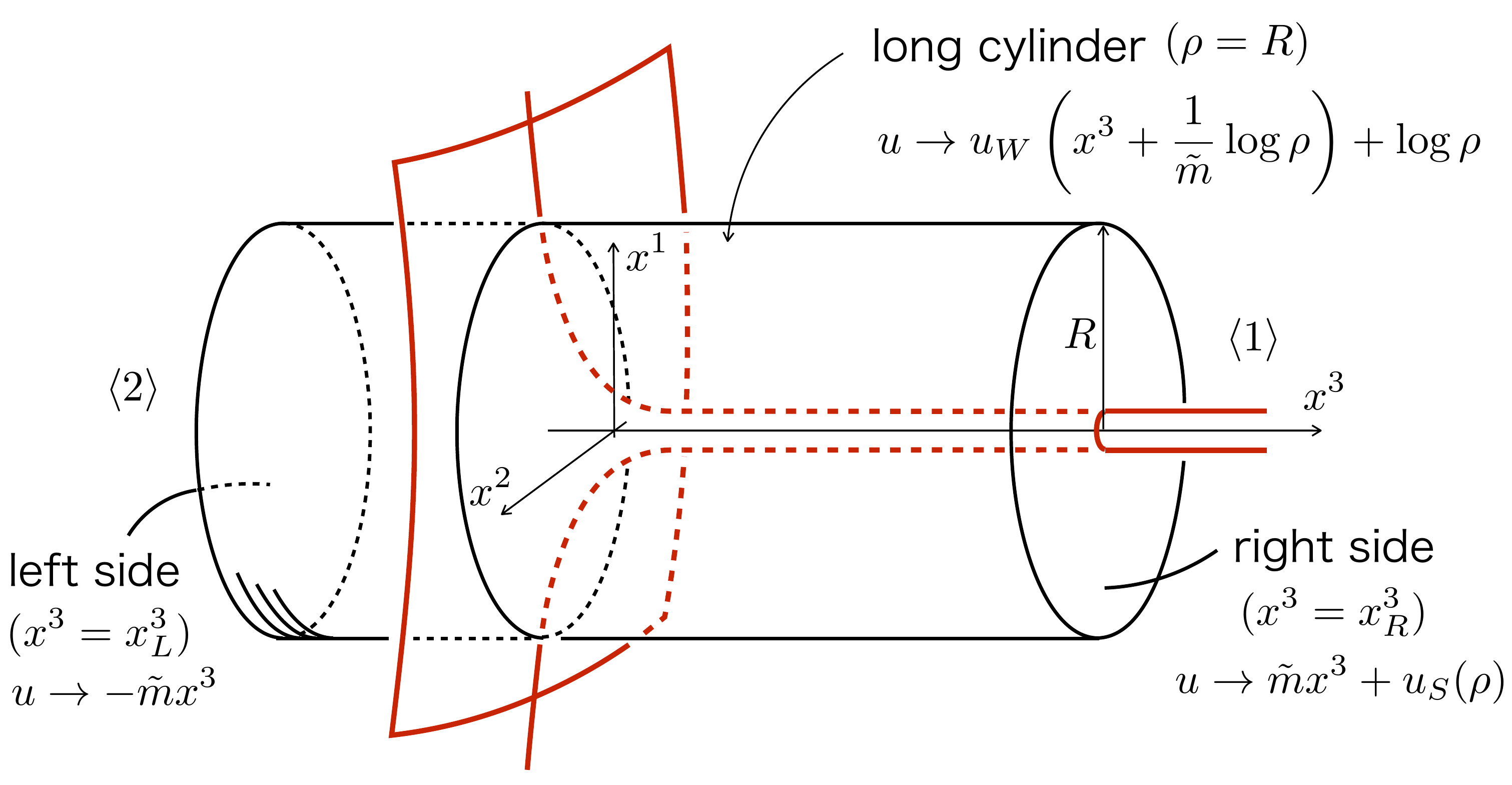}
\caption{The vortex trumpet:  vortex string ending on a logarithmically curved domain wall.
The asymptotic behavior of the function $u(\rho,x^3)$ is shown on each part of the boundary (the `long cylinder').}
\label{fig:schematic_wv}
\end{center}
\end{figure}

Let us now address how to get the simplest 1/4 BPS solutions defined by
the moduli matrix $H_0 = (z,1)$.
The corresponding gradient flow equation
is given by
\begin{eqnarray}
\left(\p_3^2 + \p_\rho^2 + \frac{1}{\rho}\p_\rho\right)U - 1 
+ \left(\rho^2 e^{\tilde m x^3} + e^{-\tilde mx^3}\right)e^{-U} = \p_t U.
\label{eq:grad_flow_1/4}
\end{eqnarray}
A significant difference from the homogeneous solitons is that
the boundaries of the 3D box parallel to the vortex string intersect with the curved domain wall
and one of the boundary parallel to the domain wall intersects with the vortex string,
see Fig.~\ref{fig:schematic_wv}.
Thus, we have to figure out an appropriate boundary condition consistent with
the curved domain wall. 

Let us first look at the region far from the vortex string axis, namely $\rho \gg 1$. 
Since $u$ depends on $x^1$ and $x^2$ only through $\log \rho$, we can drop $\partial_\rho^2 +(1/\rho) \partial_\rho$ 
term from
the master equation 
\begin{eqnarray}
\p_3^2 u - 1 + \left(\rho^2 e^{\tilde m x^3} + e^{-\tilde mx^3}\right)e^{-u} = 0. \label{eq:quater-BPS}
\end{eqnarray}
This can be rewritten as
\begin{eqnarray}
\p_3^2 (u-\log\rho) - 1 + \left(e^{\tilde m (x^3+\frac{1}{\tilde m}\log \rho)} 
+ e^{-\tilde m (x^3+\frac{1}{\tilde m}\log \rho)}\right)e^{-u+\log\rho} = 0.
\end{eqnarray}
This is nothing but the master equation for the domain wall if we regard $u-\log \rho$ as $u_W(x^3) = \lim_{t\to\infty}
U_W(x^3,t)$, so we find
\begin{eqnarray}
u \to u_W\left(x^3+\frac{1}{\tilde m}\log \rho\right) + \log \rho,\quad \left(\rho \to \infty\right).
\label{eq:u_large_rho}
\end{eqnarray}
The shift $(\log \rho)/\tilde m$ in the argument of $u_W$ reflects the fact that the domain wall asymptotically bends as
\begin{eqnarray}
x^3 \simeq - \frac{1}{\tilde m}\log \rho.
\label{eq:DW_curve}
\end{eqnarray}

Next, going far from the domain wall, $u$ approaches to the vortex string $u \to \tilde mx^3 +  u_S$ for $x^3 \gg 0$ while
becomes close to the vacuum $u \to -\tilde m x^3$ for $x^3 \ll 0$.
\begin{eqnarray}
u(\rho,x^3) \to \left\{
\begin{array}{lcccl}
\tilde mx^3 +  u_S(\rho)& \quad&\text{for} & \quad & x^3\to\infty\,,\\
-\tilde mx^3 & &\text{for} & \quad & x^3\to-\infty\,,
\end{array}
\right.\,.
\label{eq:u_large_x3}
\end{eqnarray}
with $u_S(\rho) = \lim_{t\to\infty}U_S(\rho,t)$.

A suitable function possessing these asymptotic behaviors can be obtained by replacing $\rho$ by  $e^{u_S/2}$
in the asymptotic function $u$ given in Eq.~(\ref{eq:u_large_rho}).
This is our choice for the initial configuration to solve the gradient flow
\begin{eqnarray}
U(\rho,x^3,t=0) =  u_W\left(x^3 + \frac{1}{2\tilde m} u_S(\rho)\right) + \frac{1}{2} u_S (\rho) \equiv {\mathcal U}(\rho, x^3) .
\label{eq:initial}
\end{eqnarray}
Since $u_S(\rho) \to \log \rho^2$ as $\rho \to \infty$, this function is consistent with
Eq.~(\ref{eq:u_large_rho}). Furthermore, since $u_W(x^3) \to \pm \tilde m x^3$ for $x^3 \to \pm \infty$
we have
${\mathcal U} \to \pm \tilde m \left(x^3 + \frac{1}{2\tilde m} u_S(\rho)\right) + \frac{1}{2} u_S (\rho)$, which is desired asymptotic
behavior of Eq.~(\ref{eq:u_large_x3}).
The boundary conditions consistent with the initial configuration are
\begin{eqnarray}
\p_\rho U(0,x^3,t) = 0,\quad U(L_\rho,x^3,t) = {\mathcal U}(L_\rho,x^3),\quad
U(\rho,\pm L_3,t ) = {\mathcal U}(\rho,\pm L_3).
\end{eqnarray}
Note that the first condition is ensured by  $\p_\rho u_S(\rho = 0) = 0$.
The initial configuration (\ref{eq:initial}) gives us an important information about  the domain wall's position 
for whole $\rho$ as
\begin{eqnarray}
x^3(\rho) = - \frac{1}{2\tilde m}u_S(\rho),
\label{eq:dw_posi_finite}
\end{eqnarray}
which is consistent with the asymptotic behavior given in Eq.~(\ref{eq:DW_curve}).
In the strong gauge coupling limit, we have the exact solution $u_S = \log \rho^2$, so the asymptotic
relation (\ref{eq:DW_curve}) becomes exact for whole $\rho$.
As shown in the left panel of Fig.~\ref{fig:vortex}, the domain wall is smooth everywhere in the finite gauge coupling,
but it gets logarithmically sharp at the origin in the strong gauge coupling limit.

\begin{figure}[t]
\begin{center}
   \includegraphics[width=15cm]{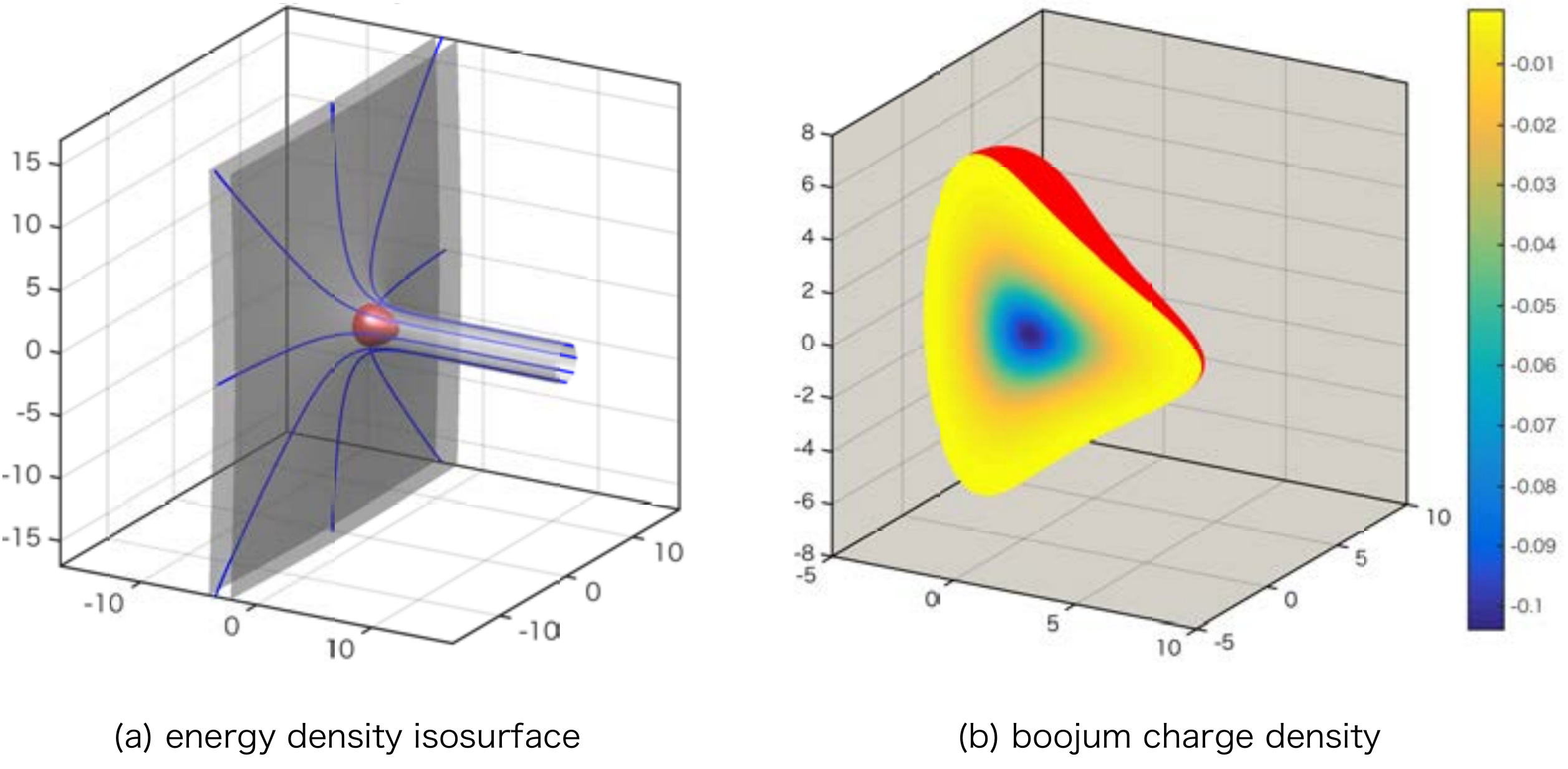}
\caption{(a) The dimensionless energy density isosurface and (b) the dimensionless boojum charge density for the finite 
vortex string ending on the bent domain wall for $\tilde m=1$. In the panel (a), the gray surface corresponds
to the dimensionless energy isosurface, the red one to the dimensionless boojum charge isosurface, and
the blue lines show the flows of the magnetic fields.}
\label{fig:simple_config}
\end{center}
\end{figure}

\subsection{Boojum at different gauge couplings}

Having the appropriate initial configuration ${\mathcal U}$, we now solve the gradient flow equation (\ref{eq:grad_flow_1/4}).
We solve it by using a standard finite difference method with the Crank-Nicolson scheme. 
A typical solution with $\tilde m=1$ is shown in Fig.~\ref{fig:simple_config}.
One can clearly see that the straight vortex string with a {\it finite} diameter ends on the logarithmically curved domain wall.
This is an obvious contrast to previously obtained solutions with singular vortex strings \cite{Portugues,Sakai1}.
Furthermore, most importantly, our solution has the boojum which contributes to the total energy
(the boojum disappears in the strong coupling limit since the boojum charge is proportional to $1/g^2$).
With our numerical solution at hand, we figure out  the correct shape of the boojum as shown in Fig.~\ref{fig:simple_config}.
So far, the shape of boojum has been only schematically realized. For example, in  Ref.~\cite{Sakai2}, 
the boojum was expressed
as just a hemisphere which is little too simple compared to our numerical solution.
The third advantage to have the numerical solution for a finite gauge coupling is that we can clearly see the distribution
of the magnetic flux which enters from the vortex string into the domain wall. In Fig.~\ref{fig:simple_config} we show several 
lines of the magnetic field (blue solid curves). They form a flux tube inside the vortex string and radially spread out once they enter the domain 
wall. This picture can correctly be obtained only for the finite gauge coupling since electromagnetic interaction disappears 
in the strong coupling limit.

\begin{figure}[t]
\begin{center}
   \includegraphics[width=15cm]{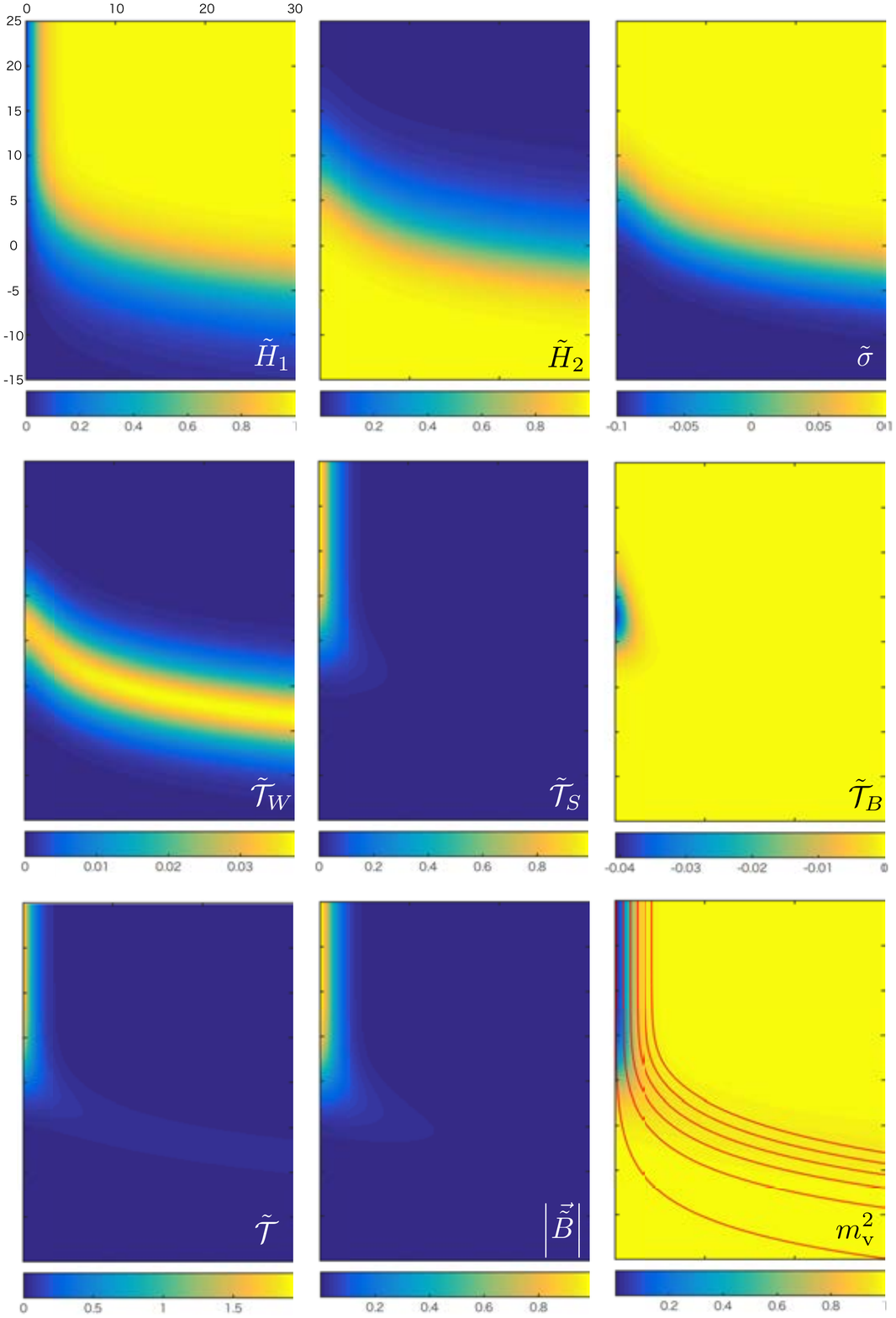}
\caption{The numerical solution for $\tilde m = 1/5$, the strong gauge coupling region
where the domain wall strongly bends.
The 1st row shows the scalar fields, and the 2nd row shows the constituent topological charge
densities. The right-bottom panel shows the effective photon mass $m_{\rm v}^2$ and the
red lines are magnetic force lines.}
\label{fig:m1o5}
\end{center}
\end{figure}

\begin{figure}[t]
\begin{center}
   \includegraphics[width=15cm]{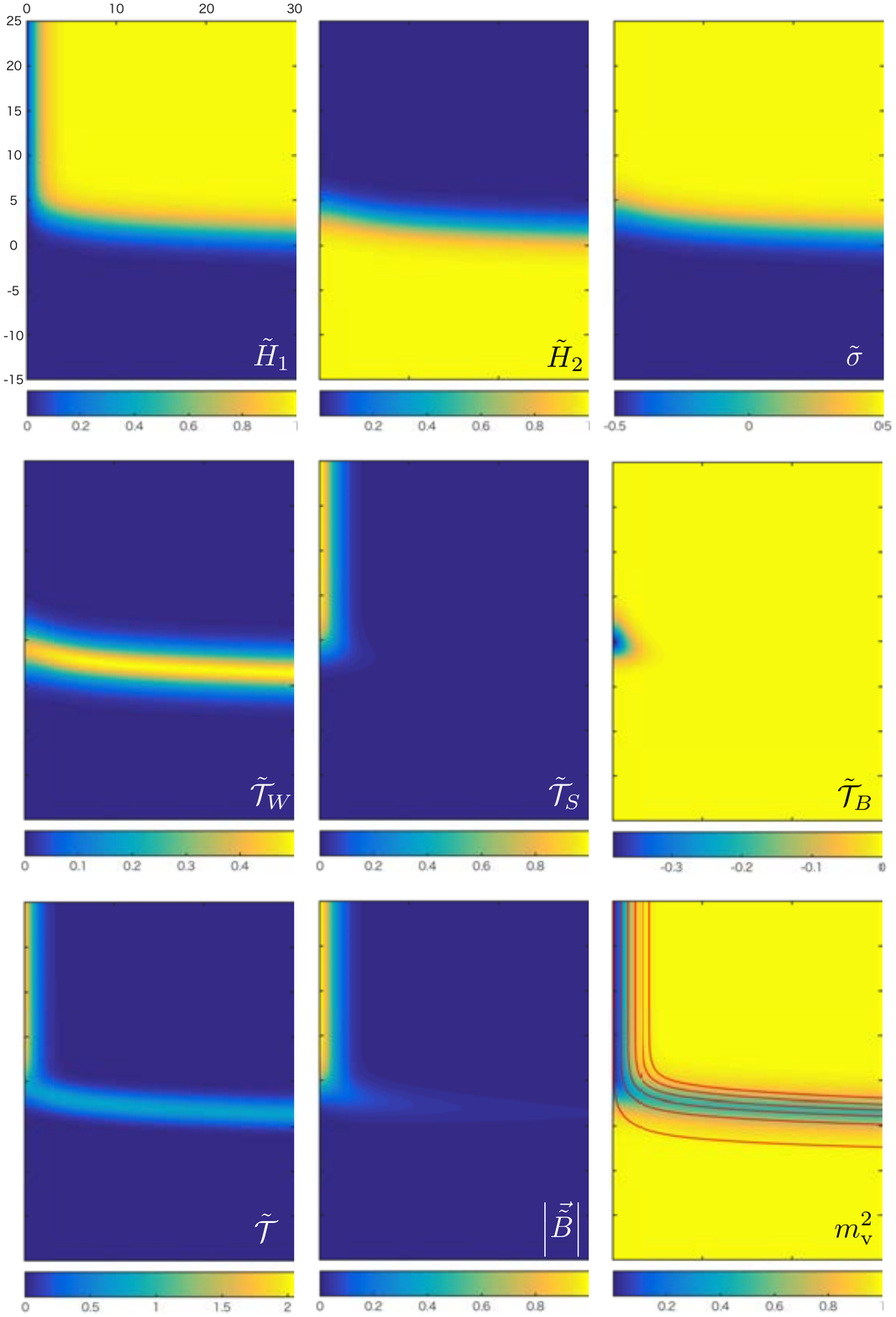}
\caption{The numerical solution for $\tilde m = 1$, an intermediate gauge coupling constant.
See also the caption of Fig.~5.}
\label{fig:m1}
\end{center}
\end{figure}

\begin{figure}[t]
\begin{center}
   \includegraphics[width=15cm]{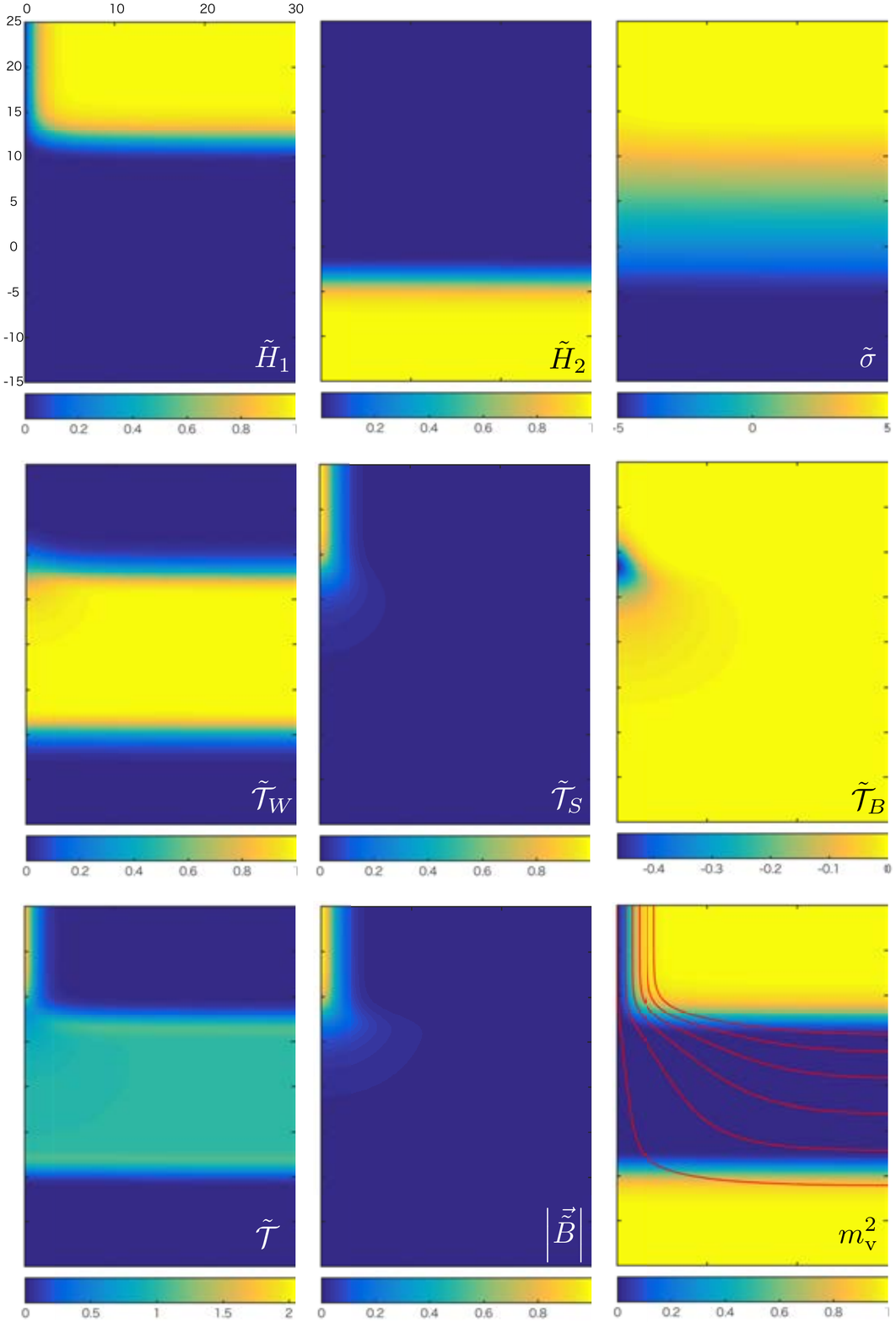}
\caption{The numerical solution for $\tilde m = 10$, the weak  gauge coupling region where
the domain wall slowly bends.
See also the caption of Fig.~5.}
\label{fig:m10}
\end{center}
\end{figure}

The numerical solutions also shed light on both qualitative and quantitative difference between
the 1/4 BPS configurations in the strong ($\tilde m \gg 1$) and the weak ($\tilde m \ll 1$) gauge coupling regions.
Especially, the vicinity of a junction point of the domain wall and vortex string is hard to correctly
understand without numerical methods. 
We show the several solutions in details; the profiles of the fields, the topological charge densities, and
the energy densities are displayed in Fig.~\ref{fig:m1o5} ($\tilde m = 1/5$),
Fig.~\ref{fig:m1} ($\tilde m = 1$) and Fig.~\ref{fig:m10} ($\tilde m = 10$).

Since we have rescaled out $gv$ dependence, the asymptotic vortex string configuration is common for
all the cases. This can be seen in regions near  the top-left edges of the middle panels of Figs.~\ref{fig:m1o5} -- \ref{fig:m10}
in which the topological charge densities of the vortex string $\tilde{\cal T}_S$ are shown.
The bottom-middle panels show the magnitude of the magnetic fields which are slightly different 
from the string topological charge density $\tilde{\cal T}_S$.

The three panels in the first row of Figs.~\ref{fig:m1o5}-\ref{fig:m10} show $\tilde H_1$, $\tilde H_2$ and $\tilde \sigma$.
The field $\tilde H_1$ develops the non-zero VEV 
in the upper region ($x^3 \gg 0$). Namely,
the light yellow regions of the top-left panels correspond to the vacuum $\left<1\right>$.
Similarly, the yellow parts of the top-middle panels show the region of the vacuum $\left<2\right>$.
Since the phase of $\tilde H_1$ has the winding number 1 for the vortex string, $\tilde H_1$ vanishes at 
the core of the vortex string (on the top-left edge $(\rho = 0, x^3 \gg 0)$). 
Since neither $\tilde H_1$ nor $\tilde H_2$ develops non-zero VEV there, the $U(1)$ gauge symmetry
is recovered. 

Shape of the bent domain wall can be seen in the middle-left panels where 
the topological charge density $\tilde{\cal T}_W$ is plotted.
The yellow-line region of $\tilde{\cal T}_W$ is consistent with the yellow-blue transit region of $\tilde\sigma$
in the top-right panels.
The domain wall strongly bends for the strong coupling ($\tilde m \ll 1$) while it seems to be almost flat for
the weak coupling ($\tilde m \gg 1$).
This difference comes from the wall to string tension ratio, which can be found by comparing the scales
of the color bars of the middle-left and middle-middle panels.

The domain wall tends to strongly bend when the tension of the vortex string is larger than that of the domain wall.
This can be also understood from Eq.~(\ref{eq:DW_curve}), which tells that the asymptotic wall curve is
$x^3 \simeq - \log \rho^{1/\tilde m}$. For example, $x^3(\rho = 10) = - \log 10$ for the case of $\tilde m = 1$. 
To get the same amount of bending in the case of $\tilde m = 10$, one needs to go to $\rho = 10^{10}$!
This is the reason why the domain wall seems almost flat for $\tilde m = 10$ in the middle-left panel of Fig.~\ref{fig:m10}.
It is also seen that the domain wall at $\rho = 0$ is not singular for any $\tilde m$.
In contrast to the inside of the vortex string, there are no topological reasons for the domain wall
core to be in the unbroken phase. Indeed, both $\tilde H_1$ and $\tilde H_2$ are non-zero inside the domain wall
with $\tilde m = 1/5$ and $1$. This is consistent with the single domain wall solutions given
in Fig.~\ref{fig:walls}. Only when the gauge coupling is very weak ($\tilde m \gg 1$), the domain wall core develops the fat inner layer of the unbroken phase.
While the diameter of the vortex string is fixed to be of order one, width of the domain wall varies 
as $\tilde m$ is changed, see Fig.~\ref{fig:walls}.
The domain wall is sharpest when $\tilde m \simeq 1$, while it becomes much fatter than the vortex string
for both weak and strong gauge coupling limits, see the bottom-left panels showing
the full energy density including the surface term.

Our prime interest is to understand the junction point of the domain wall and the vortex string.
Especially in the weak coupling limit, since  the cores of the vortex string and domain wall are in the unbroken phase.
The top-left and top-middle panels of Fig.~\ref{fig:m10} clearly show that the two unbroken regions
are just smoothly connected.
We are also interested in the boojum, the binding energy appearing at the junction point.
The middle-right panels exhibit a variety of shapes of boojum depending on $\tilde m$.
It is basically a smoothed cone which becomes thinner for $\tilde m \ll 1$ and fatter for $\tilde m \gg 1$.

Finally, in the bottom-right panels we plot the mass square of the gauge field defined by
\begin{eqnarray}
m_{\rm v}^2 \equiv \left| \tilde H_1 \right|^2 + \left| \tilde H_2 \right|^2.
\end{eqnarray}
The gauge symmetry is recovered in the deep-blue region, while is strongly broken in the light-yellow region.
In addition, we also plot flow of the magnetic field (magnetic force lines). In general, the flow tends to
be concentrated into the region where the gauge symmetry is unbroken or weakly broken.
In the case of the strong gauge coupling ($\tilde m = 1/5$ given in Fig.~\ref{fig:m1o5}), 
the incoming magnetic force lines through the vortex string are forced to bend when they encounter the
domain wall. They scatter and then spread out along the domain wall because the breaking of the 
gauge symmetry is milder inside the domain wall. However, the lines running near the core of the vortex 
string are pushed out of the domain wall. This happens because the energy cost to bend the lines all the way into 
the core of domain wall 
is larger  than that to have
mildly bending lines going through the bulk.
In contrast, in the intermediate case with $\tilde m = 1$ given in Fig.~\ref{fig:m1}, 
the magnetic force lines go along the
domain wall. This is because the mass of the gauge field inside the domain wall is sufficiently small
so that the magnetic force lines bend and tend to go inside the domain wall.
The weak coupling case with $\tilde m = 10$ shown in Fig.~\ref{fig:m10} has totally different structure.
Inside the fat domain wall is the unbroken phase. The incoming magnetic force lines spread out {\it linearly} in
the domain wall until they encounter the other side of the domain wall beyond which is the broken phase.

\clearpage

\subsection{Boojum mass}

Finally, let us  compute the tensions of the topological objects:
The domain wall tension is everywhere constant in the $x^1$--$x^2$ plane and it is given as
\begin{eqnarray}
\tilde T_W(\rho) &=& \int^\infty_{-\infty} d x^3\ \tilde{\cal T}_W(\rho,x^3) \nonumber\\
&=& \partial_3 u( x^3 \to \infty) - \partial_3 u( x^3\to -\infty) \nonumber\\
&=& 2\tilde m.
\end{eqnarray}
The string tension is a constant along the $x^3$ direction 
\begin{eqnarray}
\tilde T_S( x^3) &=& \int d x^1d x^2\ \tilde {\cal T}_S(\rho, x^3) \nonumber\\ 
&=&2\pi \int_0^\infty d \rho \ \partial_{\rho} \left( \rho \partial_{\rho}u\right) \nonumber\\
&=& 2\pi  \left[\frac{1}{\tilde m} u_W'\left(x^3+\frac{1}{\tilde m}\log \rho\right)_{ \rho \to \infty} + 1\right]\nonumber\\
&=& 4\pi,
\end{eqnarray}
where we used the asymptotic behavior given in Eq.~(\ref{eq:u_large_rho}) and 
$u_W(x^3) \to \tilde mx^3$ for $x^3 \gg 0$.
In terms of the original variables, these are rewritten as
\begin{eqnarray}
T_W &=& \frac{gv^3}{\sqrt2} \times 2\tilde m = \sqrt2\,gv^3 \frac{m}{\sqrt2\,gv} = m v^2\,\\
T_S &=& \frac{v^2}{2} \times 4\pi = 2\pi v^2.
\end{eqnarray}

One should be careful that the (dimensionless) string tension is always $4\pi$ even in the vacuum $\left<2\right>$ side
of the domain wall ($x^3 \ll 0$) where we do not put the vortex string.
This is simply a consequence of the flux conservation. 
The magnetic flux cannot invade the vacuum $\left<2\right>$, so that it flows along the surface of 
the domain wall which logarithmically bends.  
Therefore, regardless of $x^3$,
we always have a total flux of $4\pi$.
In other words, the vortex string {\it is not} chopped by the domain wall but
exponentially inflates. This observation can also be justified by
seeing the profile of the scalar field $H_1$.
The amplitude of the scalar field whose phase has the winding number must necessarily vanish
somewhere by a topological reason.
Therefore, we can think of the deep-blue region in the top-left panels of Figs.~\ref{fig:m1o5} -- \ref{fig:m10}
as the {\it inside} the vortex trumpet.

The total magnetic flux outgoing through a sufficiently long cylinder (the right side is at $x^3_R \gg \tilde m$ and 
the left side is at $x^3_L \ll -\tilde m$) of sufficiently large radius $R$ $(\gg 1)$
surrounding the vortex trumpet, see Fig.~\ref{fig:schematic_wv}, can be calculated
as follows
\begin{eqnarray}
\tilde \Phi_C &=& \int_{\rm cylinder} dS_i\ \frac{1}{2}\epsilon_{ijk}\tilde F_{jk} \nonumber\\
&=& 2\pi R \int^{x^3_R}_{x^3_L} dx^3\ \tilde F_{23}\big|_{\rho = R,\ \theta=0} \nonumber\\
&=& 2\pi R \int^{x^3_R}_{x^3_L} dx^3\ \p_3\p_\rho u\big|_{\rho = R} \nonumber\\
&=& 2\pi R \int^{x^3_R}_{x^3_L} dx^3\ \p_3\p_\rho u_W\left(x^3 + \frac{1}{\tilde m}\log \rho\right)_{\rho =R} \nonumber\\
&=& \frac{2\pi}{\tilde m} \int^{x^3_R}_{x^3_L} dx^3\ \p_3^2 u_W\left(x^3 + \frac{1}{\tilde m}\log R\right) \nonumber\\
&=& \frac{2\pi}{\tilde m} \left[\p_3 u_W\left(x^3 + \frac{1}{\tilde m}\log R\right)\right]^{x^3=x^3_R}_{x^3=x^3_L}\nonumber\\
&=& 4\pi,
\end{eqnarray}
where we used Eqs.~(\ref{eq:asm_wall}) and (\ref{eq:u_large_rho}). 
The incoming magnetic flux passing the right side of the cylinder 
can be similarly obtained as
\begin{eqnarray}
\tilde \Phi_R &=& \int  dx^1dx^2\ \tilde F_{12}\big|_{x^3 = x^3_R} \nonumber\\
&=& \int  dx^1dx^2\ -\left(\p_\rho^2 + \frac{1}{\rho}\p_\rho\right)u\big|_{x^3 = x^3_R} \nonumber\\
&=& -2\pi \int d\rho\ \rho \left(\p_\rho^2 + \frac{1}{\rho}\p_\rho\right)\left(\tilde m x^3 + u_S(\rho)\right) \nonumber\\
&=& -2\pi \int d\rho\ \p_\rho\left(\rho \p_\rho u_S\right) \nonumber\\
&=& - 4\pi,
\end{eqnarray}
where we have used the boundary conditions, $\p_\rho u_S \to 0$ for $\rho \to 0$ and 
$u_S \to \log \rho^2$ for $\rho \to \infty$.
As expected, we have $\tilde \Phi_C + \tilde \Phi_R = 0$ and $\tilde T_S = - \tilde \Phi_R$
due to the flux conservation.
This conservation implies there is no magnetic flux passing the left side of the cylinder, namely 
$\tilde \Phi_L =0$.
In order to express this in a conventional way, we need to use the original variables and 
rescale $A_\mu \to g A_\mu$. Then we have 
\begin{eqnarray}
\Phi_C = \frac{1}{2g}\tilde \Phi_C = \frac{2\pi}{g},\quad
\Phi_R = \frac{1}{2g}\tilde \Phi_R = -\frac{2\pi}{g}.
\end{eqnarray}

Similarly, the boojum mass is calculated as follows.
\begin{eqnarray}
\tilde T_B &=& 2\int d^3x\ \epsilon_{ijk} \p_i\left(\tilde \sigma \tilde F_{jk}\right) \nonumber\\
&=& 2 \int_{\rm cylinder} \!\!\!\!\!\!\!\! dS_i\ \epsilon_{ijk} \left(\tilde \sigma \tilde F_{jk}\right)  + 4\int dx^1dx^2\ \tilde \sigma \tilde F_{12}\big|_{x^3 =x^3_R}\nonumber\\
&=& 2 \tilde m \tilde \Phi_R \nonumber\\
&=& - 8\pi \tilde m,
\label{eq:mass_boojum_cal1}
\end{eqnarray}
where we used $\tilde \sigma \to \tilde m/2$ for $x^3 = x^3_R \gg \tilde m$ and the following identity
\begin{eqnarray}
\int_{\rm cylinder} \!\!\!\!\!\!\!\!\!\!\!\! dS_i\ \epsilon_{ijk} \left(\tilde \sigma \tilde F_{jk}\right)
&=& 4\pi R \int^{x^3_R}_{x^3_L} dx^3\ \tilde \sigma \tilde F_{23}\big|_{\rho = R,\ \theta = 0} \nonumber\\
&=& 2\pi R \int^{x^3_R}_{x^3_L} dx^3\ \p_3 u \p_3\p_\rho u\big|_{\rho =R} \nonumber\\
&=& 2\pi R \int^{x^3_R}_{x^3_L} dx^3\ \p_3u_W\!\!\left(x^3+\frac{1}{\tilde m}\log\rho\right)
\p_3\p_\rho u_W\!\!\left(x^3+\frac{1}{\tilde m}\log\rho\right)\bigg|_{\rho =R} \nonumber\\
&=& \frac{2\pi}{\tilde m}  \int^{x^3_R}_{x^3_L} dx^3\ \p_3u_W \p_3^2 u_W \nonumber\\
&=& \frac{\pi}{\tilde m} \left[(\p_3u_W)^2\right]^{x^3_R}_{x^3_L} \nonumber\\
&=& 0.
\label{eq:mass_boojum_cal2}
\end{eqnarray}
Note that the above result reflects an accidental $Z_2$ symmetry for our specific choice of the mass matrix
$\tilde M = {\rm diag}(\tilde m/2,-\tilde m/2)$. For generic case with
$\tilde M = {\rm diag}(\tilde m_1,\tilde m_2)$ ($\tilde m_1 > \tilde m_2$), the asymptotic behavior of $u$
at a larger $\rho$ becomes 
$u \to u_W(x^3+\frac{1}{\tilde m_1-\tilde m_2}\log\rho) + \log\rho + (\tilde m_1 + \tilde m_2)x^3$
with $u_W \to (\tilde m_1 -\tilde m_2)x^3$ for $x^3 \gg \tilde m_1 - \tilde m_2$ while
$u_W \to -(\tilde m_1 -\tilde m_2) x^3$ for $x^3 \ll -(\tilde m_1 - \tilde m_2)$.
It is straightforward to verify the following identity
\begin{eqnarray}
\int_{\rm cylinder} \!\!\!\!\!\!\!\!\!\!\!\! dS_i\ \epsilon_{ijk} \left(\tilde \sigma \tilde F_{jk}\right)
= 4\pi\left(\tilde m_1 + \tilde m_2\right).
\end{eqnarray}
Thus, we find the generic form
\begin{eqnarray}
\tilde T_B = 8\pi(\tilde m_1 + \tilde m_2) -16 \pi \tilde m_1 = -8\pi (\tilde m_1 - \tilde m_2).
\end{eqnarray}
In terms of the original variables, this can be expressed as
\begin{eqnarray}\label{eq:boojumcharge}
T_B = \frac{v}{2\sqrt{2} g}\tilde T_B = \frac{-4\pi v}{\sqrt{2} g} (\tilde m_1 - \tilde m_2) = 
- \frac{2\pi}{g^2} (m_1 - m_2).
\end{eqnarray}
This is consistent with the previously obtained value in Ref.~\cite{Sakai2}.

As a closing comment for this subsection, 
we would like to emphasize that we were able to rigorously calculate the charges
thanks to the asymptotic behavior given in Eq.~(\ref{eq:u_large_rho}). In particular, we calculated
the charges by surrounding them with the {\it finite} size cylinder as shown in Fig.~\ref{fig:schematic_wv}.  
Since we have not touched the spatial infinity, the above computations do not suffer from 
any complications coming from logarithmic bending at all.
In the literature, several attempts to avoid this problem have been done. 
In Ref.~\cite{Sakai1}, the mass of two coaxial boojums was calculated for the flat domain wall on which 
one coaxial vortex string ends on from both sides and found $2T_B = -4\pi(m_1-m_2)/g^2$.
Since the domain wall is flat in this case, there are no anxieties.
Then, in the proceeding work \cite{Auzzi}, it was proposed that the mass of the single boojum can 
be obtained by just dividing the mass of the two coaxial boojums
by 2. In the same paper \cite{Auzzi}, they also discussed that there is an uncertainty for determining the mass of the
single boojum in the case of logarithmically bent domain wall due to a geometrical ambiguity.
Ref.~\cite{Sakai2} is the first work which showed $T_B = -2\pi(m_1-m_2)/g^2$ for the isolated single boojum
in the logarithmically curved domain wall.
They reached this value by integrating the mass density ${\cal T}_B$ at $x^3 = \pm \infty$
as $T_B = \int dx^1dx^2\ {\cal T}_B\big|_{x^3=\infty} - \int dx^1dx^2\ {\cal T}_B\big|_{x^3=-\infty}$
under the assumption that no flux goes through the boundary transverse to the vortex string axis.
The first term unambiguously equals to $-2\pi m_1/g^2$. The second term is somewhat tricky since
one needs to consider the intersection of the logarithmically bent domain wall with the boundary at $x^3 = -\infty$.
In Ref.~\cite{Sakai2}, $\sigma$ was simply replaced by a mean value $(m_1+m_2)/2$ and made use of
the flux conservation, then the second term was found to be $-\pi(m_1+m_2)/g^2$.
Summing up the two contribution gives us $T_B = -2\pi(m_1-m_2)/g^2$.
This way of computing the boojum mass is plausible, but it might be better to confirm it by a 
more rigorous way.
We believe that our calculations above contribute to this point. For further confirmation,
we perform the integrations numerically and find  $-\tilde T_B/8\pi\tilde m =  0.976, 0.996, 1.010, 0.997$ 
for $\tilde m = 1, 10, 20,30$, respectively.
In conclusion, we agree with \cite{Sakai2} that the boojum mass can be cleanly separated 
from the semi-infinite vortex string and the logarithmically bent domain wall.


\clearpage

\section{Non-axially symmetric solutions}\label{nu}
\label{sec:nonaxial_Nf2}

In the previous section, we have concentrated on studying the
1/4 BPS states which are all axially symmetric about the vortex-string axis.
This symmetry reduces the problem from being three-dimensional to two-dimensional. 
In this section we will show many numerical solutions, which require full three-dimensional treatment. 
We can again use the moduli matrix formalism introduced in Sec.~(\ref{sec:mmf}) to reduce the 1/4 BPS equations  (\ref{eq:bpsws1})--(\ref{eq:bpsws4}) to the master equation (\ref{eq:quatermaster}) for a single scalar function $u(x^k)$.


Let us first look at the 1/4 BPS solutions in $N_F=2$ case with $\tilde M = (\tilde m/2,-\tilde m/2)$,
which has two distinct vacua
$\left<1\right>$ and $\left<2\right>$, and one domain wall separating those vacua.
The moduli matrix for $n_1$ ($n_2$) vortex strings in the vacuum $\left<1\right>$ ($\left<2\right>$)
is given by
\begin{eqnarray}
H_0 = (P_{n_1}(z),\ P_{n_2}(z)),
\label{eq:mm_Nf2_gene}
\end{eqnarray}
where $P_n(z)$ stands for an arbitrary polynomial of $n$-th power in $z$.
For this moduli matrix, we solve the master equation (\ref{eq:master_dimless}) written
in terms of the dimensionless parameters (\ref{eq:dimless}).
As before, we upgrade the master equation to the gradient flow (\ref{eq:GF}) 
for the real scalar function $U(x^k,t)$,
\begin{eqnarray}
\p_k^2 U - 1 + \left(|P_{n_1}|^2 e^{\tilde m x^3}  + |P_{n_2}|^2 e^{-\tilde mx^3}\right) e^{-U} = \p_t U.
\label{eq:gf_two_vor}
\end{eqnarray}
We need an appropriate initial function for solving this. It is given by
\begin{eqnarray}
{\mathcal U}(x^k) = u_W\left(x^3+\frac{u_S^{(n_1)}-u_S^{(n_2)}}{2\tilde m}\right)+\frac{u_S^{(n_1)}+u_S^{(n_2)}}{2},
\label{eq:U0_two_vor}
\end{eqnarray}
where $u_W(x^3)$ is the domain wall solution to the master equation (\ref{eq:master_dimless_wall})
and $u_S^{(n)}(x^1,x^2)$ is the $n$ vortex string solution to the master equation 
(\ref{eq:master_dimless_vortex}).
One can easily check that this correctly reproduces the initial functions given in Sec.~\ref{sec:details_axially_symmetric}
for $n_{1,2} = 0$ or $1$.

\begin{figure}[h]
 \begin{minipage}{0.5\hsize}
  \begin{center}
   \includegraphics[width=7cm]{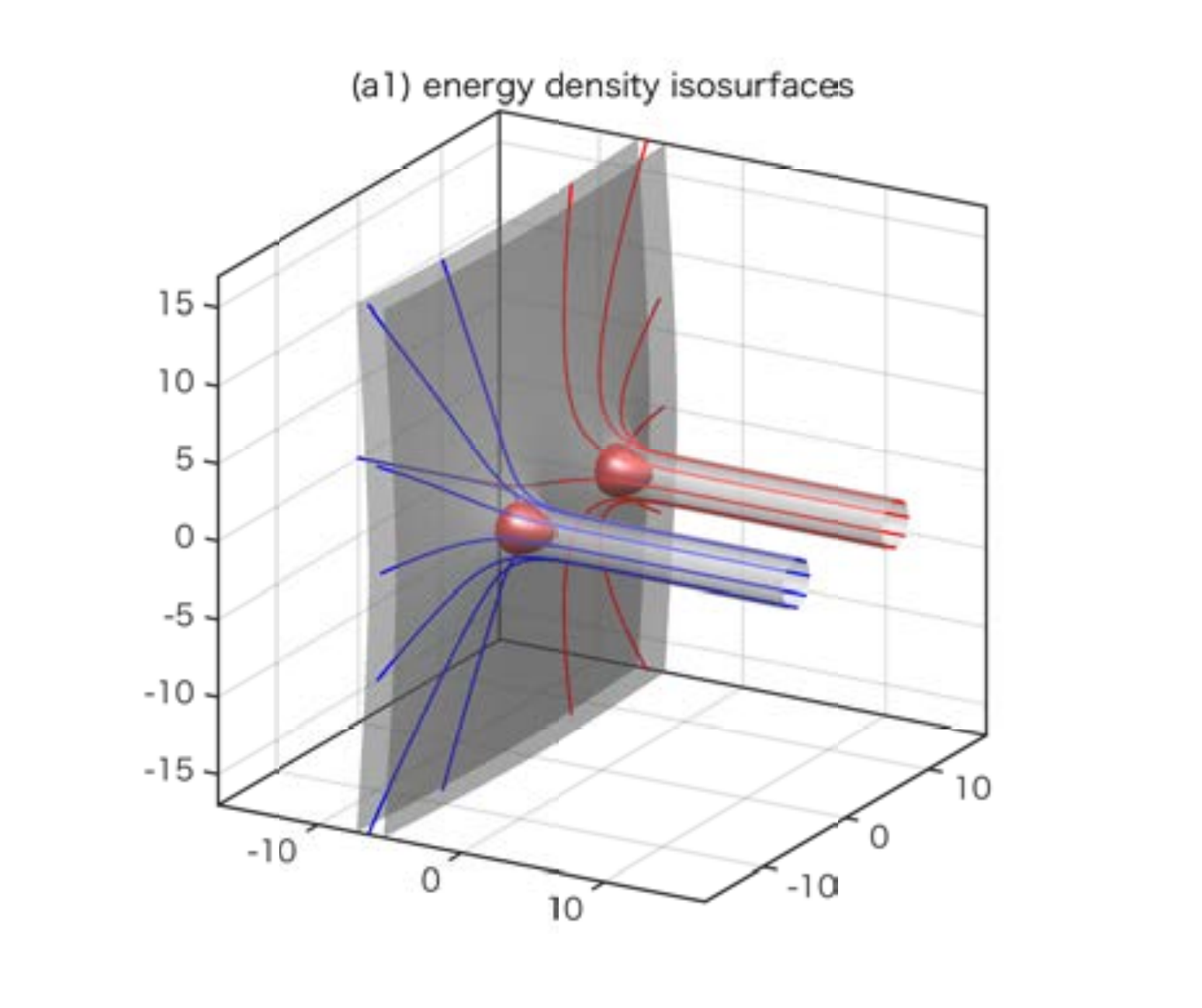}
  \end{center}
 \end{minipage}
  \begin{minipage}{0.5\hsize}
  \begin{center}
   \includegraphics[width=7cm]{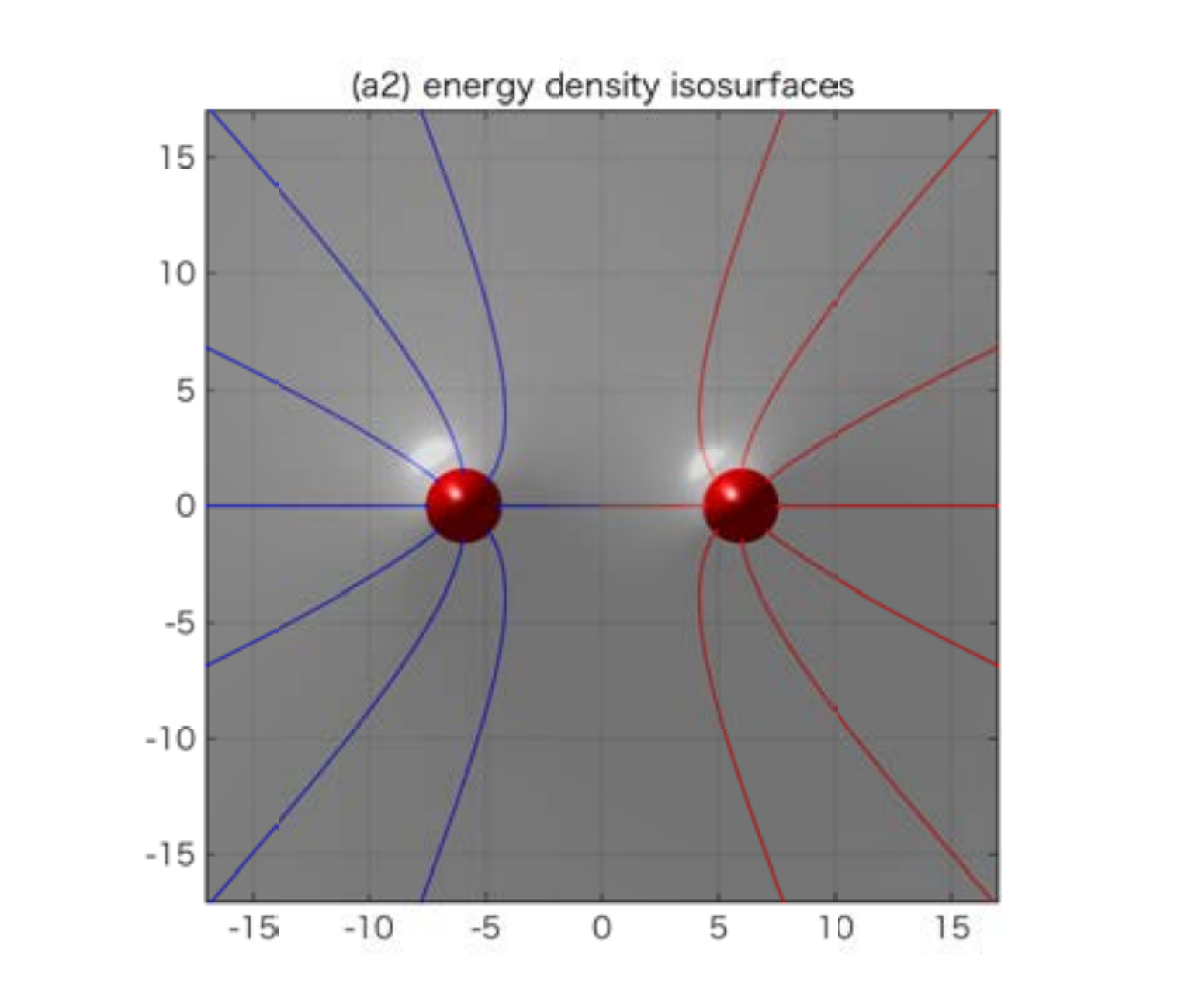}
  \end{center}
 \end{minipage}\\
 \begin{minipage}{0.5\hsize}
  \begin{center}
   \includegraphics[width=7cm]{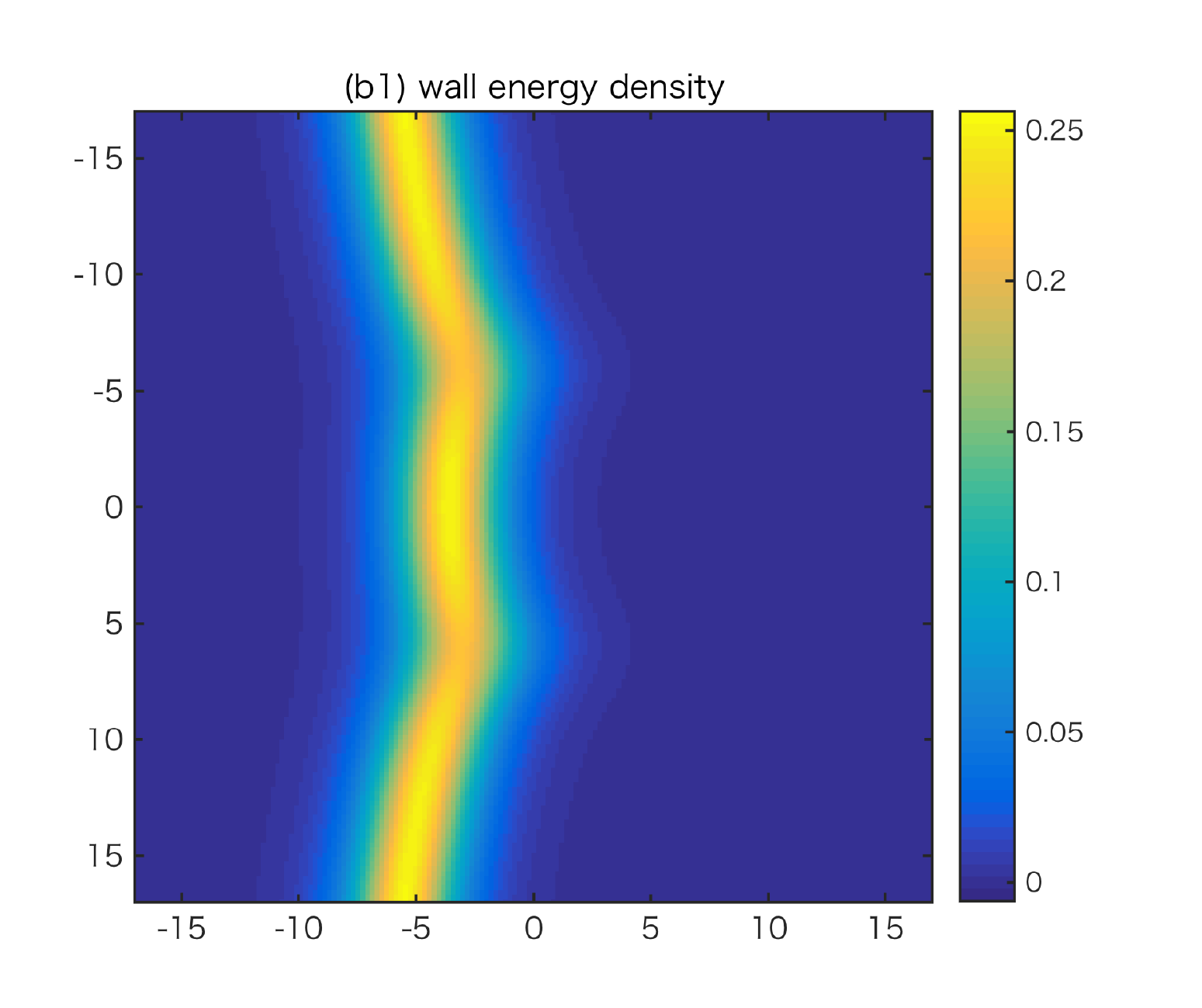}
  \end{center}
 \end{minipage}
 \begin{minipage}{0.5\hsize}
  \begin{center}
   \includegraphics[width=7cm]{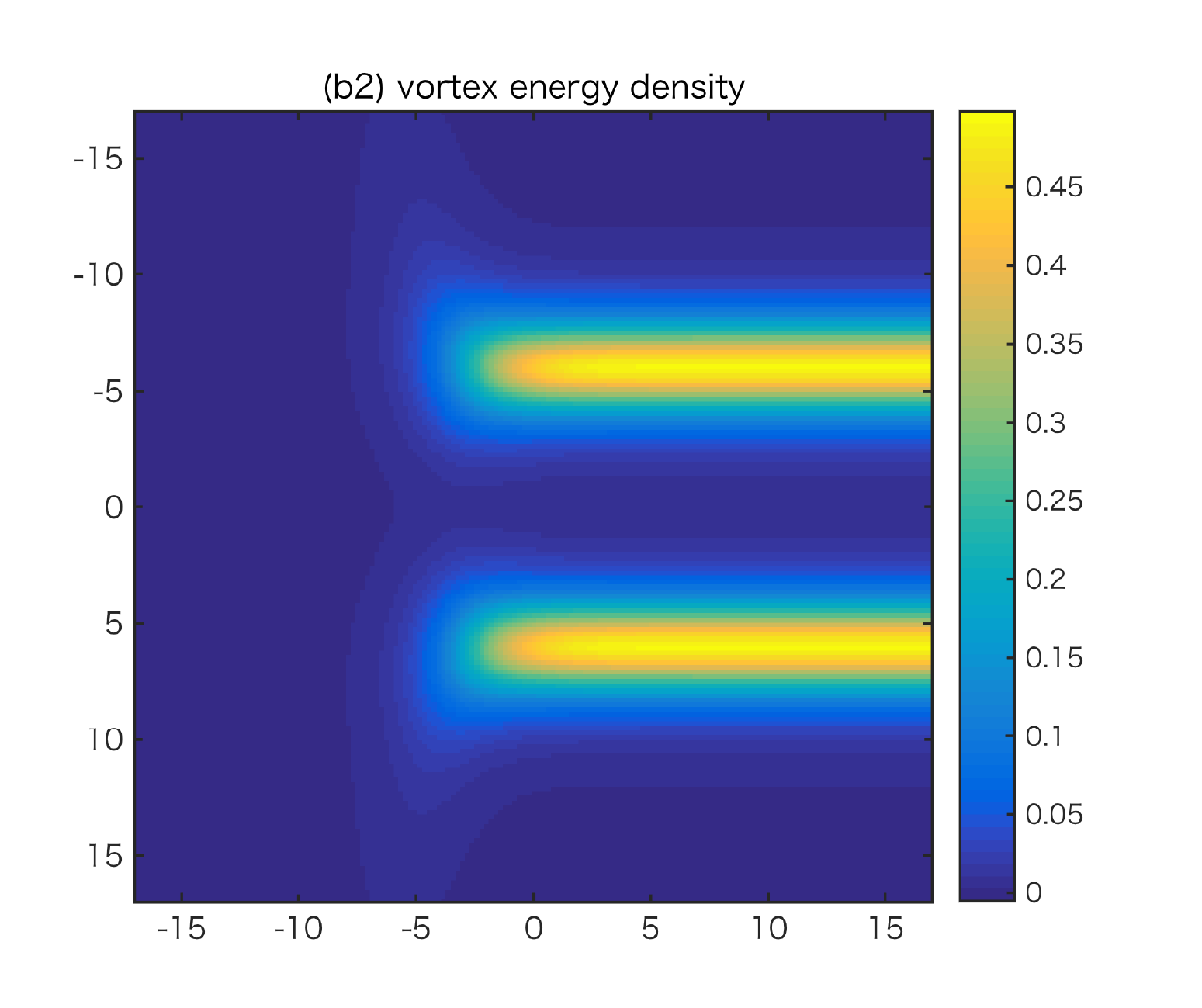}
  \end{center}
 \end{minipage}\\
  \begin{minipage}{0.5\hsize}
  \begin{center}
   \includegraphics[width=7cm]{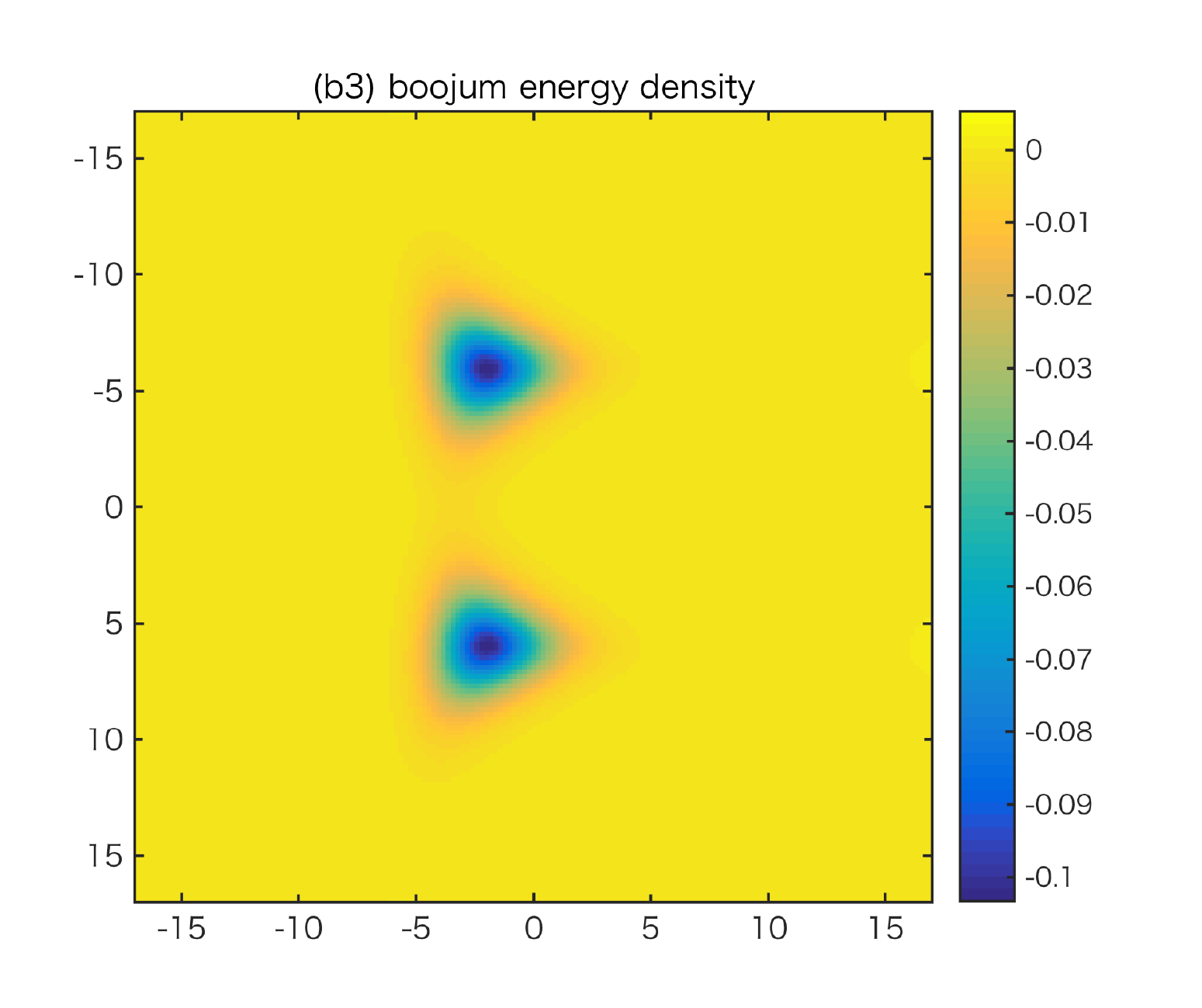}
  \end{center}
 \end{minipage}
 \begin{minipage}{0.5\hsize}
  \begin{center}
   \includegraphics[width=7cm]{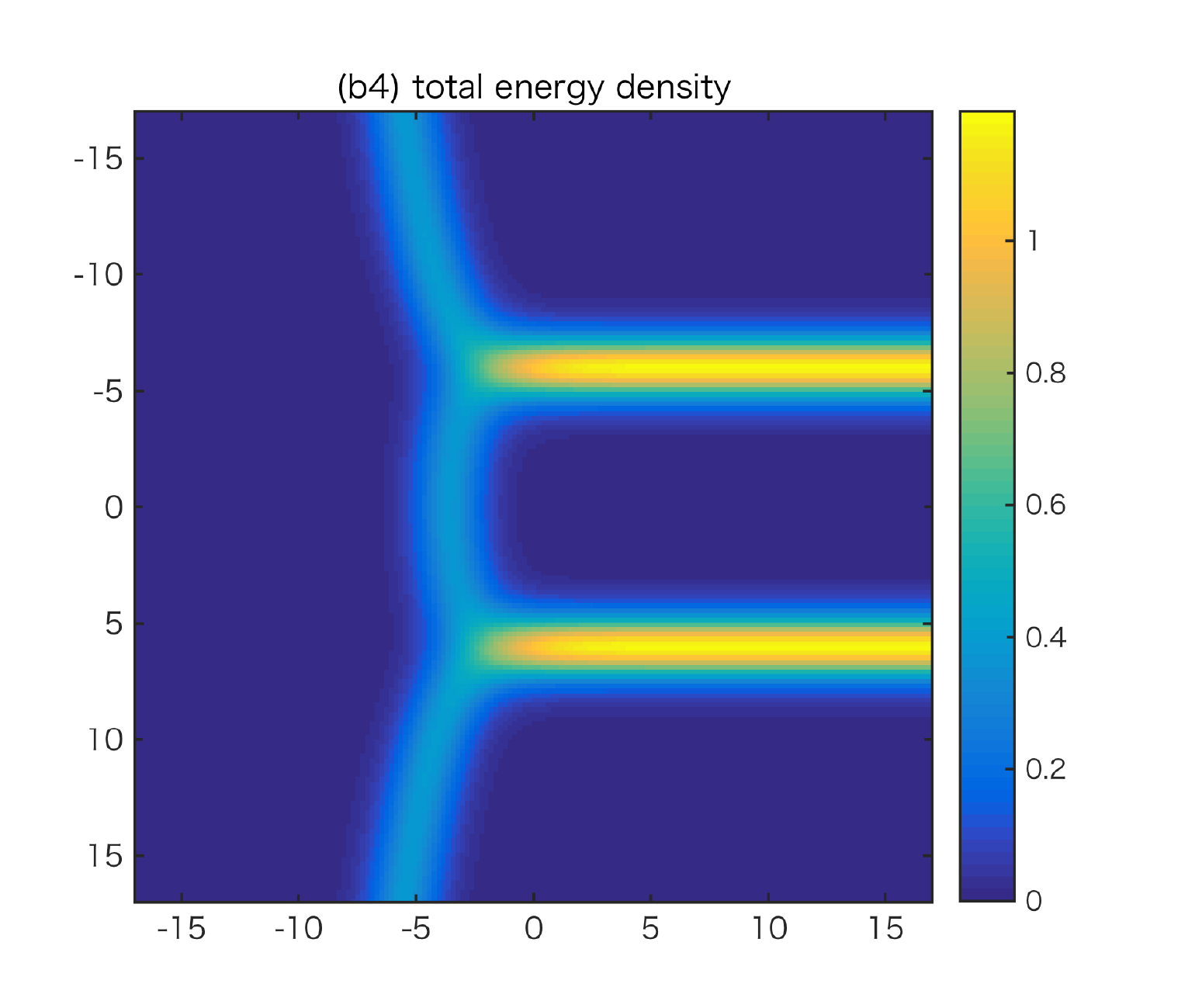}
  \end{center}
 \end{minipage}
 \caption{The plots show the energy density isosurfaces of two vortices ending on one wall (a1, a2), where the blue and the red curves show magnetic flux, the wall energy density (b1), the vortex energy density (b2), the boojum energy density (b3) and the total energy density (b4) with the distance between two vortices $L=6$. The mass is set to be $\tilde m = 1$.}
\label{fig:2vor_wall_1side_A}
\end{figure}

Two vortex strings in the first vacuum $\left<1\right>$ ending on the domain wall can be generated by
\begin{eqnarray}
P_{n_1=2}(z) = (z-L)(z+L),\quad P_{n_2=0}(z) = 1.
\label{eq:mm_two_string_one_wall}
\end{eqnarray}
The vortex strings are asymptotically parallel and located at $z=\pm L$.
In Fig.~\ref{fig:2vor_wall_1side_A} we show a numerical solution with $L=6$ for $\tilde m=1$. The gray surfaces in the panels (a1) and (a2) on Fig.~\ref{fig:2vor_wall_1side_A} are the energy density isosurfaces on which $\tilde {\cal T}_W + \tilde {\cal T}_S + \tilde {\cal T}_B$
takes one half of its maximum.

\clearpage

\clearpage

\begin{figure}[h]
 \begin{minipage}{0.5\hsize}
  \begin{center}
   \includegraphics[width=8cm]{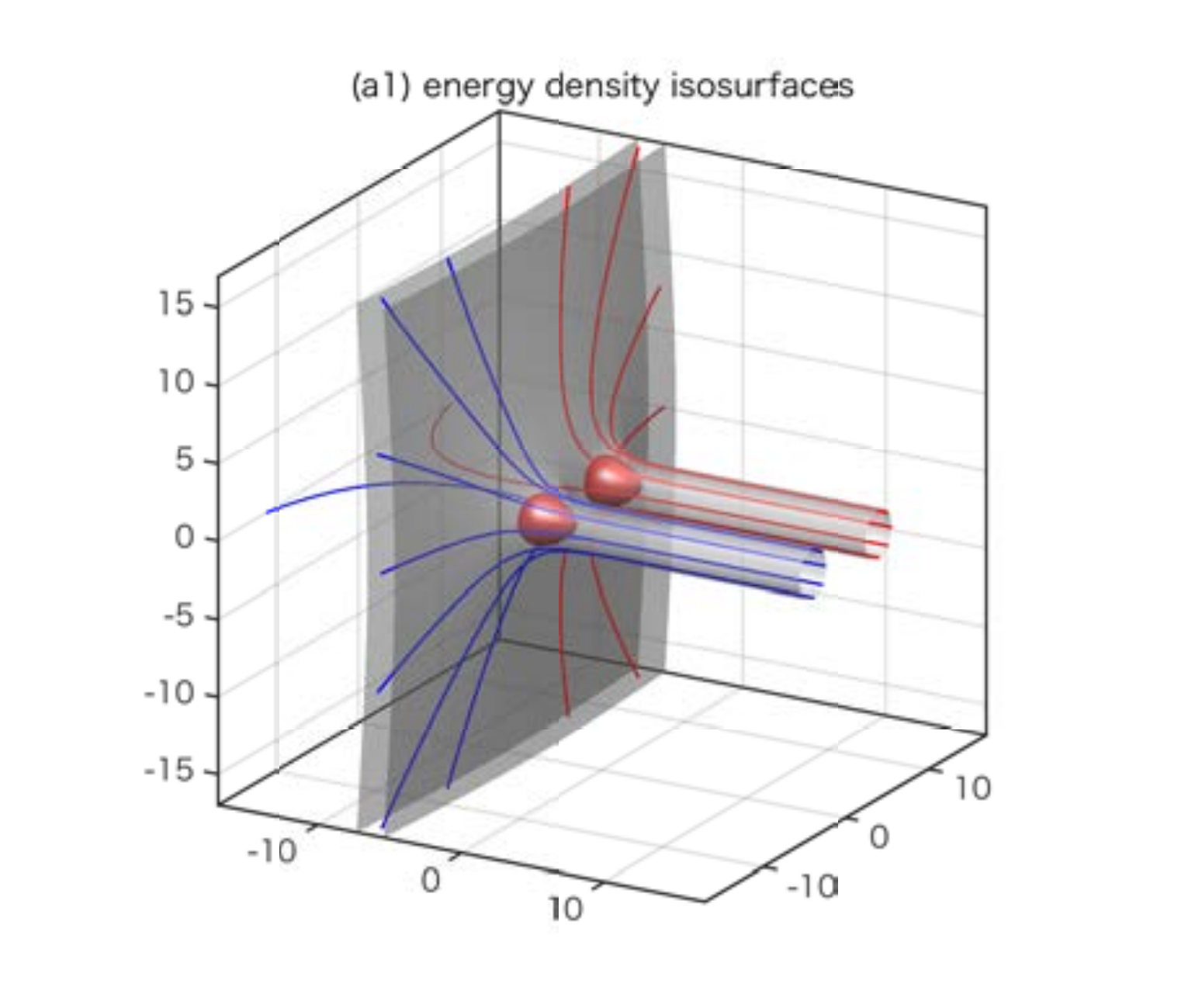}
  \end{center}
 \end{minipage}
  \begin{minipage}{0.5\hsize}
  \begin{center}
   \includegraphics[width=8cm]{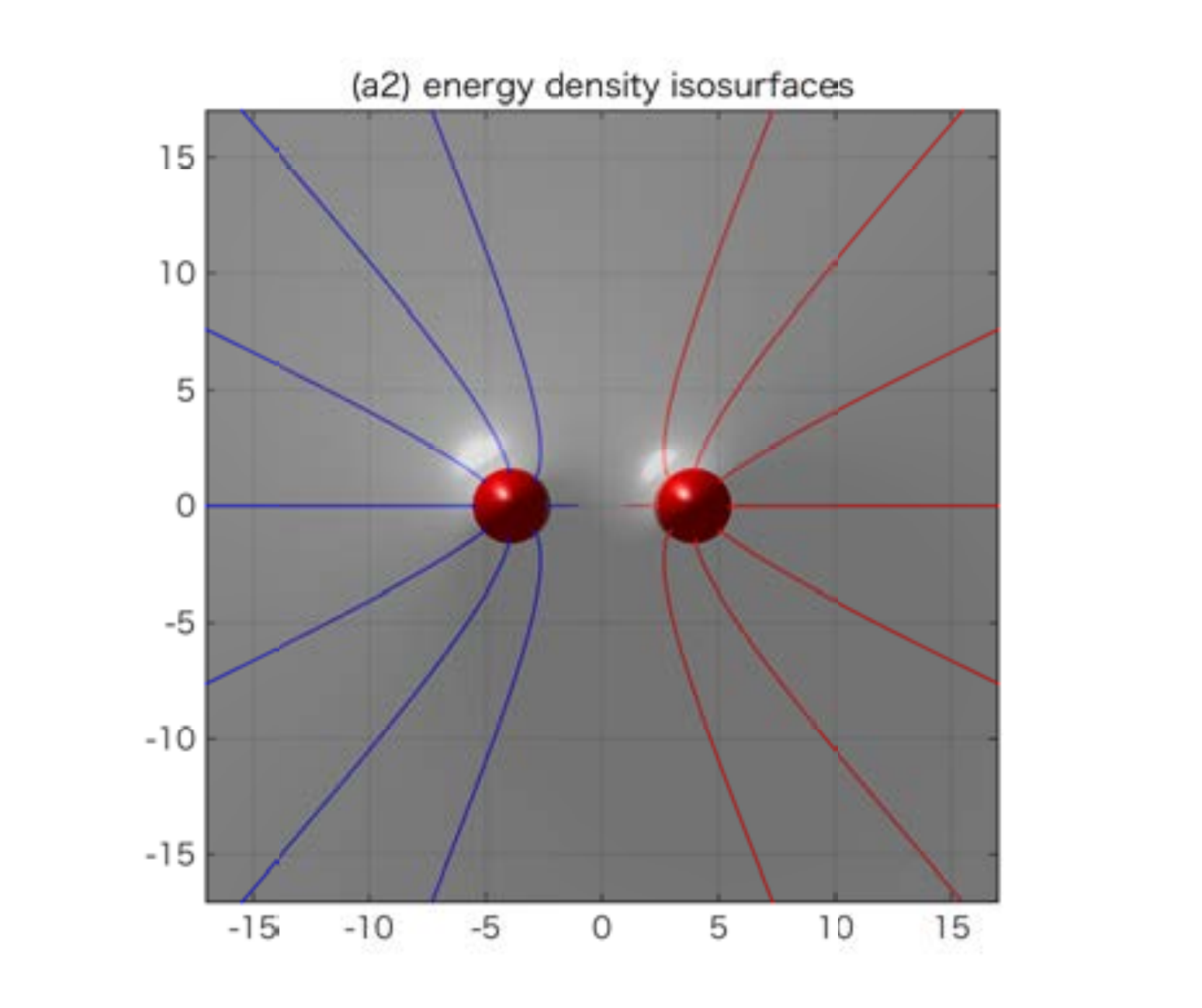}
  \end{center}
 \end{minipage}\\
 \begin{minipage}{0.5\hsize}
  \begin{center}
   \includegraphics[width=8cm]{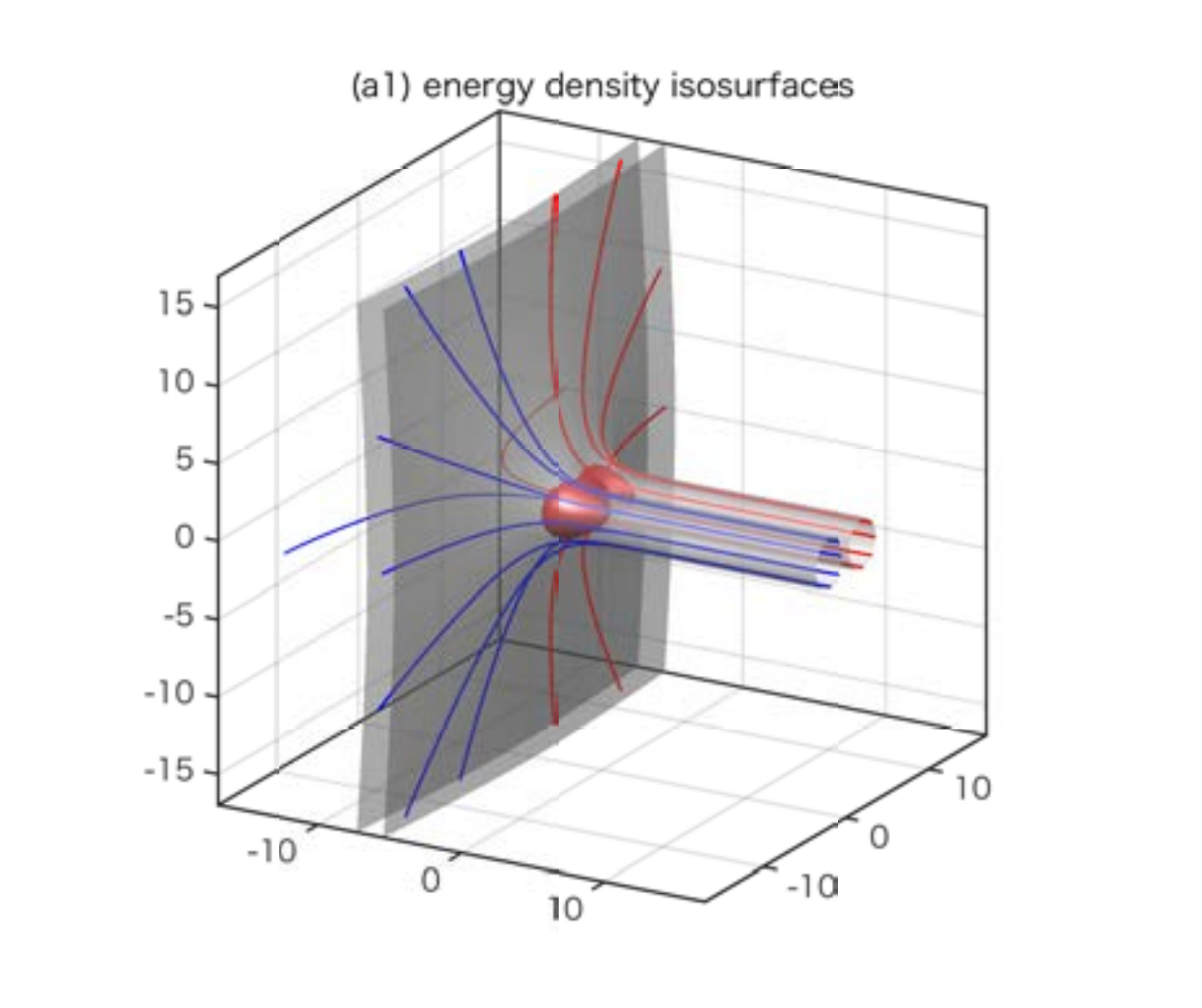}
  \end{center}
 \end{minipage}
 \begin{minipage}{0.5\hsize}
  \begin{center}
   \includegraphics[width=8cm]{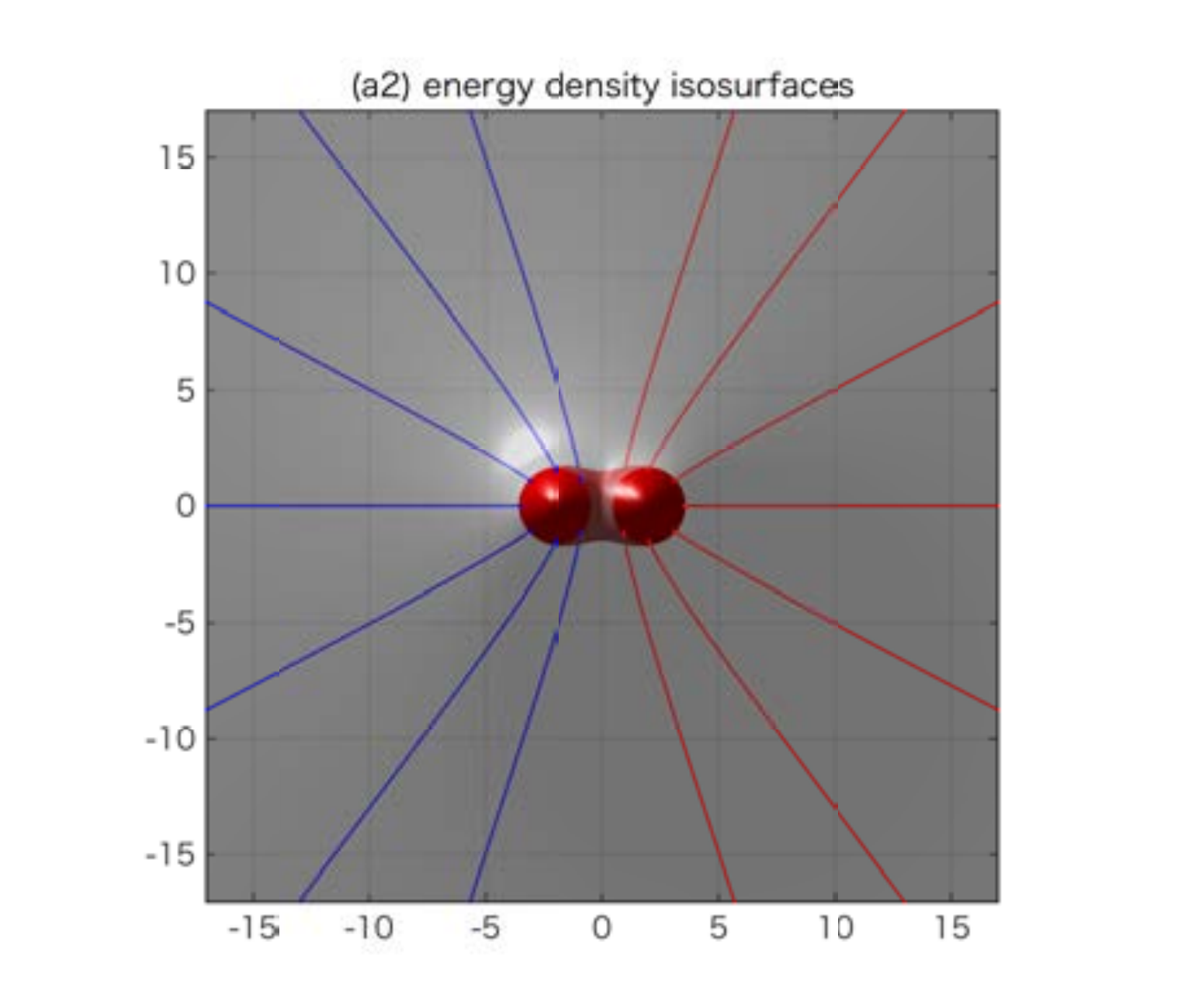}
  \end{center}
 \end{minipage}\\
  \begin{minipage}{0.5\hsize}
  \begin{center}
   \includegraphics[width=8cm]{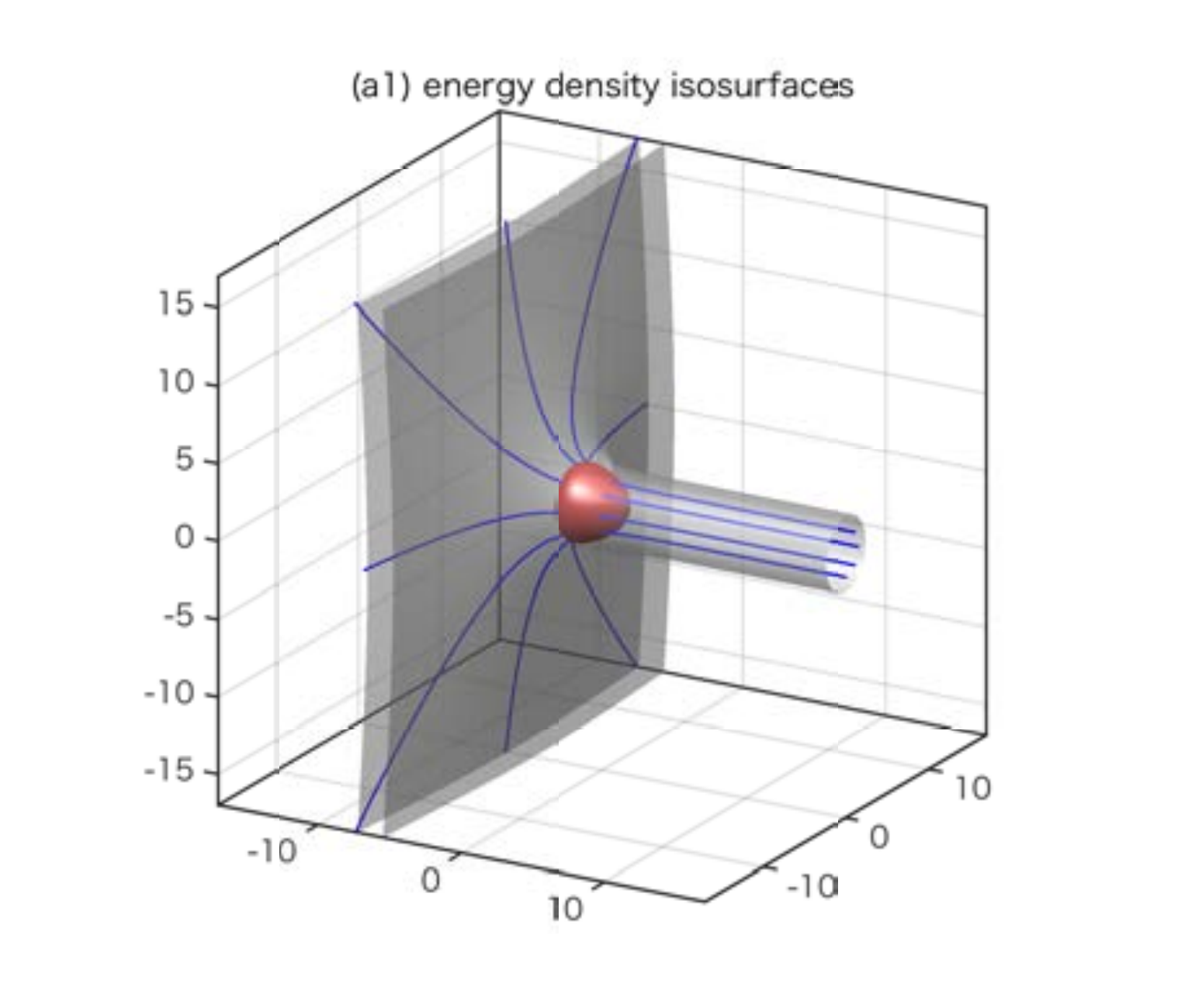}
  \end{center}
 \end{minipage}
 \begin{minipage}{0.5\hsize}
  \begin{center}
   \includegraphics[width=8cm]{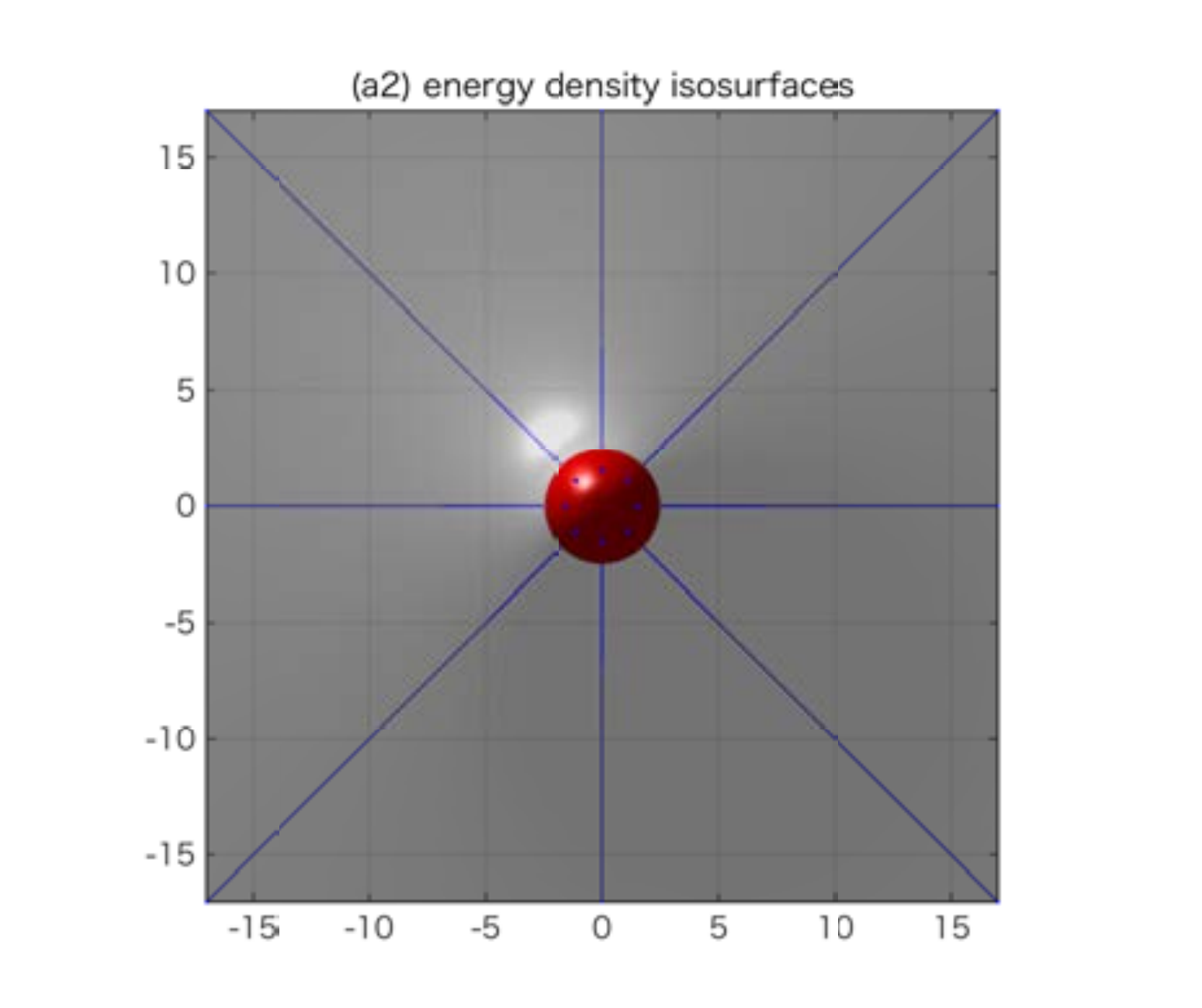}
  \end{center}
 \end{minipage}
\caption{The plots show the energy density isosurfaces of two vortices ending on one wall. The distance between two vortices is taken to be $L=4, 2, 0$ from top to bottom.}
\label{fig:2vor_wall_1side_B}
\end{figure}

\clearpage
\noindent
The red lumps in the panels (a1) and (a2) of Fig.~\ref{fig:2vor_wall_1side_A}
show the boojum density isosurface on which $\tilde {\cal T}_B$ takes one-half
of its minimum (a negative value) and the red and blue curves correspond to the magnetic force lines going through
the vortex strings. In the panels (b1), (b2), and (b3) we show the topological charge densities $\tilde{\cal T}_W$,
$\tilde{\cal T}_S$, and $\tilde{\cal T}_B$ on the cross section which passes the centers of vortex strings, respectively.
The panel (b4) of Fig.~\ref{fig:2vor_wall_1side_A} depicts the total energy density $\tilde {\cal T}$ including the 
surface terms on the same cross section.

One can easily change the distance of two vortex strings by varying $L$ in Eq.~(\ref{eq:mm_two_string_one_wall}).
The numerical solutions for $L=4,2,0$ are shown in Fig.~\ref{fig:2vor_wall_1side_B}.
It is clearly seen that the two individual boojums coalesce into one large boojum as the distance between 
two vortex strings becomes small.

\begin{figure}[h]
 \begin{minipage}{0.5\hsize}
  \begin{center}
   \includegraphics[width=7cm]{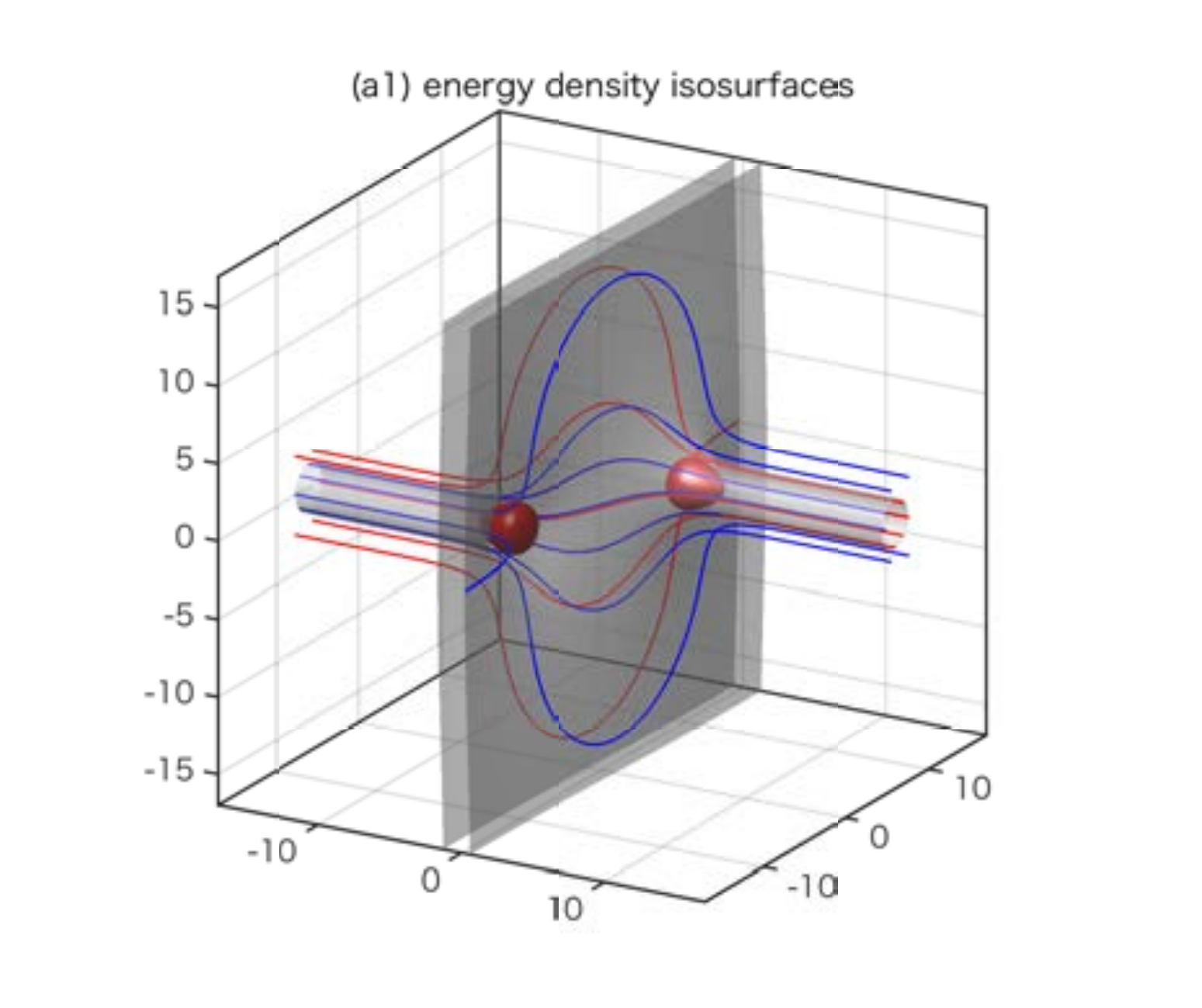}
  \end{center}
 \end{minipage}
  \begin{minipage}{0.5\hsize}
  \begin{center}
   \includegraphics[width=7cm]{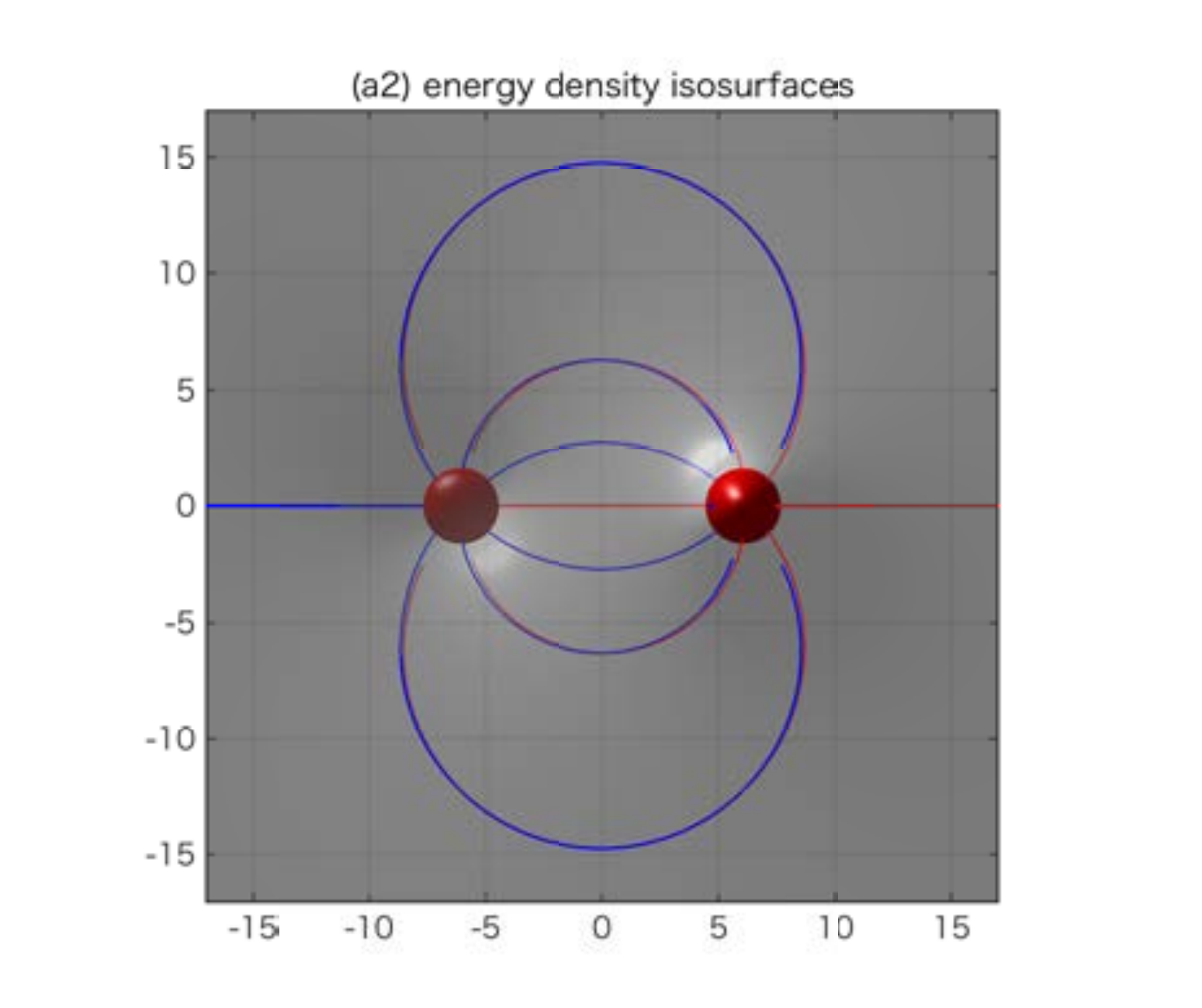}
  \end{center}
 \end{minipage}\\
 \begin{minipage}{0.5\hsize}
  \begin{center}
   \includegraphics[width=7cm]{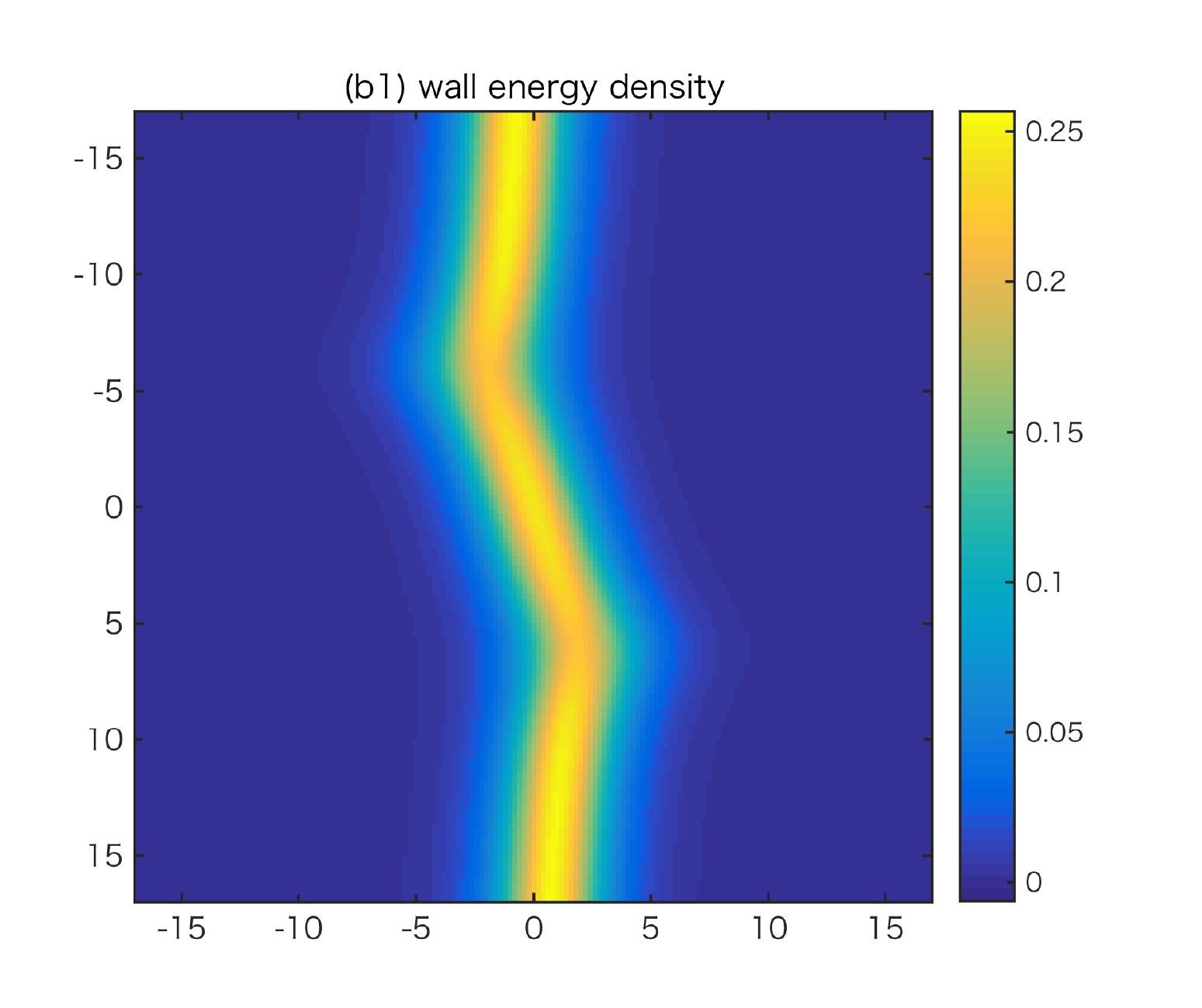}
  \end{center}
 \end{minipage}
 \begin{minipage}{0.5\hsize}
  \begin{center}
   \includegraphics[width=7cm]{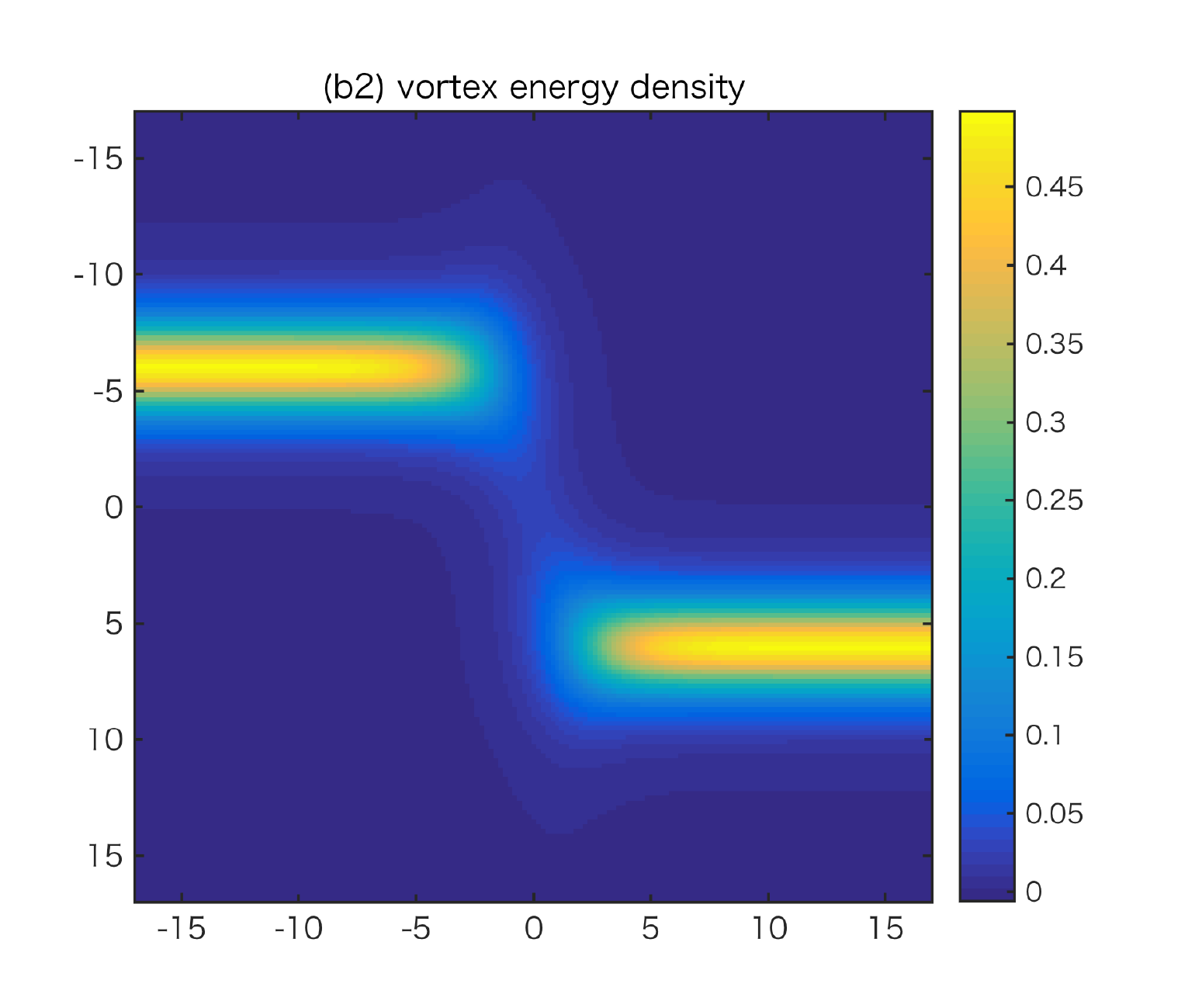}
  \end{center}
 \end{minipage}\\
  \begin{minipage}{0.5\hsize}
  \begin{center}
   \includegraphics[width=7cm]{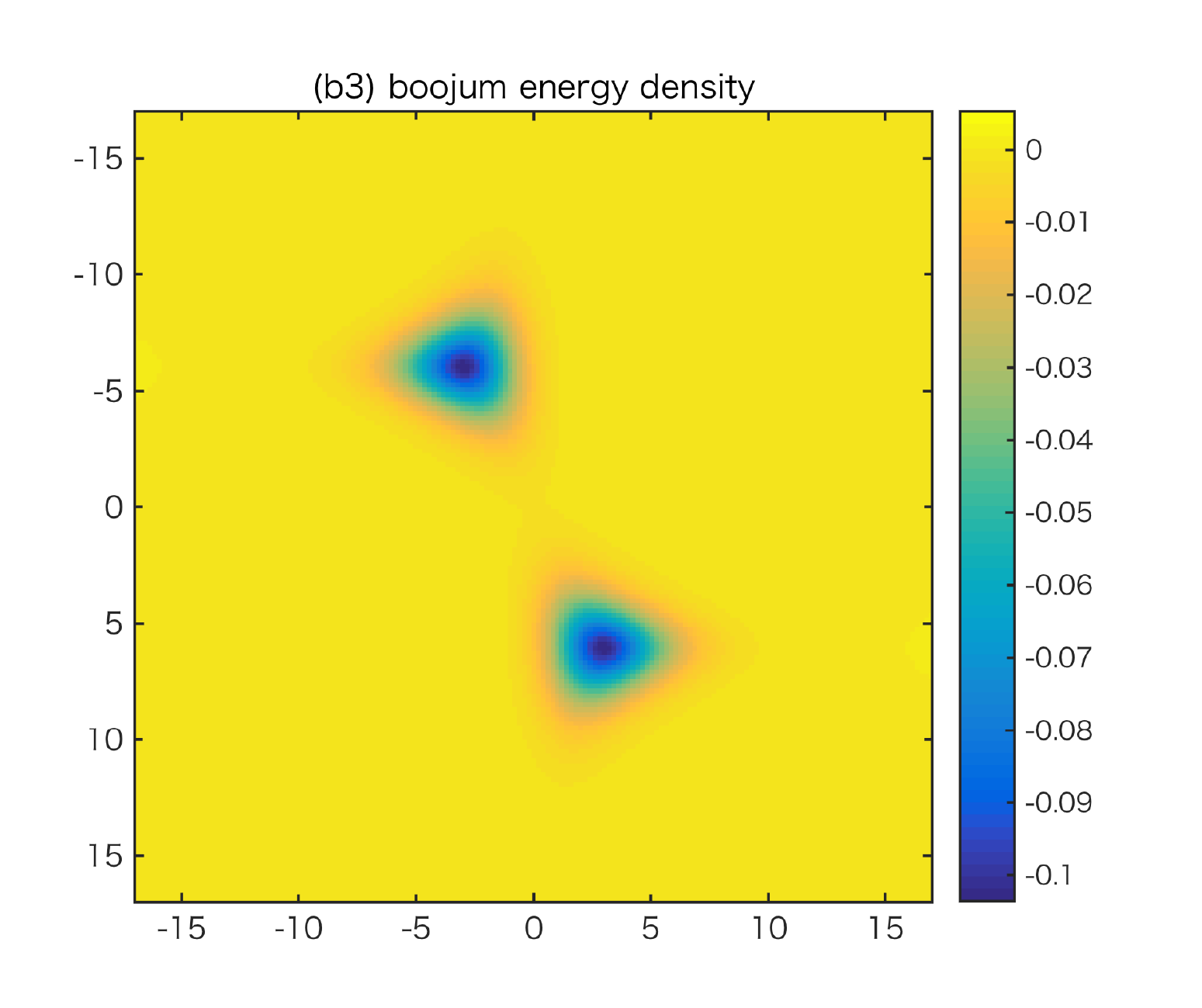}
  \end{center}
 \end{minipage}
 \begin{minipage}{0.5\hsize}
  \begin{center}
   \includegraphics[width=7cm]{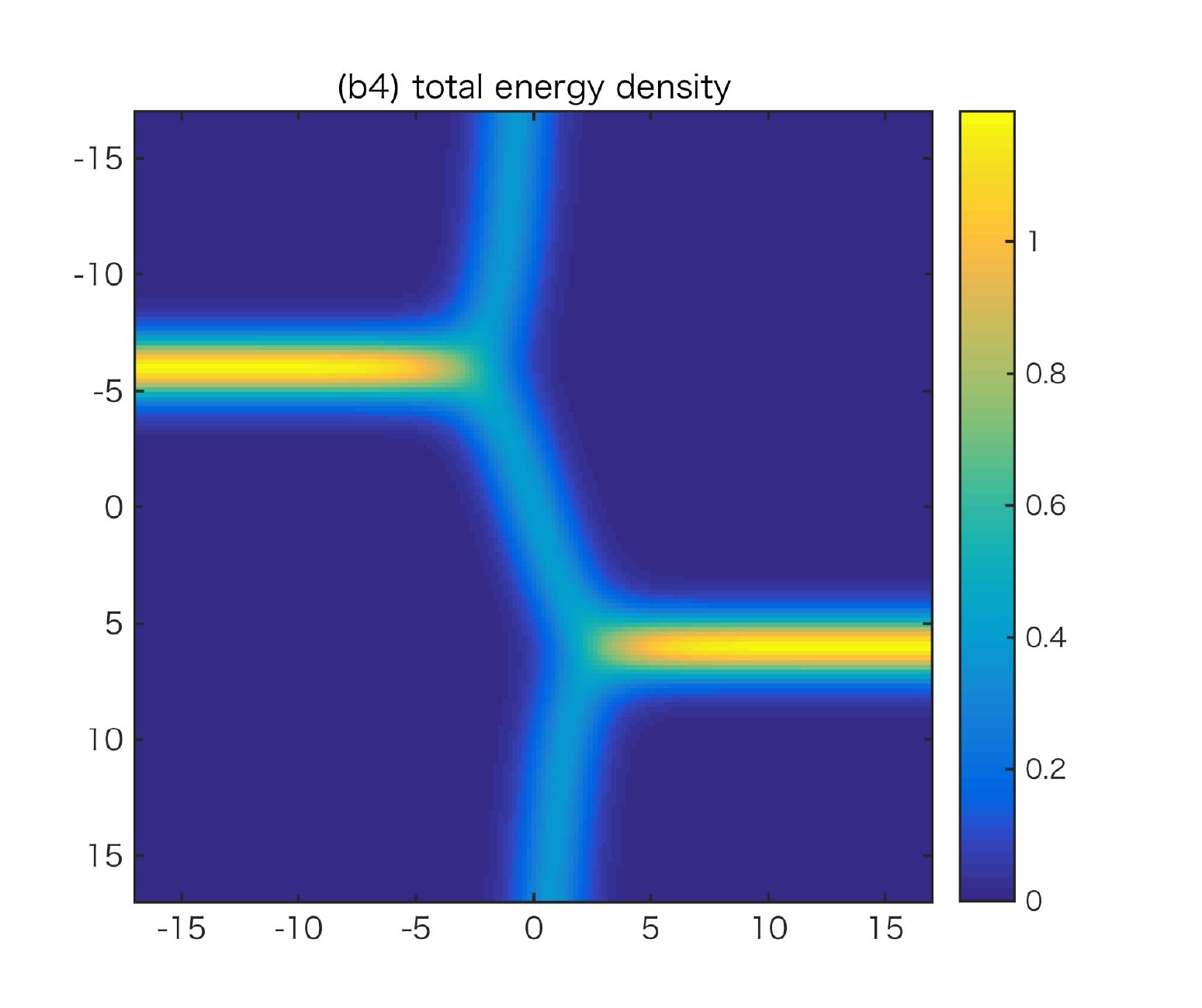}
  \end{center}
 \end{minipage}
  \caption{The plots show the energy density isosurfaces of two vortices ending on one wall from two sides (a1, a2), where the blue and the red curves show magnetic fluxes, the wall energy density (b1), the vortex energy density (b2), the boojum energy density (b3) and the total energy density (b4) with the distance between two vortices $L=6$.}
\label{fig:11_A}
\end{figure}

\begin{figure}[h]
 \begin{minipage}{0.5\hsize}
  \begin{center}
   \includegraphics[width=8cm]{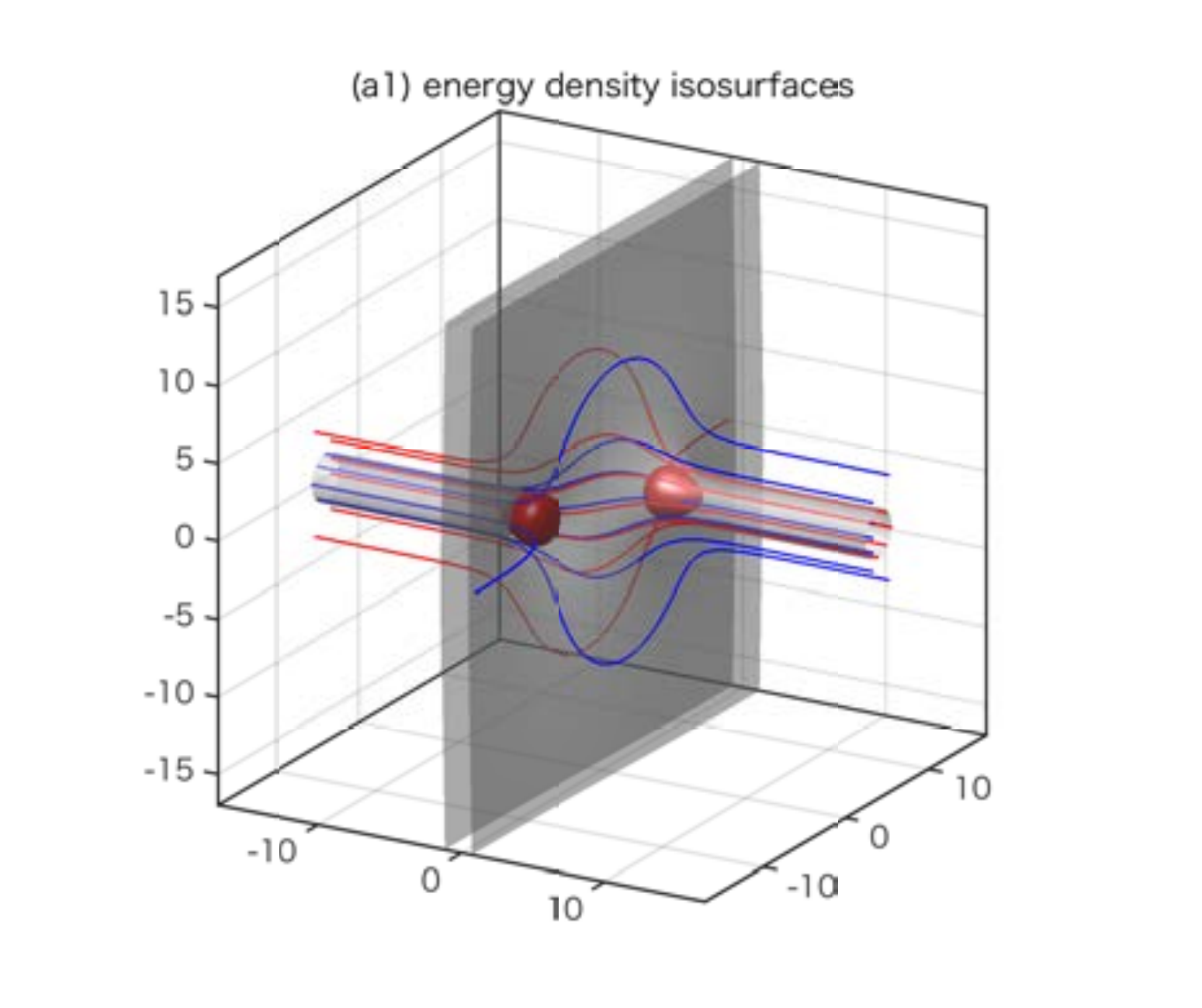}
  \end{center}
 \end{minipage}
  \begin{minipage}{0.5\hsize}
  \begin{center}
   \includegraphics[width=8cm]{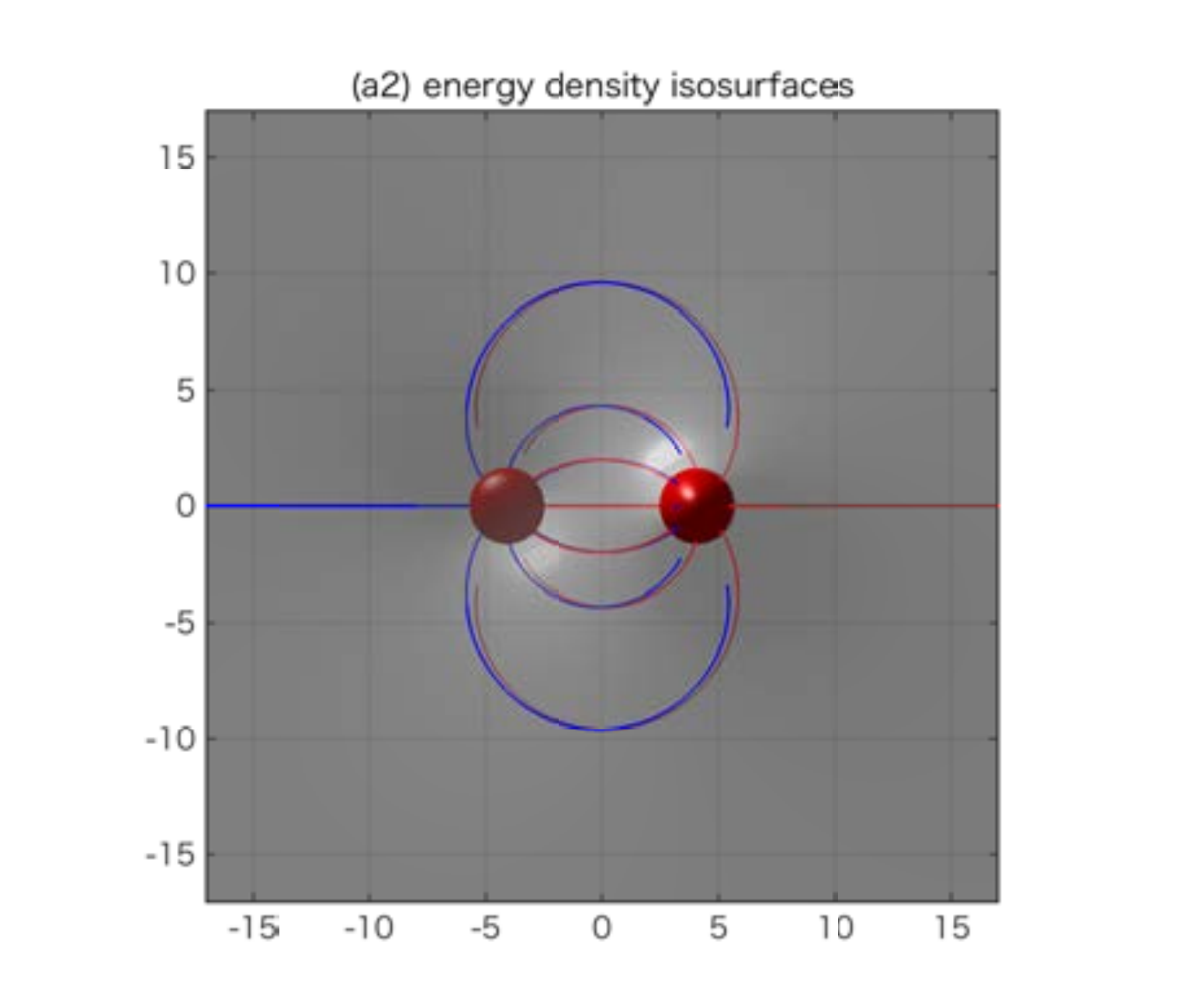}
  \end{center}
 \end{minipage}\\
 \begin{minipage}{0.5\hsize}
  \begin{center}
   \includegraphics[width=8cm]{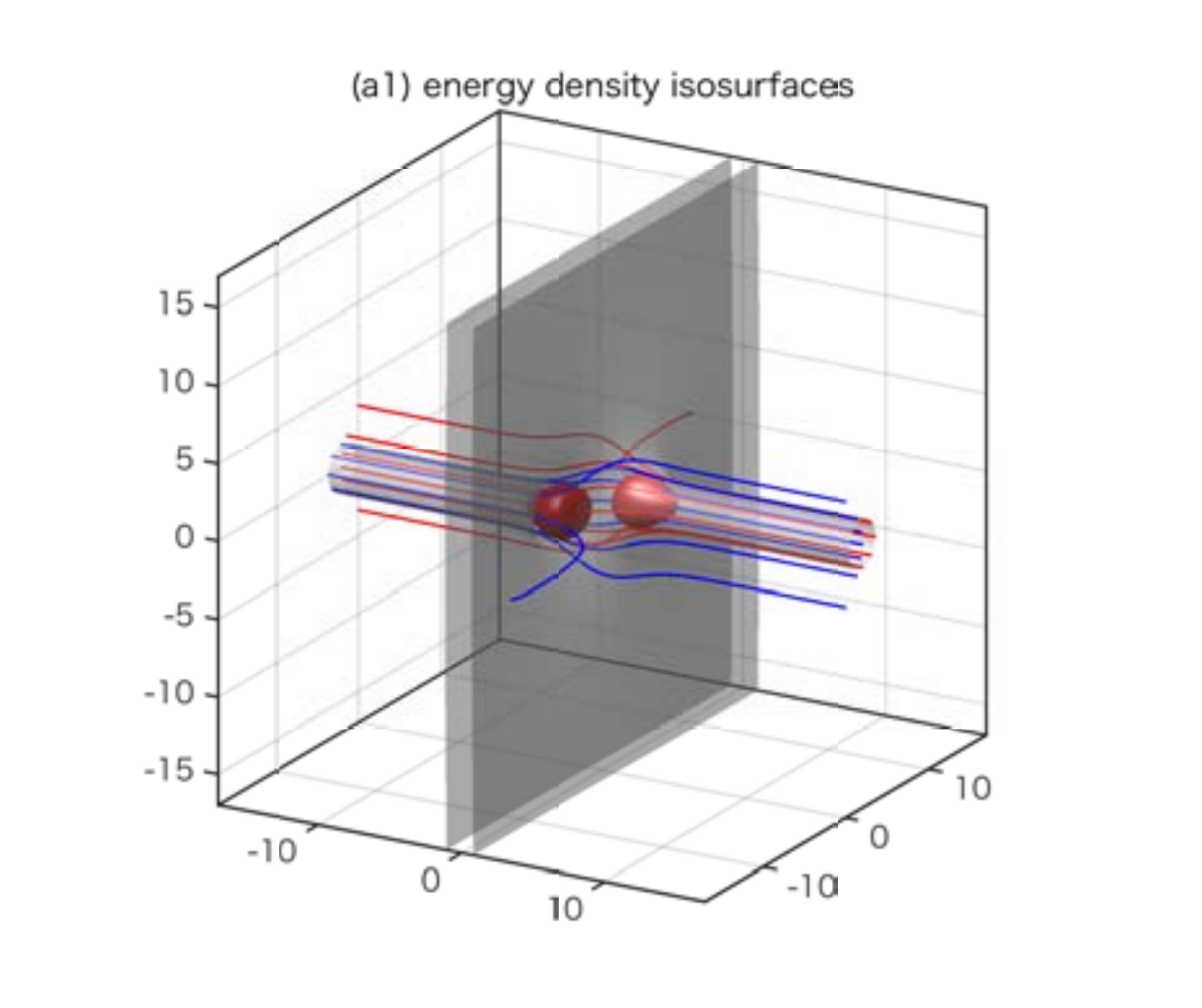}
  \end{center}
 \end{minipage}
 \begin{minipage}{0.5\hsize}
  \begin{center}
   \includegraphics[width=8cm]{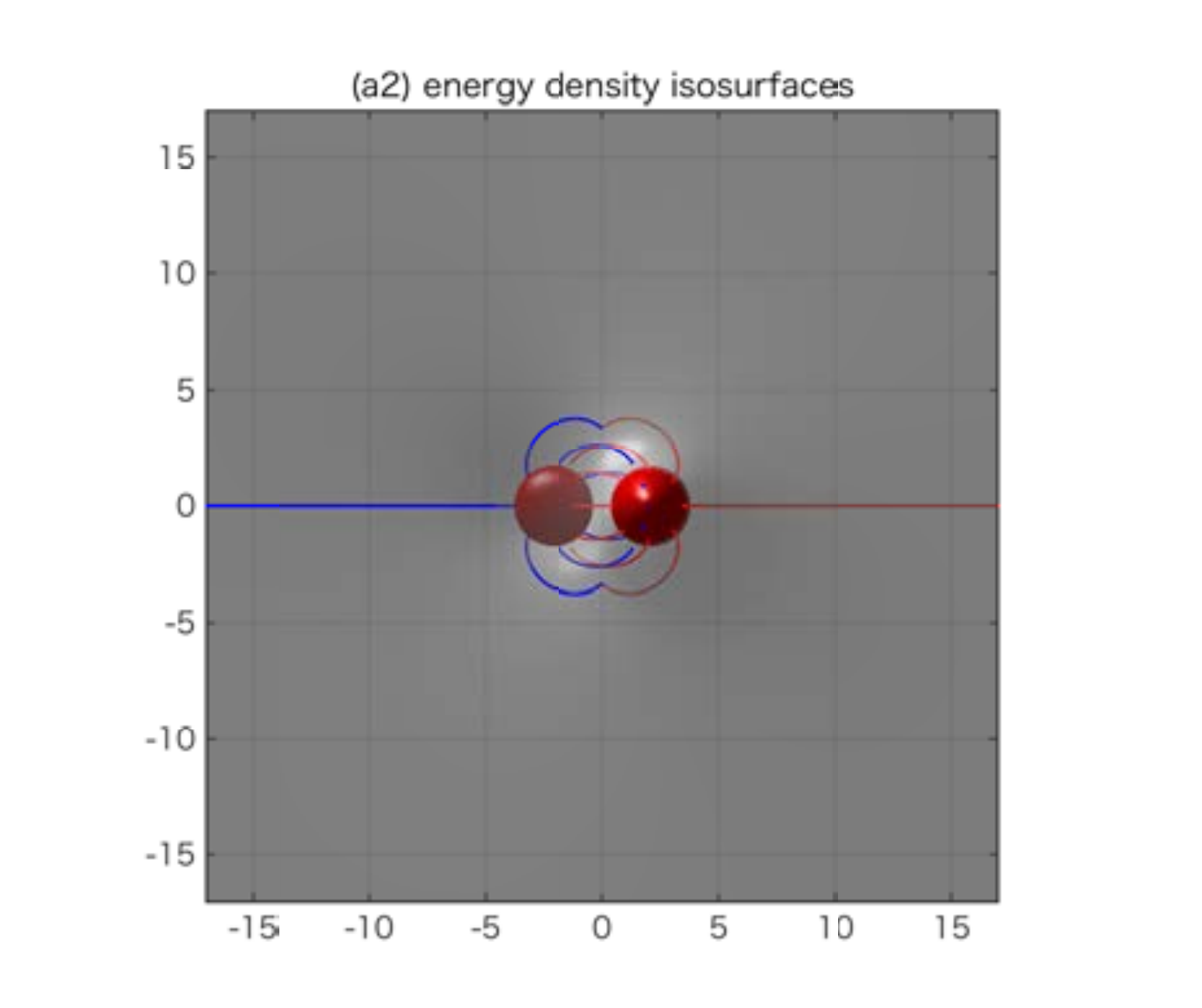}
  \end{center}
 \end{minipage}\\
  \begin{minipage}{0.5\hsize}
  \begin{center}
   \includegraphics[width=8cm]{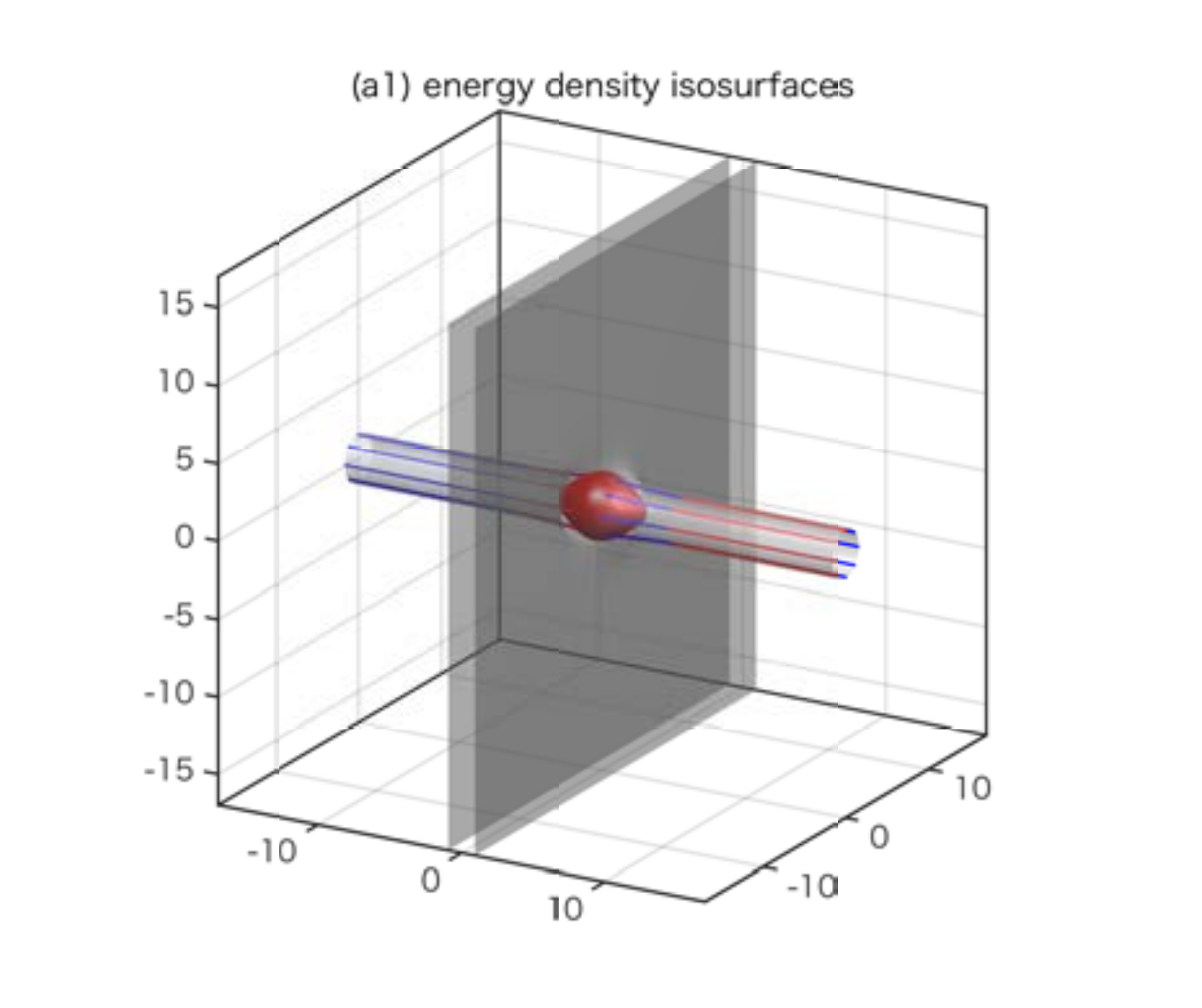}
  \end{center}
 \end{minipage}
 \begin{minipage}{0.5\hsize}
  \begin{center}
   \includegraphics[width=8cm]{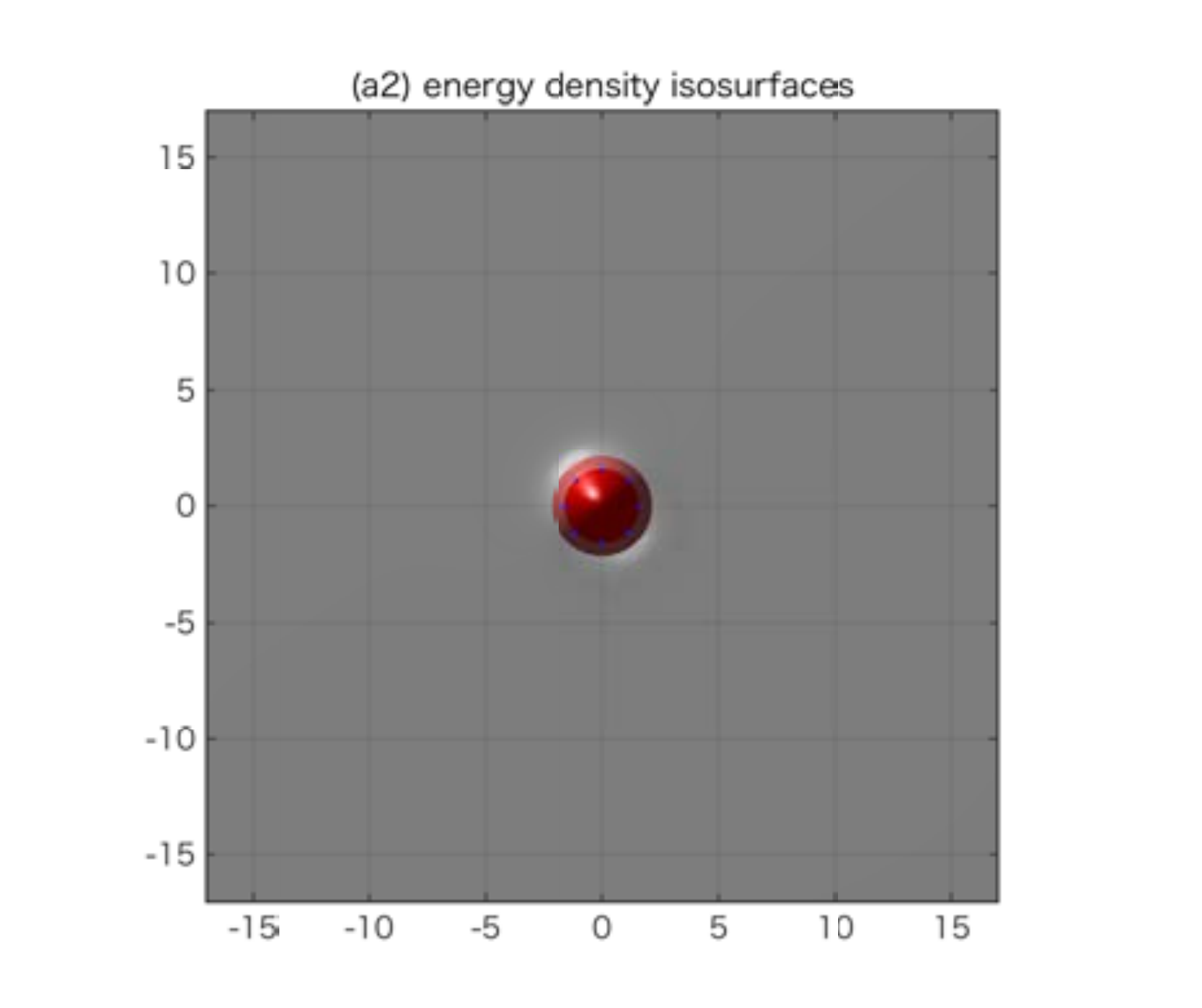}
  \end{center}
 \end{minipage}
 \caption{The plots show the energy density isosurfaces of two vortices ending on one wall from two sides. The distance between two vortices is taken to be $L=4, 2, 0$ from top to bottom.}
\label{fig:11_B}
\end{figure}

Let us next consider the asymptotically flat domain wall on which a vortex strings end from both sides.
The moduli matrix is
\begin{eqnarray}
P_{n_1=1}(z) = z-L,\quad
P_{n_2=1}(z) = z+L.
\end{eqnarray}
The parameters $z= \pm L$ correspond to positions of the string endpoints. 
We show a typical solution with $L=6$ for $\tilde m=1$ case in Fig.~\ref{fig:11_A}.
The six panels in Fig.~\ref{fig:11_A} show the same quantities plotted in the panels
in Fig.~\ref{fig:2vor_wall_1side_A}.
The magnetic force lines incoming from the vortex string on the positive $x^3$ side go into
the other vortex string on the negative $x^3$ side. Seen from the $x^3$ axis, the distribution of the magnetic
force lines is like a magnetic dipole, see the panel (a2) of Fig.~\ref{fig:11_A}.
This is a remarkable contrast to the two vortex strings ending on the domain wall from one side.
The former can be thought of as the dipole with opposite magnetic charges and the latter as the dipole with
the same magnetic charges. As we reduce the separation $L$ of the two strings, the region in which the magnetic
flux expands gets smaller, see Fig.~\ref{fig:11_B} where we plot the numerical solutions for $L=4,2,0$.
Note that when $L=0$ the two vortex strings become collinear and the magnetic charges in the $2+1$ dimensional
sense exactly cancel. Even in the limit $L\to0$, the boojums do not disappear because the domain wall has 
the finite width and the boojums are separate in $x^3$ direction.

\clearpage

\clearpage

\clearpage



\begin{figure}[t]
 \begin{minipage}{0.31\hsize}
  \begin{center}
   \includegraphics[width=6cm]{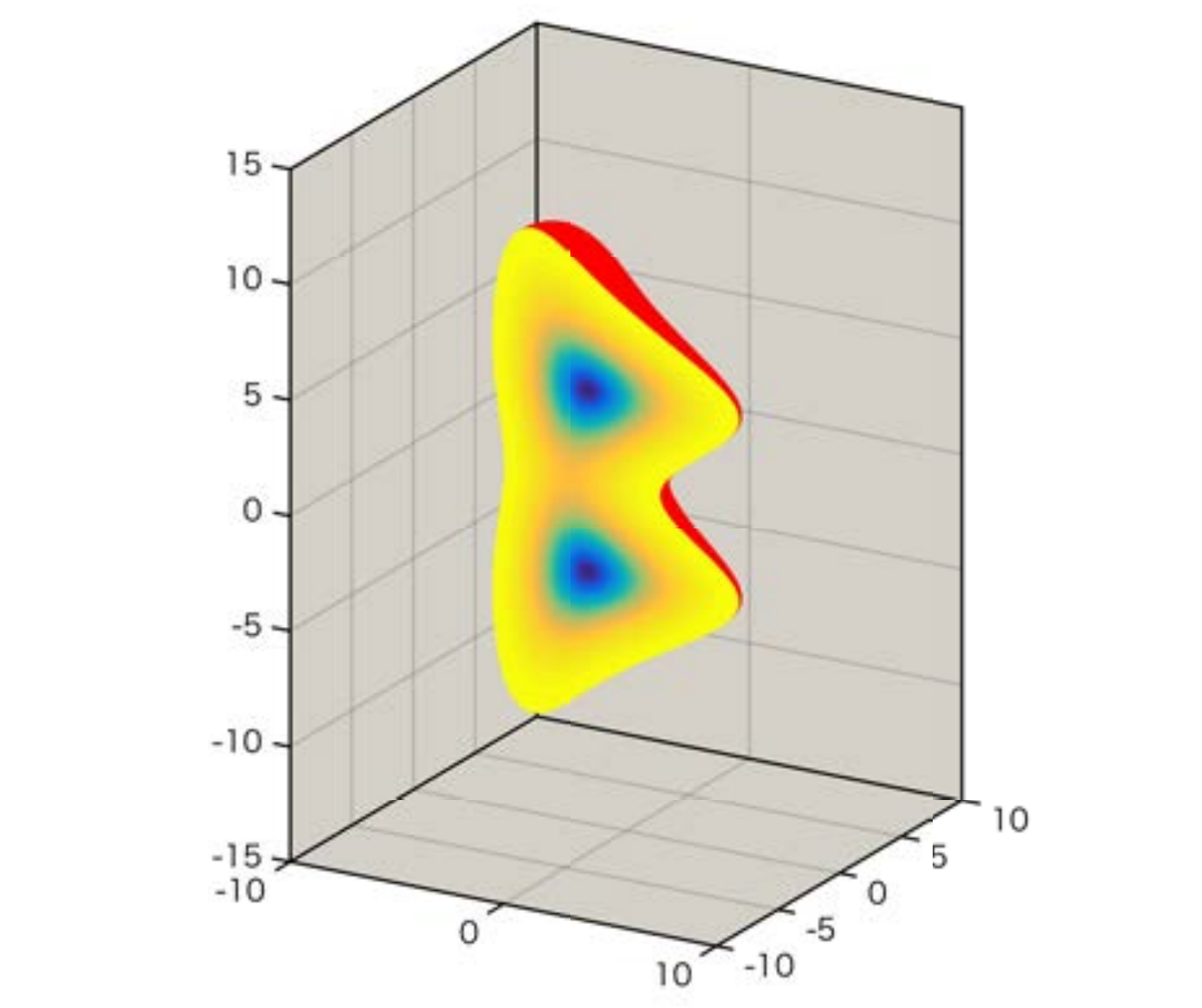}
  \end{center}
 \end{minipage}
  \begin{minipage}{0.31\hsize}
  \begin{center}
   \includegraphics[width=6cm]{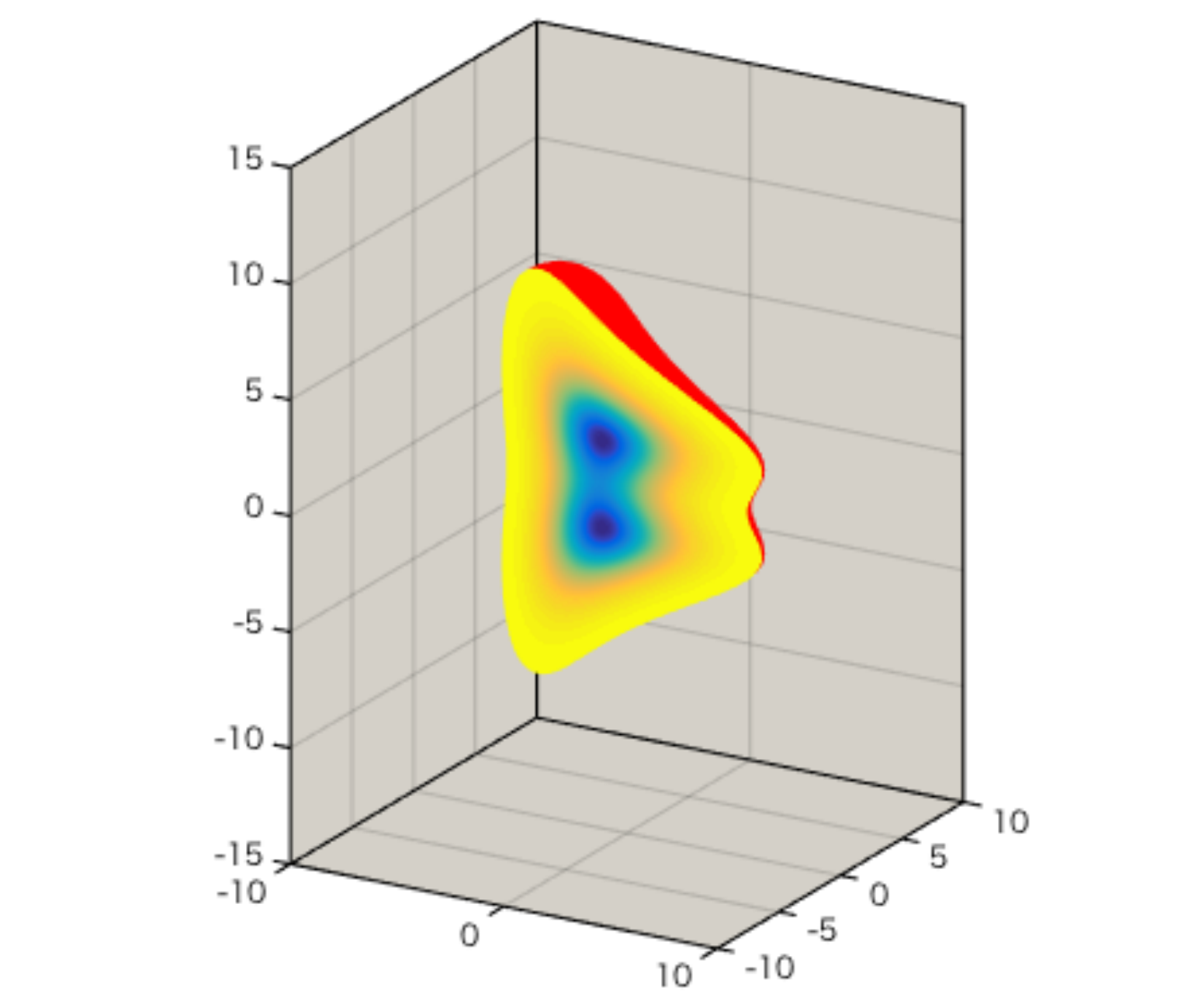}
  \end{center}
 \end{minipage}
 \begin{minipage}{0.31\hsize}
  \begin{center}
   \includegraphics[width=6cm]{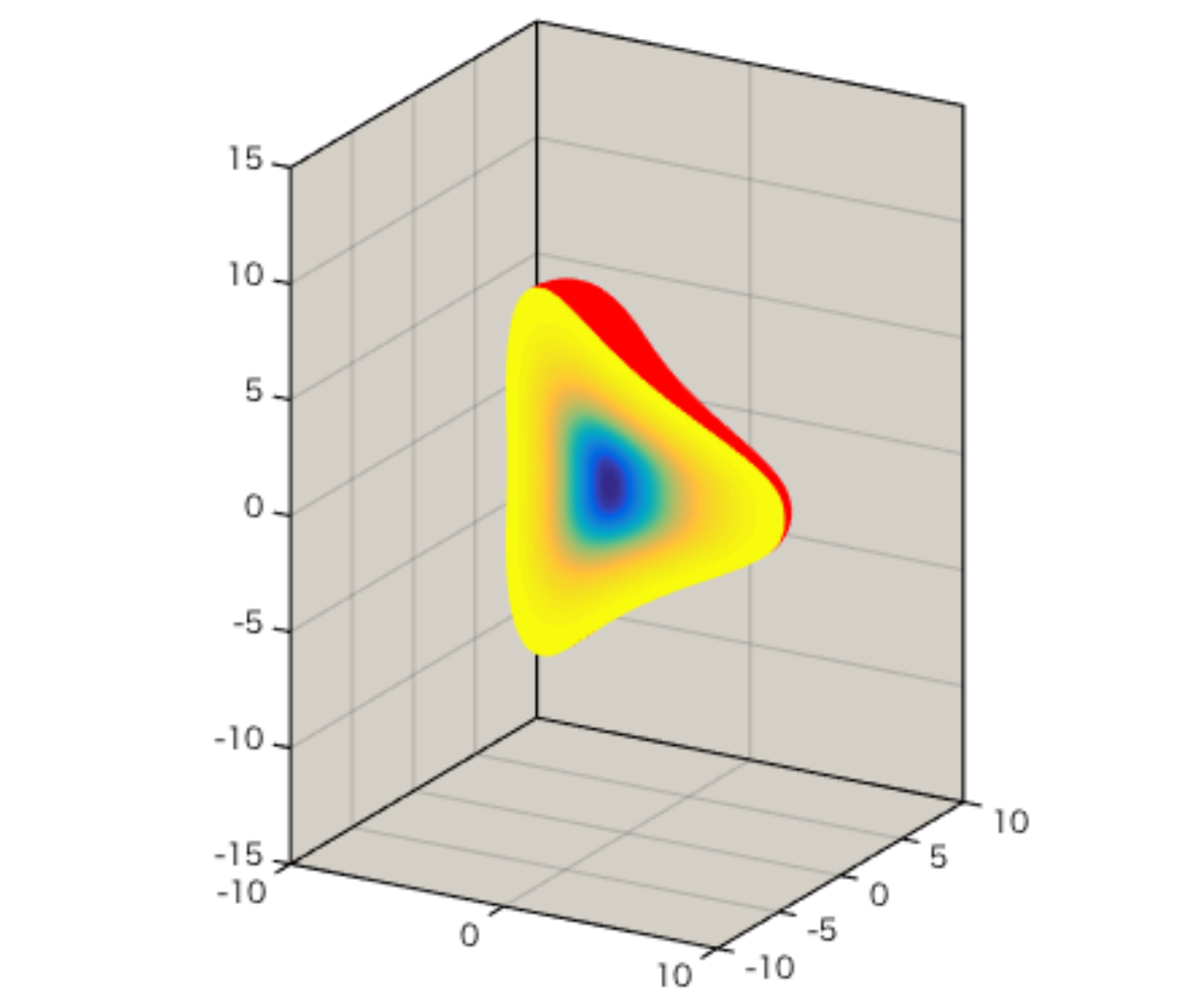}
  \end{center}
 \end{minipage}
\caption{The plots of the boojum energy density for the case that two vortices ending on one wall. The distances of two vortices are $L=4, 2, 0$ from left to right.}
\label{fig:boojum_a}
\end{figure}

\begin{figure}[t]
 \begin{minipage}{0.31\hsize}
  \begin{center}
   \includegraphics[width=6cm]{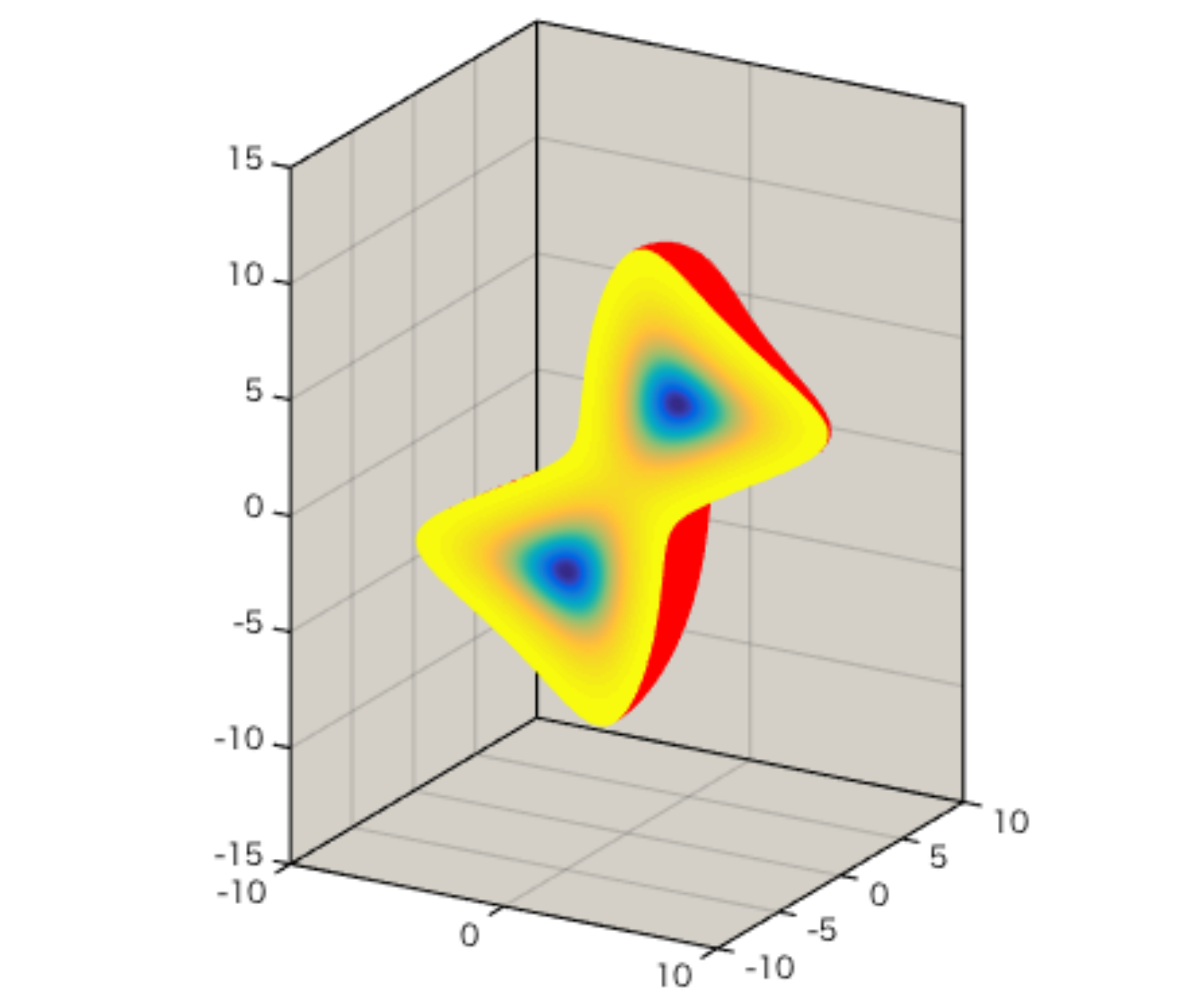}
  \end{center}
 \end{minipage}
  \begin{minipage}{0.31\hsize}
  \begin{center}
   \includegraphics[width=6cm]{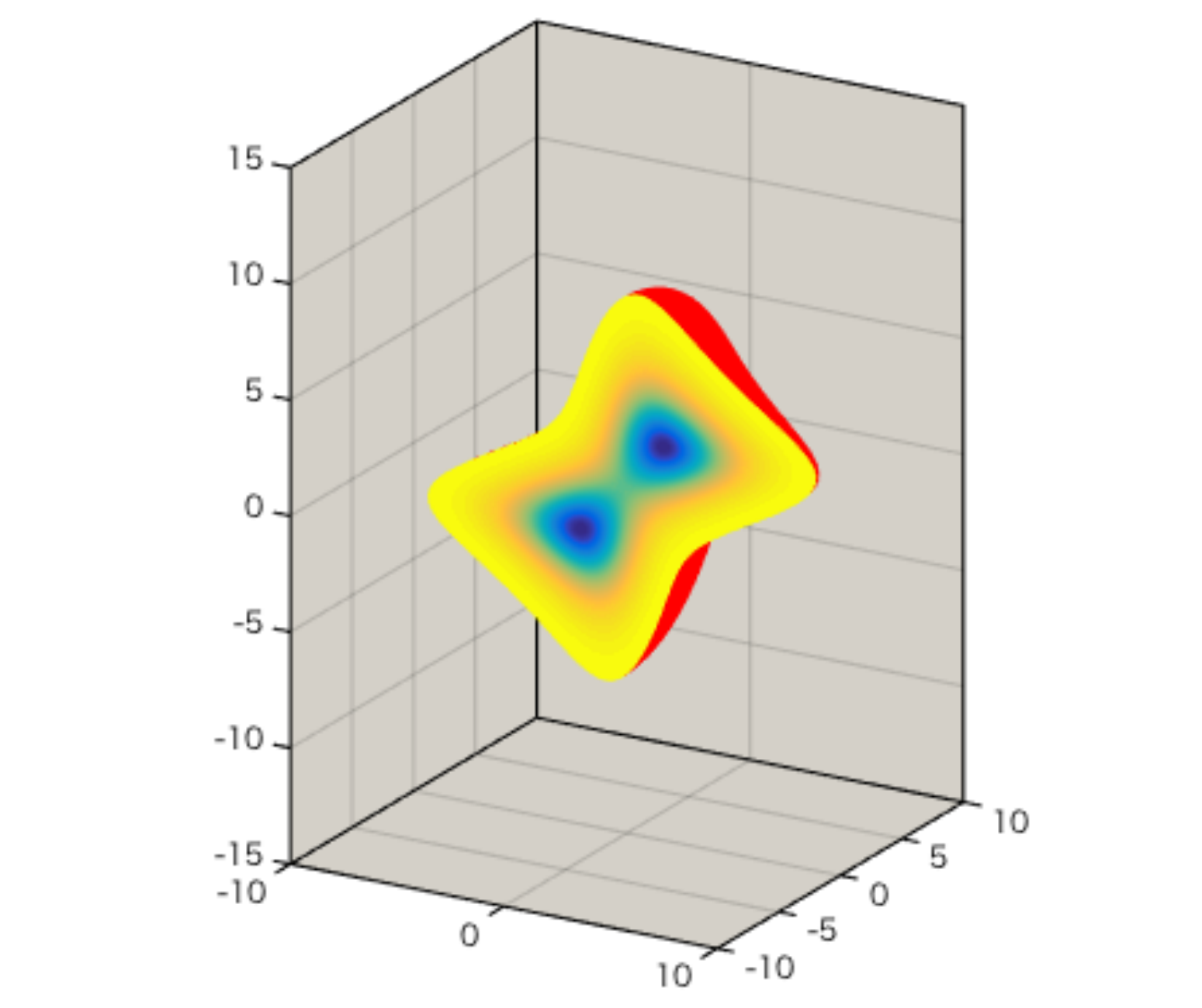}
  \end{center}
 \end{minipage}
 \begin{minipage}{0.31\hsize}
  \begin{center}
   \includegraphics[width=6cm]{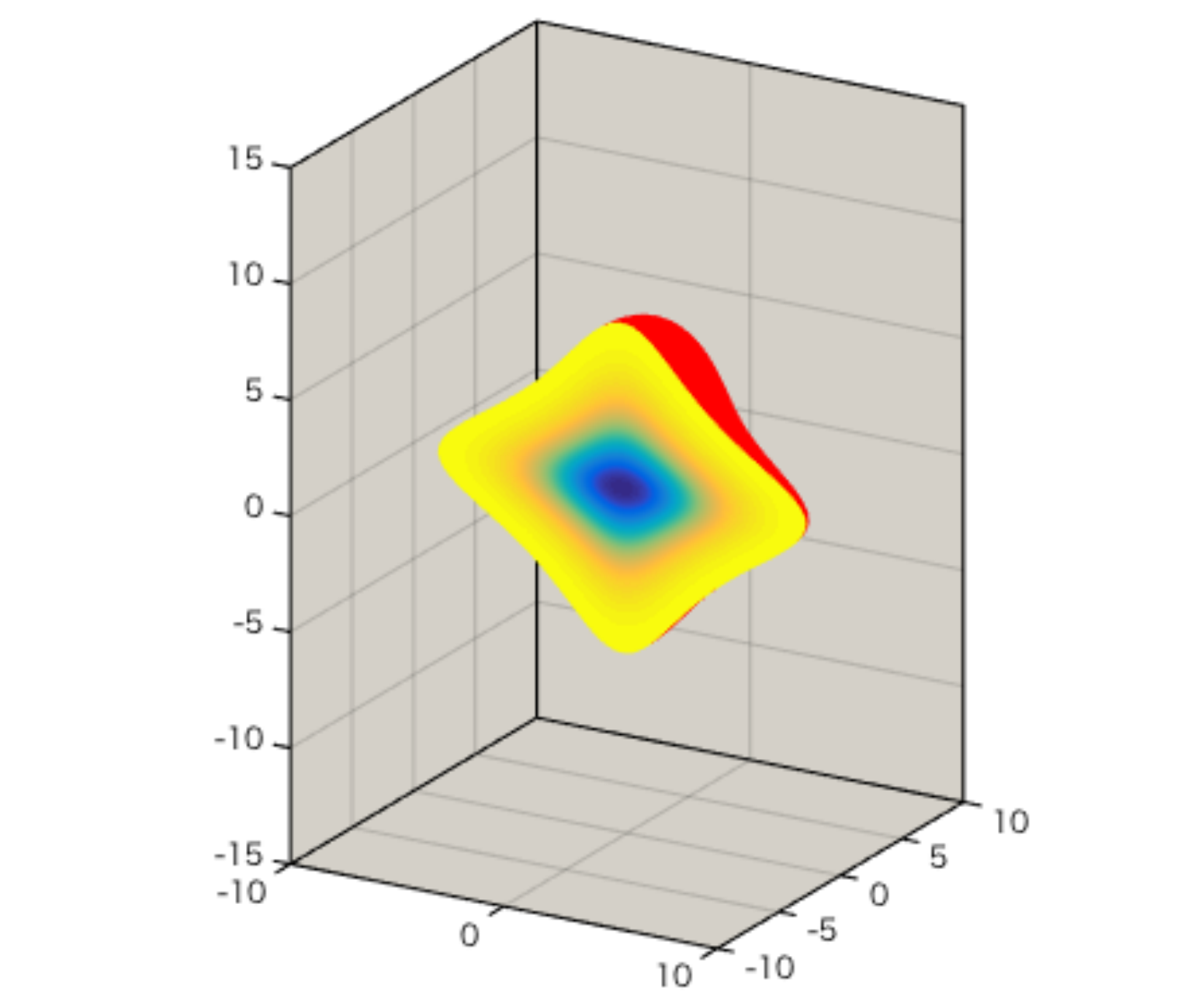}
  \end{center}
 \end{minipage}
\caption{The plots of the boojum energy density for the case that two vortices ending on one wall from two sides. The distance of two vortices is $L=4, 2, 0$ from left to right.}
\label{fig:boojum_b}
\end{figure}

Let us make a comment on the shapes of the boojums. When they are well separated,
the shape of the individual boojum is approximately the same as that drawn in the right panel of  Fig.~\ref{fig:simple_config}.
When the vortex strings are closer to each other than the vortex string size, the boojums merge.
One may recall that rich 3D structure appears when several BPS 't Hooft-Polyakov monopoles 
in $SU(2)$ gauge theory get close. For the boojums in the Abelian-Higgs theory, such drastic change in the shape
is not observed, see Figs.~\ref{fig:boojum_a} and \ref{fig:boojum_b}.


\begin{figure}[t]
  \begin{center}
   \includegraphics[width=7cm]{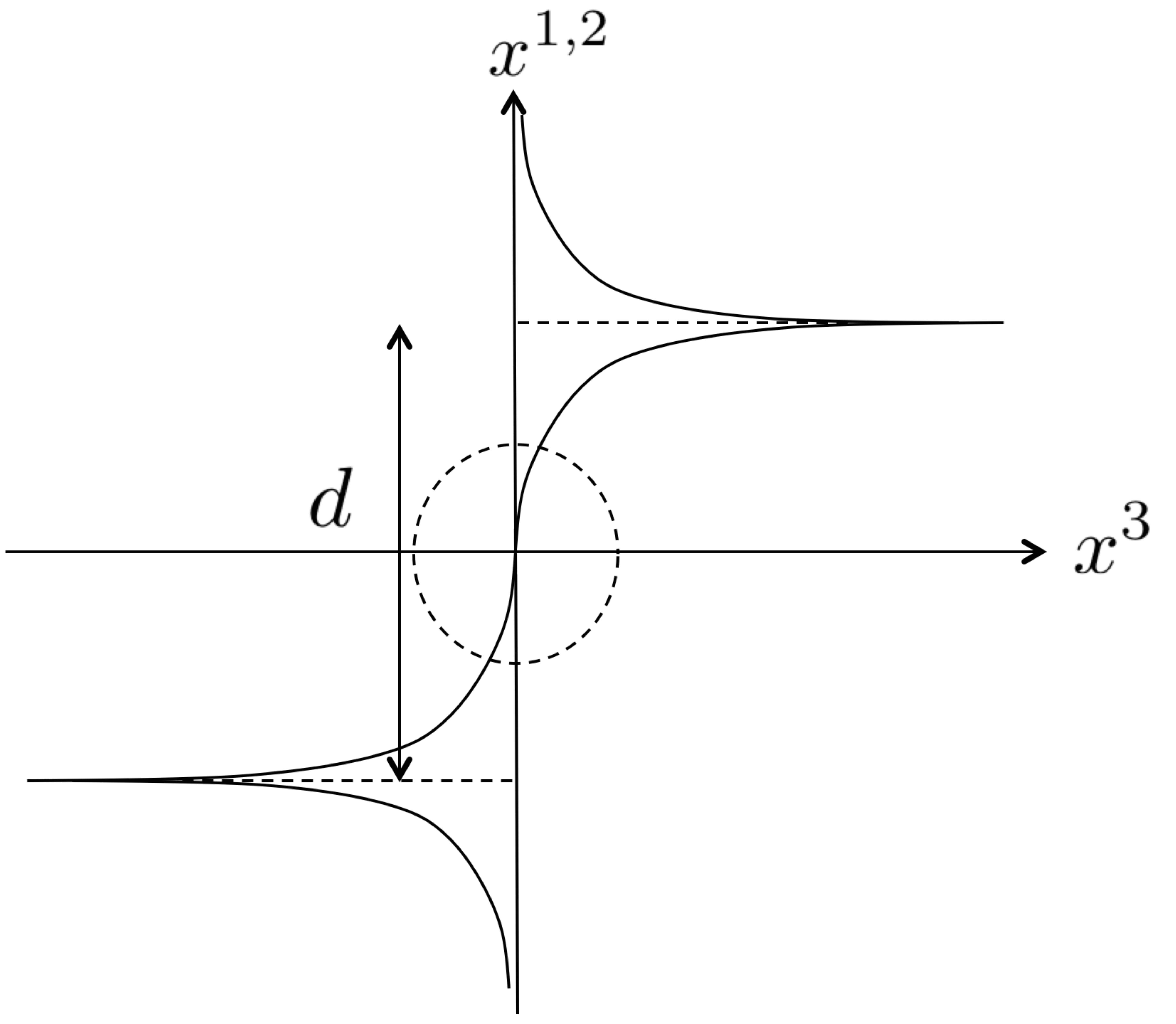}
  \end{center}
 \caption{Schematic picture of the configuration of two vortices attached to the wall from both sided.}\label{fig:vs}
\end{figure}

Our last comment here is about the question of the definition of the binding energy raised in \cite{Auzzi}. Consider two vortex-strings attached to the domain wall from both sides, see Fig. \ref{fig:vs}. Two vortex-strings are separated by the distance $d$. For this configuration, the authors of Ref.~\cite{Auzzi} argued that there are two possibilities where the binding energy is located. The first possibility is that the binding energy only localizes around the junction points. The second one is that it localizes not only around the junction points but also near the origin (dotted-circled domain in the right figure in Fig. \ref{fig:vs}). Our numerical solutions, for example Fig.~\ref{fig:11_A}, 
strongly supports that the former is true.

Lastly, in Appendix~A we present more complicated configurations with two or more domain walls.

\clearpage


\section{Approximate
solutions to 1/4 BPS Abelian master equations}\label{sec:APP}

In this paper, we have analyzed various 1/4 BPS configurations 
by numerically solving the corresponding master equations. However, in doing so, the need for asymptotically well-behaving initial configurations 
led us to the
novel method, how to approximate composite solitons. 
In this section, we will collect our findings and present a general formula for an arbitrary configuration of vortex-strings interacting with up to two domain walls. The extension of this formula to three or more domain walls is, however, a straightforward task.
  
Let us illustrate our method by recalling the simplest 1/4 BPS configuration, that is a vortex string attached to the domain wall, which we have studied in Sec.~\ref{sec:details_axially_symmetric}. In moduli matrix formalism, the corresponding master equation in dimensionless coordinates reads
\begin{equation} \label{eq:appmaster1}
\p_3^2 u - 1 + \left(\rho^2 e^{\tilde m x^3} + e^{-\tilde mx^3}\right)e^{-u} = 0.
\end{equation}
In Sec.~\ref{sec:details_axially_symmetric} we argued that 
an 
approximate solution, which asymptotically approaches 
the true solution in all directions is 
\begin{equation}\label{eq:analapp}
{\mathcal{U}} \equiv u_W\left(x^3 + \frac{1}{2\tilde m} u_S(\rho)\right) + \frac{1}{2} u_S (\rho),
\end{equation}
where $u_W(x^3)$ is a solution to a domain wall part and $u_S(\rho)$ a solution to the vortex string part of the composite soliton. In other words, these functions are solutions to 1/2 BPS master equations 
\begin{gather}\label{eq:appwall}
\partial_{3}^2 u_W  = 1-\Bigl(e^{\tilde m x^3}+e^{-\tilde m x^3}\Bigr)e^{-u_W}, \\ \label{eq:appstring}
\partial_\rho^2 u_S + \frac{1}{\rho}\partial_\rho u_S = 1- \rho^2 e^{-u_S}. 
\end{gather}
The function ${\mathcal{U}} $ is designed to solve Eq.~\refer{eq:appmaster1} in regions far away from vortex string, where $\abs{\partial_\rho {\mathcal{U}} } \ll 1$. Indeed, using asymptotic behavior of the string solution $u_S(\rho) \sim \log \rho^2$ we can easily see that the function
\begin{equation} 
u_W\left(x^3+\log \rho/\tilde m\right)+ \log \rho,
\end{equation} 
solves Eq.~\refer{eq:appmaster1}, if we neglect derivatives with respect to $\rho$. On the other hand, if we look at region dominated by the vortex string, that is $x^3 \gg 1$, we see from the asymptotic property of the domain wall part $u_W(x^3) \sim \tilde m x^3$ that the approximate solution takes the form
\begin{equation}
{\mathcal{U}}  \sim u_S(\rho) +\tilde m x^3, \hspace{5mm} x^3 \gg 1\,.
\end{equation}
This indeed solves Eq.~\refer{eq:appmaster1}, if we neglect the term $e^{-\tilde m x^3}$ on the right hand side. The only place where ${\mathcal{U}} $ fails to solve the master equation is the junction point of the vortex string and the domain wall -- the boojum. Thus, ${\mathcal{U}} $ approximates the true solution \emph{globaly}, which is not the case for the commonly used infinite gauge coupling approximation
$ {\mathcal{U}}_{\infty}  \equiv \log\left(\rho^2 e^{\tilde m x^3}+e^{-\tilde m x^3}\right) $. Indeed, if we evaluate the left hand side of the master equation \refer{eq:appmaster1} along the domian wall's position $x^3 = -\log \rho/\tilde m$ we obtain
\begin{equation}
\partial_k^2  {\mathcal{U}}_{\infty}  (\rho,x^3)\Big|_{x^3 = -\log\rho/\tilde m} = \tilde m^2+\frac{1}{\rho^2}.
\end{equation}
This shows that region of discrepancy between the true solution and $ {\mathcal{U}}_{\infty}$ is not bound to a finite region like in the case of $ {\mathcal{U}}_{\infty} $ . 

Amongst other advantages, the approximation ${\mathcal{U}}$ also enabled us in Sec.~\ref{sec:details_axially_symmetric} to calculate all topological charges \emph{exactly} and helped us to refine the arguments, which lead to the derivation of the general formula for the boojum charge in Eq.~\refer{eq:boojumcharge}.

More importantly, the same strategy can be used to create approximations to virtually any 1/4 BPS configuration. 
For example,
a collinear vortex strings attached from both sides of the domain wall is approximated as:
\begin{gather}
\partial_k^2 u = 1- \rho^2\left( e^{\tilde m x^3}+e^{-\tilde m x^3}\right)e^{-u}, \\
{\mathcal{U}} =u_W(x^3) + u_S(\rho).
\end{gather}
Again, the reasoning is the same. Far away from the strings the $\partial_\rho$ terms can be neglected and we are left with the master equation for a single domain wall, which is solved as
\begin{equation}
u \approx u_W(x^3) + \log(\rho^2)\,. 
\end{equation}
On the other hand, far away from the domain wall $\abs{x^3}\gg 1$ we can neglect $\partial_3$ derivatives and solve the master equation approximately as
\begin{equation}
u \approx \log\Bigl(e^{\tilde m x^3}+e^{-\tilde m x^3}\Bigr)+u_S(\rho)
\end{equation}
with the first term on the right-hand side being an asymptotic form of the domain wall solution $u_W(x^3)$.
The global approximation for two collinear vortex strings attaching the domain wall from both sides is thus $u \approx {\mathcal{U}} =u_W(x^3) + u_S(\rho)$ as claimed.

Let us now provide a short list of other 1/4 BPS configurations and their global approximations. Their derivation follows the same arguments as presented above and we will not repeat them for brevity. 

A semi-local vortex string  of the size $a$ attached to the wall is approximated as:
\begin{gather}
\partial_k^2 u = 1-\left((\rho^2+a^2)e^{\tilde m x^3}+e^{-\tilde m x^3}\right)e^{-u}, \\
{\mathcal{U}}= u_W\left(x^3 +\frac{1}{2\tilde m}u_{SLS}(a)\right)+\frac{u_{SLS}(a)}{2},
\end{gather}
where $u_{SLS}(a)$ is a solution to semi-local vortex master equation
\begin{equation}
\partial_\rho^2 u_{SLS}+\frac{1}{\rho}\partial_\rho u_{SLS} = 1-(\rho^2+a^2)e^{-u_{SLS}}.
\end{equation}
Two semi-local vortex strings of sizes $a_1$ and $a_2$ attached from both sides can be approximated as 
\begin{gather}
\partial_k^2 u = 1-\left((\rho^2+a_1^2)e^{\tilde m x^3}+(\rho^2+a_2^2)e^{-\tilde m x^3}\right)e^{-u}, \\
{\mathcal{U}} =u_W\left(x^3 +\frac{u_{SLS}(a_1)-u_{SLS}(a_2)}{2\tilde m}\right)+\frac{u_{SLS}(a_1)+u_{SLS}(a_2)}{2}.
\end{gather}
A general formula for approximate solution for arbitrary configuration of $n_1$ vortices attached from the right and $n_2$ vortices attached from the left is
\begin{gather}
\partial_k^2 u = 1-\left(\abs{P_{n_1}}^2e^{\tilde m x^3}+\abs{P_{n_2}}^2e^{-\tilde m x^3}\right)e^{-u}, \\
{\mathcal{U}}=u_W\left(x^3 +\frac{u_{S}^{(n_1)}-u_{S}^{(n_2)}}{2\tilde m}\right)+\frac{u_{S}^{(n_1)}+u_{S}^{(n_2)}}{2},
\end{gather}
where the string solutions obey
\begin{align}
\partial_\rho^2 u_{S}^{(n)}+\frac{1}{\rho}\partial_\rho u_{S}^{(n)} = 1-\abs{P_n}^2e^{-u_{S}^{(n)}}.
\end{align}
Generalizing further to a configuration of two domain walls with $n_1$ vortex strings attached to the right wall, $n_2$ vortex strings in the middle and $n_3$ vortex strings attached to the left wall, we find the approximation in the form:
\begin{gather}\label{eq:analapp2}
\partial_k^2 u = 1-\left(\abs{P_{n_1}}^2e^{\tilde m x^3}+\delta^2 \abs{P_{n_2}}^2 +\abs{P_{n_3}}^2e^{-\tilde m x^3}\right)e^{-u}, \\
{\mathcal{U}}=u_W\left(x^3 +\frac{u_{S}^{(n_1)}-u_{S}^{(n_3)}}{2\tilde m}; \delta \frac{\abs{P_{n_2}}}{\sqrt{\abs{P_{n_1}P_{n_3}}}}\right)+\frac{u_{S}^{(n_1)}+u_{S}^{(n_3)}}{2}.
\end{gather}
Here, we must supply double wall solution $u_W(x^3;\delta)$
\begin{equation}
\partial_3^2 u_W(x^3; \delta) = 1-\left(e^{\tilde m x^3}+\delta^2 + e^{-\tilde m x^3}\right)e^{-u_W(x^3; \delta)}.
\end{equation}
Arbitrary configuration of semi-local vortex strings attached from the right and arbitrary configuration of  vortex strings attached from the left is given as
\begin{gather}
\partial_k^2 u = 1-\left(\left(\abs{P_{n_1}}^2+\abs{Q_{n_1-1}}^2\right)e^{\tilde m x^3}+\abs{P_{n_2}}^2e^{-\tilde m x^3}\right)e^{-u}, \\
{\mathcal{U}}=u_W\left(x^3 +\frac{u_{SLS}^{(n_1)}-u_{S}^{(n_2)}}{2\tilde m}\right)+\frac{u_{SLS}^{(n_1)}+u_{S}^{(n_2)}}{2},
\end{gather}
where
\begin{equation}
\partial_\rho^2 u_{SLS}^{(n)}+\frac{1}{\rho}\partial_\rho u_{SLS}^{(n)} = 1-\left(\abs{P_n}^2+\abs{Q_{n-1}}^2\right)e^{-u_{SLS}^{(n)}}.
\end{equation}
And lastly, vortex string attached to a domain wall under the angle $\tan\alpha = 2\eta/\tilde m$ is approximated as
\begin{gather}
\partial_k^2 u = 1-\left(e^{\tilde m x^3}+e^{4\eta x^1}e^{-\tilde m x^3}\right)e^{-u}, \\
{\mathcal{U}}= u_W\left(x^3-\frac{2\eta}{\tilde m}x^1+\frac{1}{2\tilde m}u_S\right)+2\eta x^1+\frac{u_S}{2}.
\end{gather} 
This ends our short list of approximate solutions, which we used at various places in this paper and in the second paper \cite{Boojum2}. As we see, the general way how to construct approximations to composite solitons is to first solve the corresponding domain wall part $u_W$ by ignoring $x^1$ and $x^2$ derivatives and then replace these coordinates inside $u_W$ by appropriate vortex string solutions. This method can be thus applied to even more complicated configurations not mentioned in this section.
As we see, to obtain fully analytical approximations we need to supply 1/2 BPS ingredients into our global approximations. We develop such analytic approximations for a single string and a single domain wall in the Appendix~\ref{sec:APP2}. 



\section{Exact solutions to 1/4 BPS solitons}
\label{sec:ESB}

In this section, we will complement both numerical and approximative analysis by several \emph{exact} solutions of 1/2 and 1/4 BPS solitons which we have found. 
    
Any composite soliton can be reduced to pure wall(s) or pure vortex-string(s) in some appropriate limit, i.e. either by shifting wall or string to infinity via the corresponding moduli parameter(s). Therefore, any exact 1/4 BPS solution potentially contains 1/2 BPS solutions as its limiting cases. Hence, before we can discuss exact composite soliton solutions we must first discuss exact domain wall and vortex string solutions.

In the following subsections, we will present a novel exact solution to a particular configuration of semi-local vortex-strings and an exact solution of a particular configuration of domain walls, out of which we construct new 1/4 BPS solutions. Throughout this section, we use dimensionful coordinates.

\subsection{$M=\mbox{diag}(m,0,-m,-m)$}

The master equation for the model with $N_F=4$ and the mass matrix given as above reads in general
\begin{equation}\label{eq:exquarter1}
\frac{1}{2g^2v^2}\Bigl(4\partial_{z}\partial_{\bar z}u+\partial_3^2 u\Bigr) = 1- \Bigl(e^{2m x^3} \abs{h_1(z)}^2+\abs{h_2(z)}^2+e^{-2m x^3}\bigl(\abs{h_3(z)}^2+\abs{h_4(z)}^2\bigr)\Bigr)e^{-u},
\end{equation}
where $H_0 = (h_1,h_2,h_3,h_4)$ are elements of the moduli matrix. The generic configuration consists of a pair of domain walls, where the right wall has vortices attached from positive direction of $x^3$ axis (complex zeros of $h_1$), left wall has semi-local vortices attached from negative direction of $x^3$ axis (complex zeros of $h_3$ and/or $h_4$) and both walls 
have
vortices stretched between them (complex zeros of $h_2$). The walls are bent if the net vorticity of incoming and outgoing  vortex-strings is not equal. Asymptotically, the right wall has the profile $x^3 \approx \tfrac{1}{m}\log\bigl(\abs{h_1}/\abs{h_2}\bigr)$, while the left wall has $x^3 \approx -\tfrac{1}{2m}\log\bigl(\abs{h_2}^2/(\abs{h_3}^2+\abs{h_4}^2)\bigr)$.

Let us first discuss an exact vortex string solution of this model. Since there are no known exact ANO vortices we set $h_1 = h_2 = 0$ and look for exact solutions over the leftmost degenerate vacuum. Getting rid of the $x^3$ dependence by shifting $u\to u-2m x^3$ in the master equation \refer{eq:exquarter1}, we are left with the master equation for semi-local vortices:
\begin{equation}
\frac{2}{g^2v^2}\partial_{z}\partial_{\bar z}u = 1- (\abs{h_3(z)}^2+\abs{h_4(z)}^2)e^{-u}.
\end{equation}
This equation has an exact solution, namely 
\begin{equation}\label{eq:exsl}
u_{SLS}^{2F} = 2\log\Bigl( \abs{z}^2+\frac{4}{g^2v^2}\Bigr),
\end{equation}
for which $\abs{h_3}^2 = \tfrac{8}{g^2v^2}\abs{z}^2$ and $\abs{h_4}^2 = \abs{z}^4$. Loosely, we could think about $u_{SLS}^{2F}$ as a (non-linear) superposition of a local vortex and a semi-local vortex of the core size $\sqrt{8}/(gv)$. The vorticity of $u_{SLS}^{2F}$ is indeed 2.

Now, let us consider domain wall solutions. We set $\abs{h_1}^2 = 1$, $\abs{h_2}^2 = e^{2R}$ and $\abs{h_3}^2 = \abs{h_4}^2 = 1$. The master equation \refer{eq:exquarter1} reduces to
\begin{equation}
\frac{1}{2g^2v^2}\partial_3^2 u = 1 -\Bigl(e^{2m x^3}+ e^{2 R} + e^{-2m x^3}\Bigr)e^{-u},
\end{equation}
where the parameter $2R/m$ can be interpreted as the separation of walls for $R \gg m$. If $R$ is close to or less than $m$, the domain walls cannot be distinguished from one another and they form a single (compressed) domain wall. In the limit $R\to -\infty$ the middle vacuum disappears completely and only a single elementary wall remains.

There are two exact solutions for this equation. The first one
\begin{equation}
u_{DW} = 2\log\Bigl(e^{m x^3}+\sqrt{e^{2R}-6}+e^{-m x^3}\Bigr)\,, \hspace{5mm}  \frac{m^2}{2g^2v^2} = 1,
\end{equation}
was originally reported in \cite{Sakai3}. Notice that $u_{DW}$ describes arbitrarily separated/com-pressed pair of walls (depending on the value of parameter $R$), but the ratio $m/(gv)$ is fixed. On the other hand, the second solution
\begin{equation}
u_{CW} = 2\log\Bigl(e^{m x^3}+e^{-m x^3}\Bigr)\,, \hspace{5mm}  \frac{m^2}{2g^2v^2} \leq \frac{1}{4},
\end{equation}
describes walls with separations up to $R =\tfrac{1}{2}\log\bigl(2- \tfrac{4m^2}{g^2v^2}\bigr) \leq \log\sqrt{2}$. In other words, walls in $u_{CW}$  are always compressed, but unlike in the previous solution, ratio $m/(gv)$ can be varied up to the upper bound $1/\sqrt{2}$. Reaching this limit, we have $R\to -\infty$ and the solution $u_{CW}$ reduces to an exact single wall solution, reported in \cite{Sakai3}, which we have denoted as $u_1$ in Appendix~\ref{sec:APP2}.

Now we have all the necessary pieces to discuss an exact composite soliton solution, which reads 
\begin{equation}\label{eq:exslw}
u_{SW}^{4F} = 2\log\Bigl(e^{m x^3}+e^{-mx^3}\bigl(\tfrac{4}{g^2v^2}+\abs{z}^2\bigr)\Bigr)\,, \hspace{5mm} \frac{m^2}{2g^2v^2} = \frac{1}{8},
\end{equation}
with 
\begin{equation}
\abs{h_1}^2 = 1\,, \hspace{5mm} \abs{h_2}^2 = \abs{z}^2\,, \hspace{5mm} \abs{h_3}^2 = \frac{8}{g^2v^2}\abs{z}^2\,, \hspace{5mm} \abs{h_4}^2 = \abs{z}^4.
\end{equation}
The interpretation of this solution is most explicit if we rewrite it as
\begin{equation}
u_{SW}^{4F} = u_{CW}\bigl(x^3 -\tfrac{1}{4m}u_{SLS}^{2F}\bigr)+\frac{1}{2}u_{SLS}^{2F}.
\end{equation}
Remarkably, this formulation is functionally identical to the analytic approximation \refer{eq:analapp} for a single vortex string attached to the domain wall (only this time, the string is attached from the left side).\footnote{The disparity of factors multiplying string solutions inside the wall solutions is due to the different choices of the mass matrix $M$ for the approximate solution \refer{eq:analapp} and for the exact solution \refer{eq:exslw}.}

The bending of the compressed domain wall can be exactly obtained by equating both terms inside the logarithm in the solution \refer{eq:exslw}, that is $x^3 = -\tfrac{2}{gv}\log\bigl(\abs{z}^2+\tfrac{4}{g^2v^2}\bigr)$. Curiously, this bending is equivalent to the bending produced by a single semi-local vortex of the core size $\tfrac{4}{g^2v^2}$. However, by comparing the solution \refer{eq:exslw} with the Eq.~\refer{eq:exsl} we recognize that the attached string is identical to $u_{SLS}^{2F}$. (The solutions become identical as $x^3 \to -\infty$.)

\begin{figure}[h]
\begin{center}
\includegraphics[width=14cm]{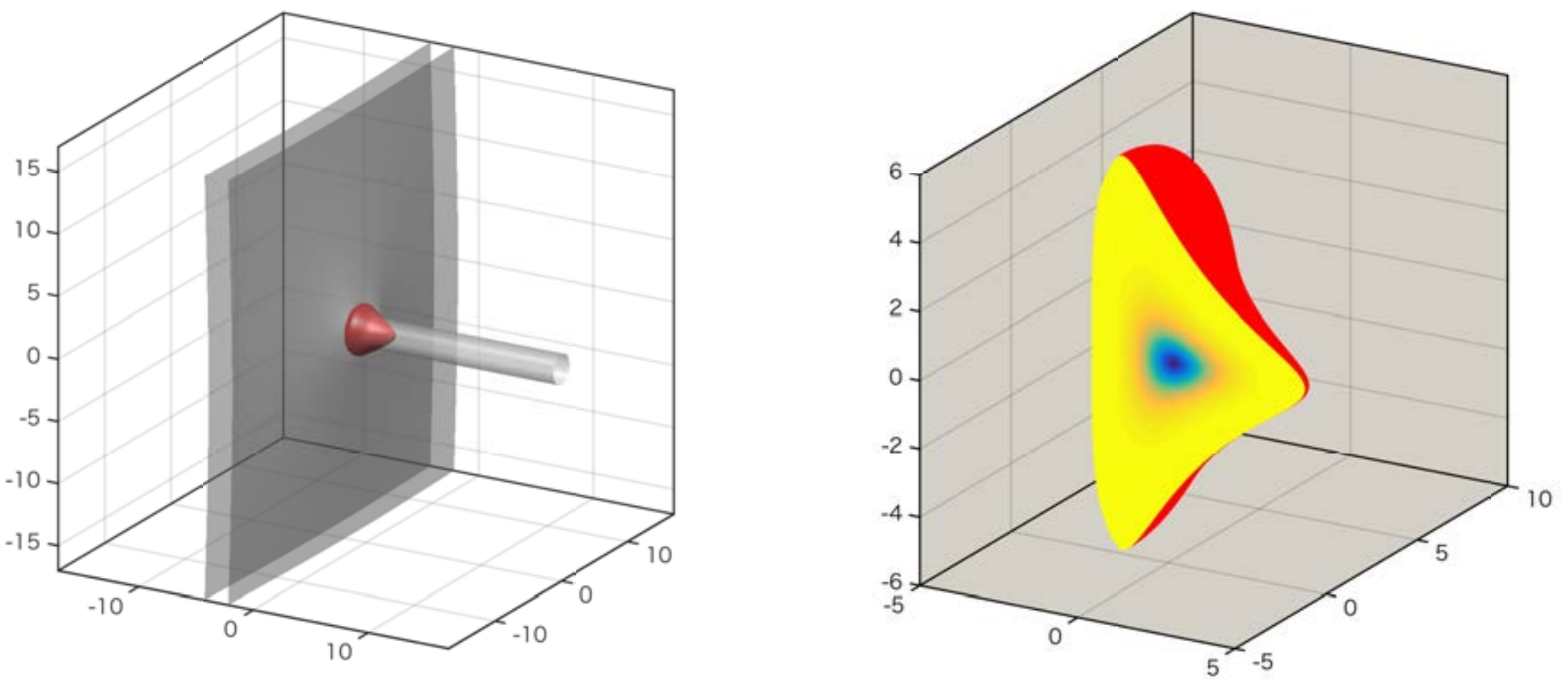}
\caption{The energy density surface (left) and the boojum density (right) for $u^{4F}_{SW}$
for $gv = 2$ and $m=-1$.}
\label{fig:exact_sol_1}
\end{center}
\end{figure}

Thus, the solution $u_{SW}^{4F}$ indeed describes a non-elementary wall $u_{CW}$ with non-elementary string $u_{SLS}^{2F}$ attached from the left. The same configuration with a string attached from the right side can be achieved by reflection $x^3 \to -x^3$. Notice that the position of the domain wall can be changed by shifting $x^3$, while the position of the string can be changed by shifting $z\to z-z_0$.
We plot the total energy density and boojum charge density on Fig.~\ref{fig:exact_sol_1}.

\subsection{$M=\mbox{diag}(2m,m,0,0,-m,-2m)/2$}

The master equation for the model with $N_F=6$ case and with the mass matrix given as above reads in general
\begin{align}\label{eq:exquarter2}
\frac{1}{2g^2v^2}\Bigl(\partial_{z}\partial_{\bar z}u+\partial_3^2 u\Bigr) = 1- \Bigl( & e^{2m x^3} \abs{h_1(z)}^2+e^{m x^3}\abs{h_2(z)}^2+\abs{h_3(z)}^2+\abs{h_4(z)}^2 \nonumber \\
& +e^{-m x^3}\abs{h_5(z)}^2+e^{-2m x^3}\abs{h_6(z)}^2\bigr)\Bigr)e^{-u},
\end{align}
where $H_0 = (h_1,h_2,h_3,h_4, h_5, h_6)$ are elements of the moduli matrix. The generic soliton configuration is composed of four domain walls. As in the previous example, the complex zeros of the moduli elements $h_i(z)$, $i=1,\ldots, 6$ determine number and positions of vortices, attached and stretched between successive walls (as well as their asymptotic profiles). 

Since we have doubly degenerate vacuum between the second and third wall, the exact vortex string solution of this model is again $u_{SLS}^{2F}$, as defined in Eq.~\refer{eq:exsl}. However, it  is
necessary to 
introduce a more general solution of the semi-local vortex string based on $N_F = 3$ model, with the master equation
\begin{equation}
\frac{2}{g^2v^2}\partial_z\partial_{\bar z} u = 1- \bigl(\abs{h_1}^2+\abs{h_2}^2+\abs{h_3}^2\bigr)e^{-u}.
\end{equation}
That solution is
\begin{equation}
u_{SLS}^{3F}(s) = 2\log\bigl(\abs{z}^2+s^2\bigr),
\end{equation}
with 
\begin{equation}
\abs{h_1}^2 = \abs{z}^4, \hspace{5mm}
\abs{h_2}^2 = 2 s^2 \abs{z}^2, \hspace{5mm}
\abs{h_3}^2 = s^2 \bigl(s^2-\tfrac{4}{g^2v^2}\bigr).
\end{equation}
The solution $u_{SLS}^{3F}$ is well-defined only if the core size $s$ is sufficiently big, namely $s\geq \tfrac{2}{gv}$. Indeed, if $s=\tfrac{2}{gv}$ this solution reduces to the previous case  $u_{SLS}^{3F}\bigl(\tfrac{2}{gv}\bigr)=u_{SLS}^{2F}$ and bellow this limit $\abs{h_3}$ is imaginary. The interpretation of $u_{SLS}^{3F}(s)$ is best seen if we rewrite the master equation as
\begin{equation}
\frac{2}{g^2v^2}\partial_z\partial_{\bar z} u = 1- \bigl(\abs{z}^2+s^2 +\tfrac{2s}{g v}\bigr)\bigl(\abs{z}^2+s^2 -\tfrac{2s}{g v}\bigr)e^{-u},
\end{equation}
from which we see that $u_{SLS}^{3F}$ describes a pair of coincident semi-local vortices with core sizes $s^2 \pm\tfrac{2s}{g v}$.

Let us now study the domain wall solution for the master equation \refer{eq:exquarter2}, which is given as
\begin{equation}
u_{DCW}(x^3;R) = 2\log\Bigl(e^{m x^3}+e^{R}+e^{-m x^3}\Bigr), \hspace{5mm} \frac{m^2}{2g^2v^2}\leq \frac{1}{4},
\end{equation}
with 
\begin{gather}
\abs{h_1}^2 = \abs{h_6}^2=1, \hspace{5mm} \abs{h_2}^2=\abs{h_5}^2= 2e^{R}\bigl(1-\tfrac{m^2}{2g^2v^2}\bigr), \\
\abs{h_3}^2+\abs{h_4}^2 = e^{2R}+2-\tfrac{8m^2}{2g^2v^2}.
\end{gather} 
Again, the parameter $2R/m$ can be interpreted as the separation of walls for $R \gg m$. However, we can rewrite $u_{DCW}$ as
\begin{align}\label{eq:exdcw2}
u_{DCW}(x^3;S) &= 2\log\Bigl(e^{m (x^3-S)/2}+e^{-m (x^3-S)/2}\Bigr)\Bigl(e^{m (x^3+S)/2}+e^{-m (x^3+S)/2}\Bigr) \nonumber \\
& = u_{CW}(x^3-S;m/2)+u_{CW}(x^3+S;m/2),
\end{align}
where 
\begin{equation}
e^{R} =2\cosh(m S). 
\end{equation}
In other words, the solution $u_{DCW}$ describes a pair of compressed walls $u_{CW}$ with separation $2S$.\footnote{Notice that the solution is defined for all values of $R$, but the parameter $S$ is only defined for $R\geq \log 2$. If $R<\log 2$, it is possible to ascribe $S$ a purely imaginary value $S = \tfrac{i}{m}\arccos(e^R/2)$, with which the solution \refer{eq:exdcw2} is still real. This confirms the intuition that the pair of wall merges at $R =\log 2$, where separation becomes zero $S=0$.}

Now we have all the pieces to study an exact 1/4 BPS solution of Eq.~\refer{eq:exquarter2}. This solution is a combination of the wall solution $u_{DCW}$ with the semi-local string solution $u_{SLS}^{3F}$ stretched between its two domain walls: 
\begin{equation}\label{eq:exwslw}
u_{SW}^{6F} = 2\log\Bigl(e^{m x^3}+ a + b \abs{z}^2+ e^{-m x^3}\bigr)\,, \hspace{5mm} \frac{1}{2}\leq \alpha^2 \leq 1\,, 
\end{equation}
with the moduli elements given as
\begin{gather}
\abs{h_1}^2 = 1\,, \hspace{5mm} \abs{h_2}^2 = 2 b \abs{z}^2 (1-\alpha^2)\,, \hspace{5mm} \abs{h_3}^2 = 2ab \abs{z}^2 \\
 \abs{h_4}^2 = b^2\abs{z}^4\,, \hspace{5mm} \abs{h_5}^2 = 2 b \abs{z}^2 (1-\alpha^2)\,, \hspace{5mm} \abs{h_6}^2 = 1\,,
\end{gather}
where we have denoted
\begin{equation}
\alpha^2 \equiv \frac{m^2}{2g^2v^2}\,, \hspace{5mm} a \equiv \sqrt{\frac{2-8\alpha^2}{1-2\alpha^2}}\,, \hspace{5mm}
b \equiv \frac{g^2v^2}{2}\bigl(1-\alpha^2\bigr)a\,.
\end{equation} 
The profile of the right compressed wall is asymptotically equal to $x^3 = \tfrac{1}{m}\log\bigl(a+b \abs{z}^2\bigr)$, while 
that of the left wall is  $x^3 = -\tfrac{1}{m}\log\bigl(a+b \abs{z}^2\bigr)$. Therefore, the minimal separation between both composite walls is roughly $ \tfrac{2}{m}\log a $. 
As in the previous case, we can rewrite this solution using its 1/2 BPS constituents:
\begin{equation}
u_{SW}^{6F} = u_{DCW}(x^3; u_{SLS}^{3F}\bigl(\tfrac{1}{gv}\sqrt{\tfrac{2}{1-\alpha^2}})+\tfrac{1}{2}\log \tfrac{a g^2v^2(1-\alpha^2)}{2}\bigr)\,.
\end{equation} 
This is functionally identical to the approximate solution for a similar configuration given in Eq.~\refer{eq:analapp2}.

Notice that the ratio of mass and gauge coupling (parameter $\alpha$) can only vary in a certain interval, outside of which the solution is not valid. The edges of the interval correspond to extreme cases. In the limit $\alpha^2 \to 1/2$ we see that $a,b\to \infty$, $a/b \to \tfrac{4}{g^2v^2}$ and the solution becomes $u_{SW}^{6F}\to u_{SLS}^{2F}$. In other words, this limit corresponds to the infinite separation of walls. In the other extreme $\alpha^2 \to 1$ we have $a\to \sqrt{6}$, $b\to 0$ and $h_2, h_3, h_4, h_5 \to 0$. As a consequence, both domain walls are infinitely compressed and the vortex string disappears. This domain wall is an exact solution, which we denoted as $u_2$ is Sec.~\ref{sec:APP}.

By shifting $x^3$ and $z$ coordinates we can change the position of the configuration as a whole, but interestingly the separation between the walls is not a modulus of the solution, but rather it is controlled by parameters of the model. As we can see, as the walls are further apart the core size of the vortex string  $\tfrac{1}{gv}\sqrt{\tfrac{2}{1-\alpha^2}}$ shrinks ultimately to  
$\tfrac{2}{g v}$. But, when the wall approaches each other, the size of the semi-local string fattens and, ultimately, diverges. This is a direct confirmation of a numerical observation which we make in Appendix A.

\begin{figure}[h]
\begin{center}
\includegraphics[width=14cm]{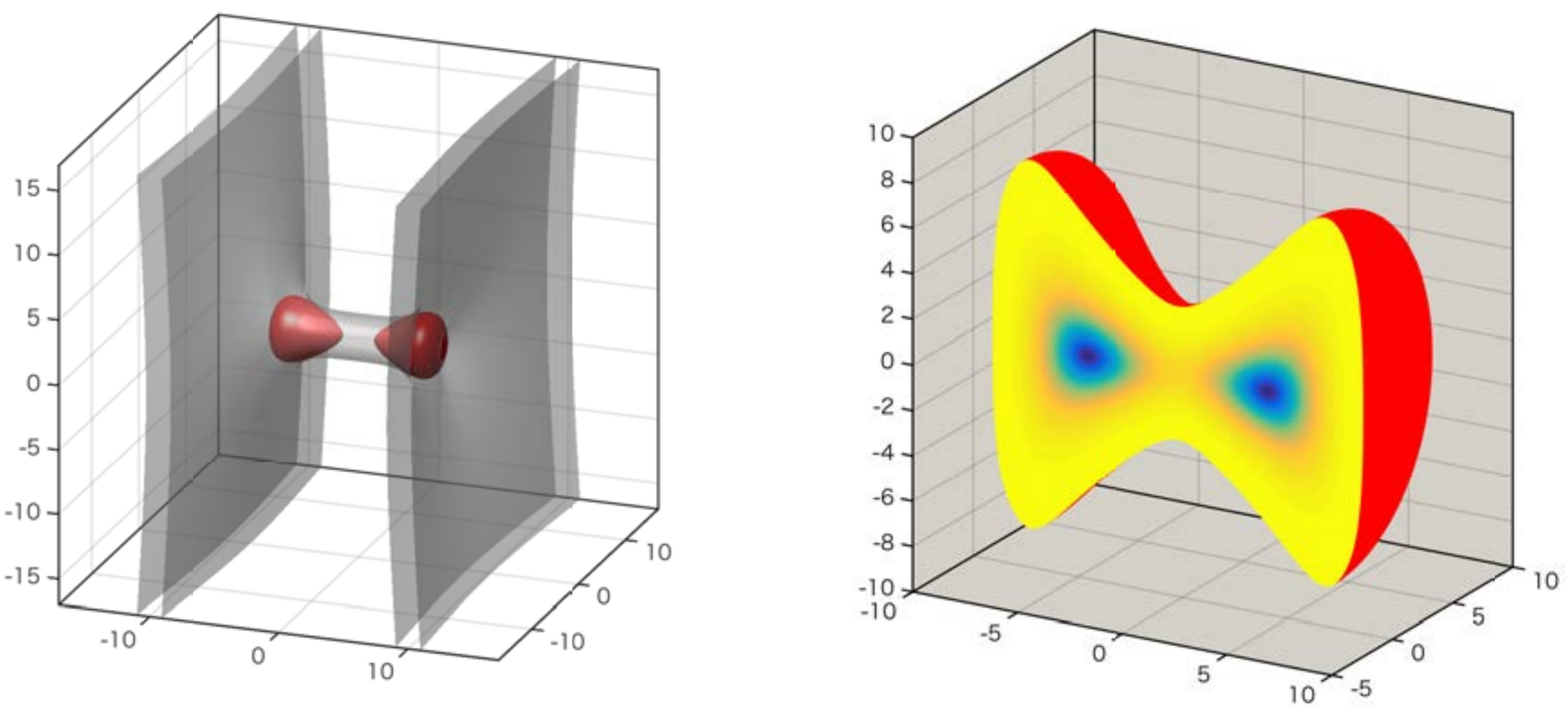}
\caption{The energy density surface (left) and the boojum density (right) for $u^{6F}_{SW}$
for $gv=1$ and $m=1.0001$.}
\label{fig:exact_sol_2}
\end{center}
\end{figure}

We plot the total energy density and boojum charge density on Fig.~\ref{fig:exact_sol_2}.

\subsection{Other exact solutions}

The two exact 1/4 BPS solitons  $u_{SW}^{4F}$ and $u_{SW}^{6F}$ are just illustrative examples of the rich hierarchy of exact solutions.
We believe that there exists 1/4 BPS exact solutions in multi-flavour models. 
Our preliminary investigations show that many new configurations of domain walls and vortices can be constructed as a suitable combination of its 1/2 BPS constituents, in the same spirit as described in this section. The list is potentially infinite if the number of flavors can be extended indefinitely. To provide an exhaustive list of exact 1/4 BPS solution up to a certain number of flavors seems to be an important task, which however lies somewhat outside the scope of this paper. Therefore, we plan to address this issue fully in a separate work. For parallel domain walls, this was already achieved by one of us \cite{exactwalls}. 

All our exact solutions presented in this section shares the same trait: they are only a special configuration of solitons. The domain wall solution $u_{CW}$ describes a pair of compressed elementary domain walls and both vortex string solutions $u_{SLS}^{2,3F}$ describes a pair of coincident semi-local vortices. In other words, some moduli are missing and we cannot provide exact results for generic configuration of solitons. Composite solitons $u_{SW}^{4F}$ and $u_{SW}^{6F}$ naturally inherit this trait.

As a closing comment of this section,
the existence of exact solutions of Abelian master equations directly implies the existence of exact solutions in non-Abelian case. This can be seen from the fact, that non-Abelian analog of the master equation can be, for special $U(1)$-factorizable solutions, decomposed into a set of independent Abelian master equations \cite{Isozumi:2004va}. Thus, any exact solution in Abelian theory can be used to construct exact solutions in non-Abelian theories. 


\section{Outlook}
\label{sec:out}

In this paper, we have studied 1/4 BPS equations in the Abelian-Higgs theory.
As written at the end of Sec.~\ref{sec:ESB}, our solutions also solve non-Abelian BPS equations
by trivial embedding. While these embedding solutions are essentially Abelian solutions, genuine non-Abelian
objects, like 't Hooft-Polyakov monopoles, may join the game in the non-Abelian theories. 
We will investigate non-Abelian extensions of 1/4 BPS solutions including monopoles, boojums, domain 
walls and vortex strings in forthcoming works.
It is also interesting to study the 1/4 BPS equations in different spacetime dimensions. 
For example, the 1/4 BPS equations for vortex sheets and instanton particles in 5 dimensions \cite{Hanany:2004ea,Eto:2004rz}, 
and those for the domain wall junctions in 3 dimensions \cite{Eto:2005cp,Eto:2005fm} are known, but only a little is known
about their solutions both in the Abelian and non-Abelian theories for the finite gauge coupling constants.
We will also proceed to obtain numerical/analytical solutions for these series of 1/4 BPS equations.
Furthermore, the technique developed in this paper may help to solve another type of 1/4 BPS equations or 1/8 BPS equations \cite{Lee:2005sv,Eto2} for which nothing has been known except for the BPS equations.


We have stumbled upon unexpected analytic results, namely exact solutions to 1/2 and 1/4 BPS master equations in models with $N_F\leq 6$ described in Sec.~\ref{sec:ESB}. As we stated, we believe that these solutions are just simplest examples of a potentially inexhaustible wealth of exact solutions in models with an ever larger number of flavors of fundamental Higgs fields.
The indirect evidence for this stems from the fact that our exact composite solitons, denoted as $u_{SW}^{4F}$ and $u_{SW}^{6F}$, have been `build' from of 1/2 BPS exact solutions for walls and strings. Therefore, it is natural to expect
that more complicated combinations should give us another 1/4 BPS solutions in higher flavor models. 
The exact rules of this `solitonic engineering' are so far unclear. However, the preliminary results show that certain restrictions on 1/2 BPS components must be in place if they are to be used in the construction of 1/4 BPS solutions. We have spotted glimpses of this in Sec.~\ref{sec:ESB}, namely that domain walls must have sufficiently low tension and cores of semi-local strings must be sufficiently large. The full analysis of this is a direction of future research.
Furthermore, for parallel 1/2 BPS domain walls, such analysis was done by one of us in the study \cite{exactwalls}, where it was shown that exact multi-wall solutions can be build up from certain single wall solutions in quite an arbitrary fashion, except that the total tension of the final domain wall configuration must be small enough. Effectively, this restriction prevents the development of an inner layer inside domain walls, where unbroken phase appears. In fact, no exact solution of domain wall with this inner layer is known at present. Finding such a solution remains as an interesting open problem.
The similar study for semi-local vortices is in preparation. By combining the findings for exact 1/2 BPS solitons we may be able to clarify, how their combination can yield exact 1/4 BPS solitons. Naturally, in doing so we may discover similar rules for other composite solitons, such as  wall-wall and string-string junctions, which we did not address in this paper.
An interesting question is, whether all exact 1/4 BPS solution can be decomposed as combinations of exact 1/2 BPS solutions, or whether some `irreducible' solutions exist.

\section*{Acknowledgements} 
This work is supported by Grant-in Aid for Scientific Research 
No.25400280 (M.\ A.\ ) and from the Ministry of Education, 
Culture, Sports, Science and Technology  (MEXT) of Japan. 
The work of M. E. is supported in part by
JSPS Grant-in-Aid for Scientic Research (KAKENHI Grant No.~26800119)
and the MEXT-Supported Program for the Strategic
Research Foundation at Private Universities ``Topological Science''
(Grant No.~S1511006).
F. B. is an international research fellow of the Japan Society for the Promotion of Science.
This work was supported by Grant-in-Aid for JSPS Fellows, Grant Number 26004750.


\appendix


\section{Non-axially symmetric solutions for $N_F\ge3$}

In this appendix, we describe configurations of two or more domain walls with a various number of vortex-strings attached. Throughout this section, we use dimensionless units.

\subsection{Non-degenerate masses }

Let us explain how to obtain the 1/4 BPS configurations with two domain walls separating three different vacua
$\left<A\right>$ ($A=1,2,3$) in which  $n_A$ vortex strings exist. The minimal model is $N_F=3$ with non-degenerate
mass matrix $\tilde M = (\tilde m_1/2,(\tilde m_2 - \tilde m_1)/2,-\tilde m_2/2)$. The tension of the domain wall interpolating
$\left<1\right>$ and $\left<2\right>$ is $v^2(2\tilde m_1-\tilde m_2)/2$ and that of the domain wall 
interpolating $\left<2\right>$ and $\left<3\right>$ is $v^2 (2\tilde m_2-\tilde m_1)/2$.
For simplicity, we will focus on the case $\tilde m_1 = \tilde m_2 = \tilde m$, namely that the two domain walls
have the same tension.

\begin{figure}[t]
\begin{center}
\includegraphics[width=15cm]{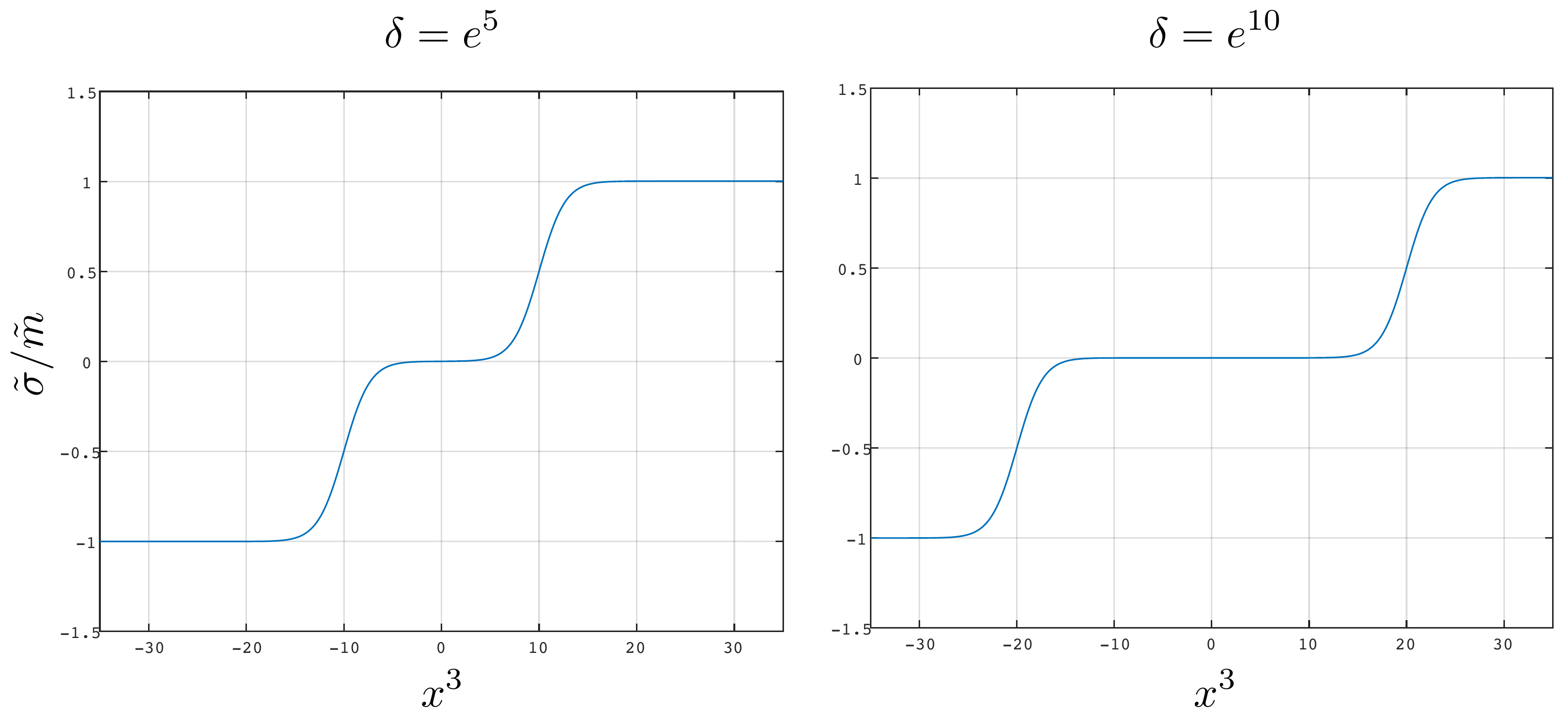}
\caption{$\tilde \sigma$ for two separated domain walls for $\tilde m_1 = \tilde m_2 = \tilde m = 1$ in $N_F=3$ case. 
The left panel is for
$\delta = e^5$ ($R=20$) and the right panel is for $\delta = e^{10}$ ($R=40$).}
\label{fig:two_walls}
\end{center}
\end{figure}

First of all, let us consider the configuration with no vortex strings. 
The moduli matrix is characterized by only one real parameter $\delta$ as
\begin{eqnarray}
H_0 = (1,\ \delta,\ 1).
\end{eqnarray}
When $\delta \gg 1$ the physical meaning of this parameter is related to the separation of walls, since the position of the walls can be estimated as 
\begin{eqnarray}
x^3 = \pm\frac{1}{\tilde m} \log \delta^2.
\end{eqnarray}
Thus the separation $R$ is
\begin{eqnarray}
R = \frac{2}{\tilde m} \log \delta^2,
\end{eqnarray}
see Fig.~\ref{fig:two_walls}. 
When $\delta$ is close to or smaller than 1, two domain walls
coalesce and $2(\log \delta^2)/\tilde m$ can no longer be understood as the distance. When $\delta \to -\infty$,
the two walls collapse into one heavier domain wall. The solution is parametrized as $u=u_W(x^3;\delta)$ which
satisfies the following reduced master equation
\begin{eqnarray}
\p_3^2 u_W(x^3;\delta) - 1 + \left(e^{\tilde m x^3} + \delta^2 + e^{-\tilde mx^3}\right)e^{-u_W(x^3;\delta)} = 0.
\label{eq:master_dw_nf3}
\end{eqnarray}

Next, we put $n_A$ vortex strings in the $\left<A\right>$ vacuum. 
The moduli matrix for this  is given by
\begin{eqnarray}
H_0 = (P_{n_1}(z),\ \delta P_{n_2}(z),\ P_{n_3}(z)),
\end{eqnarray}
where $P_{n_A}(z)$ stands for a monomial of $n_A$-th degree.
The corresponding gradient flow equation is 
\begin{eqnarray}
\p_k^2 U - 1 + \left(|P_{n_1}|^2 e^{\tilde mx^3} + \delta^2 |P_{n_2}|^2 + |P_{n_3}|^2 e^{-\tilde mx^3}\right) e^{-U} = \p_t U.
\end{eqnarray}
One can rewrite this as
{\small
\begin{eqnarray}
\p_k^2 U - 1 + 
\left(e^{\tilde mx^3 + \log \left|\frac{P_{n_1}}{P_{n_3}}\right|} +  \frac{\delta^2|P_{n_2}|^2}{|P_{n_1}P_{n_3}|} +
 e^{-\tilde mx^3-\log \left|\frac{P_{n_1}}{P_{n_3}}\right|}\right) e^{-U+\log|P_{n_1}P_{n_3}|} = \p_t U.
\end{eqnarray}}
Comparing this with Eq.~(\ref{eq:master_dw_nf3}), one finds
a suitable initial configuration for solving the three-dimensional master equation 
\begin{eqnarray}
{\mathcal U}(x^k) = u_W\left(x^3+\frac{u_S^{(n_1)} - u_S^{(n_3)}}{2\tilde m}; \delta \frac{|P_{n_2}|}{\sqrt{|P_{n_1}P_{n_3}|}}\right) + \frac{u_S^{(n_1)}+u_S^{(n_3)}}{2}.
\end{eqnarray}
Validity of this initial configuration is ensured by the asymptotic behavior $u_S^{(n)} \to \log |P_n|^2$ for $\rho \gg 1$. 
The positions of the two domain walls are estimated as
\begin{eqnarray}
x^3\big|_{\left<1\right>\leftrightarrow\left<2\right>} 
&=& \frac{1}{\tilde m} 
\left(\log \delta^2 + u_S^{(n_2)} - u_S^{(n_1)} \right),
\label{eq:mag_pot_left}\\
x^3\big|_{\left<2\right>\leftrightarrow\left<3\right>} 
&=& -\frac{1}{\tilde m} 
\left(\log \delta^2 + u_S^{(n_2)} - u_S^{(n_3)} \right).
\label{eq:mag_pot_right}
\end{eqnarray}
Then the separation of the two domain walls is
\begin{eqnarray}
R &=& \frac{2}{\tilde m}\left(\log \delta^2 + u_S^{(n_2)} - \frac{u_S^{(n_1)} + u_S^{(n_3)}}{2}\right),\\
&\simeq& \frac{2}{\tilde m} \log \delta^2 \frac{|P_{n_2}|^2}{|P_{n_1}P_{n_3}|}, \nonumber\\
&\to& \frac{2}{\tilde m}\log \delta^2 \rho^{2n_2-n_1-n_3}
\quad (\rho \to \infty).\nonumber
\end{eqnarray}
The domain walls are asymptotically parallel when $2n_2 = n_1 + n_3$ and are asymptotically flat only
when $n_1 = n_2 = n_3$.

\begin{figure}[h]
 \begin{minipage}{0.5\hsize}
  \begin{center}
   \includegraphics[width=7cm]{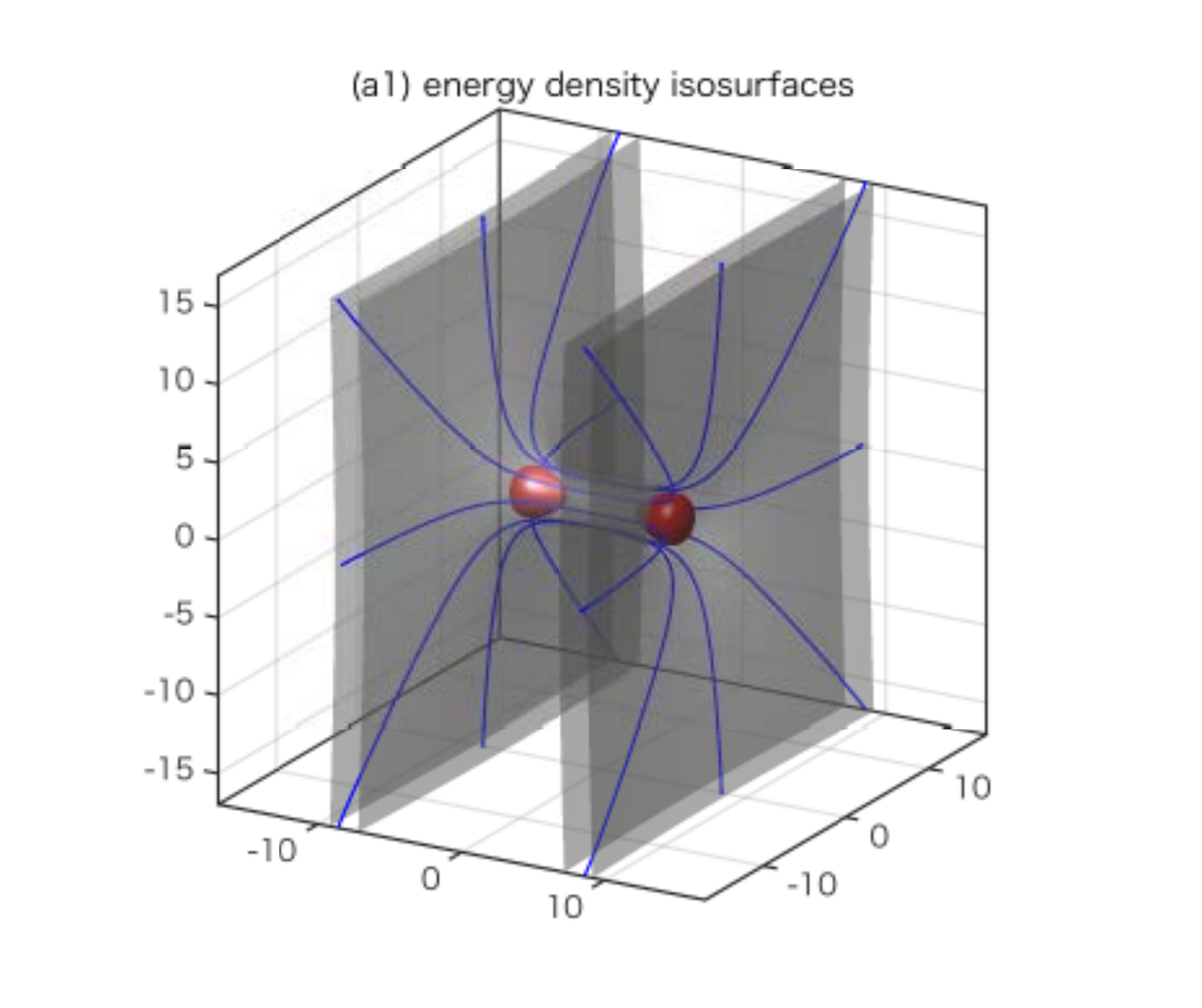}
  \end{center}
 \end{minipage}
  \begin{minipage}{0.5\hsize}
  \begin{center}
   \includegraphics[width=7cm]{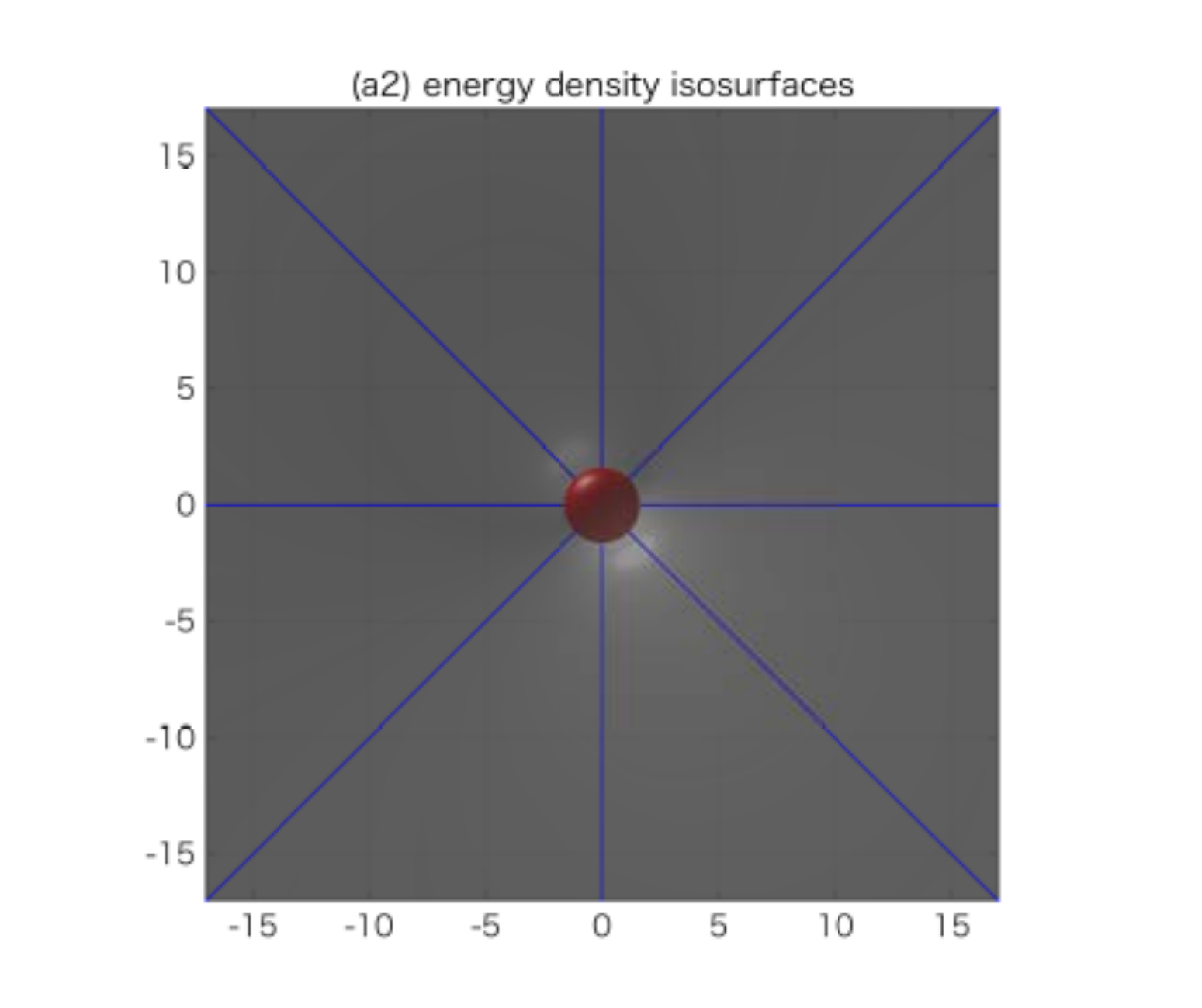}
  \end{center}
 \end{minipage}\\
 \begin{minipage}{0.5\hsize}
  \begin{center}
   \includegraphics[width=7cm]{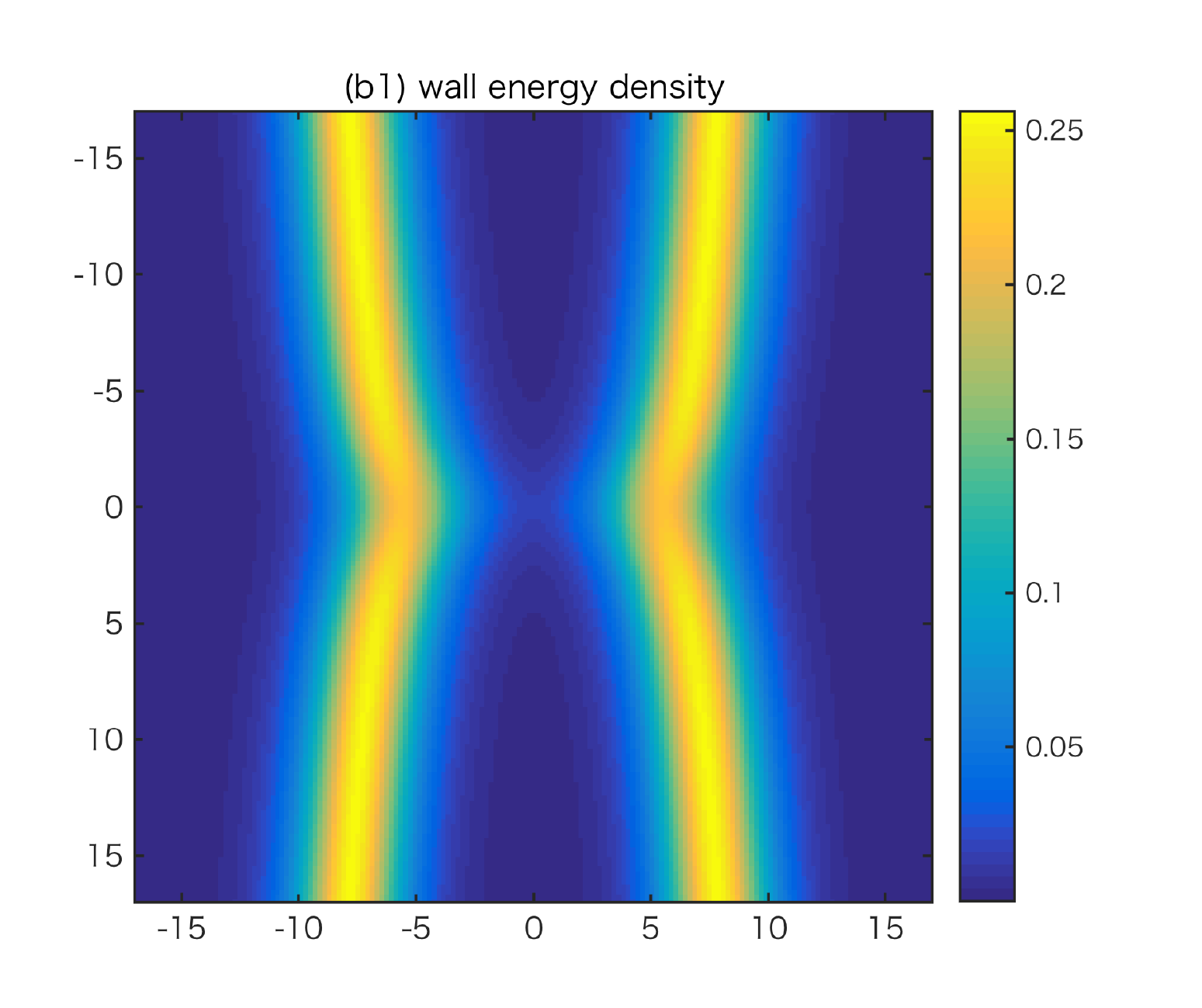}
  \end{center}
 \end{minipage}
 \begin{minipage}{0.5\hsize}
  \begin{center}
   \includegraphics[width=7cm]{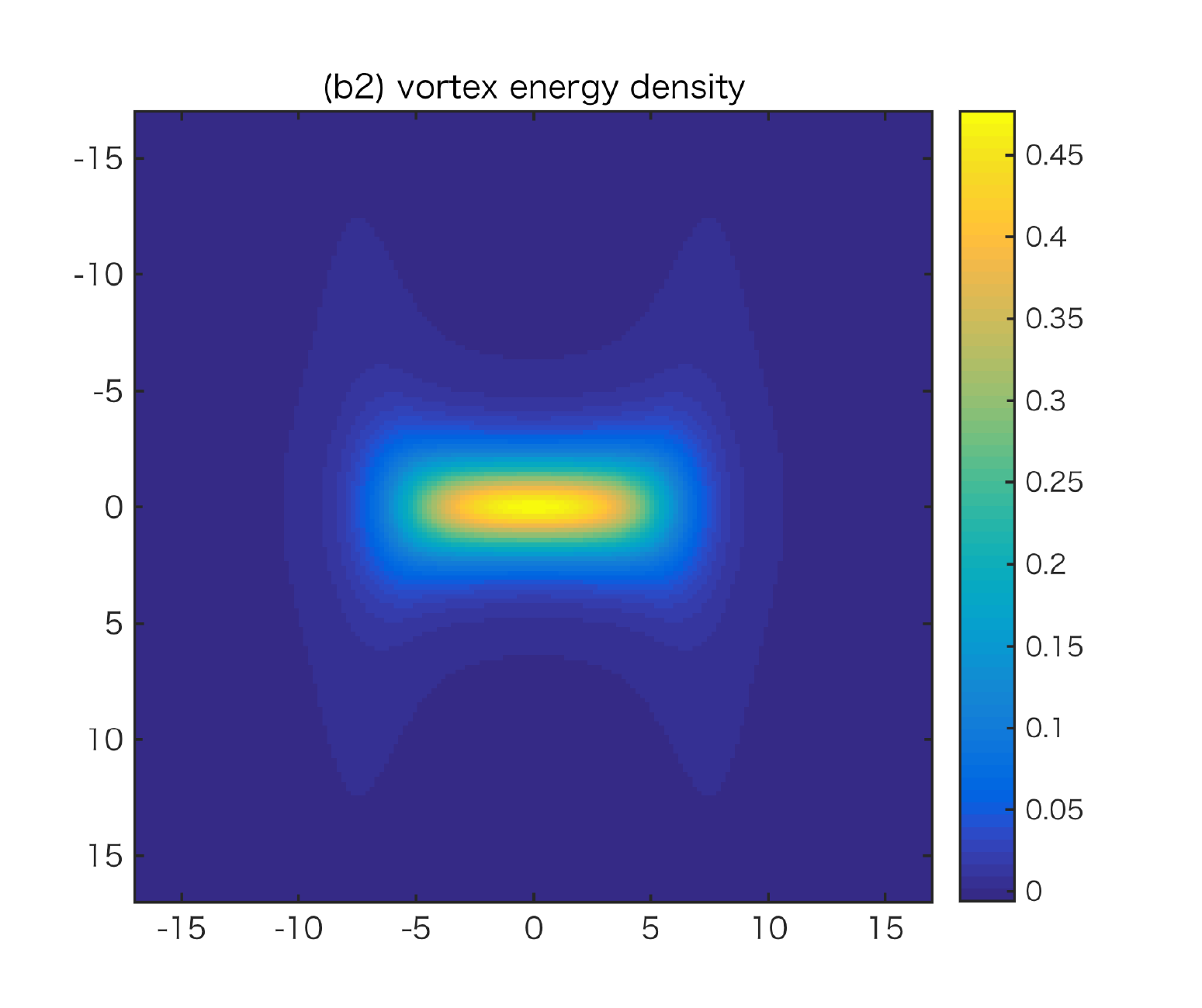}
  \end{center}
 \end{minipage}\\
  \begin{minipage}{0.5\hsize}
  \begin{center}
   \includegraphics[width=7cm]{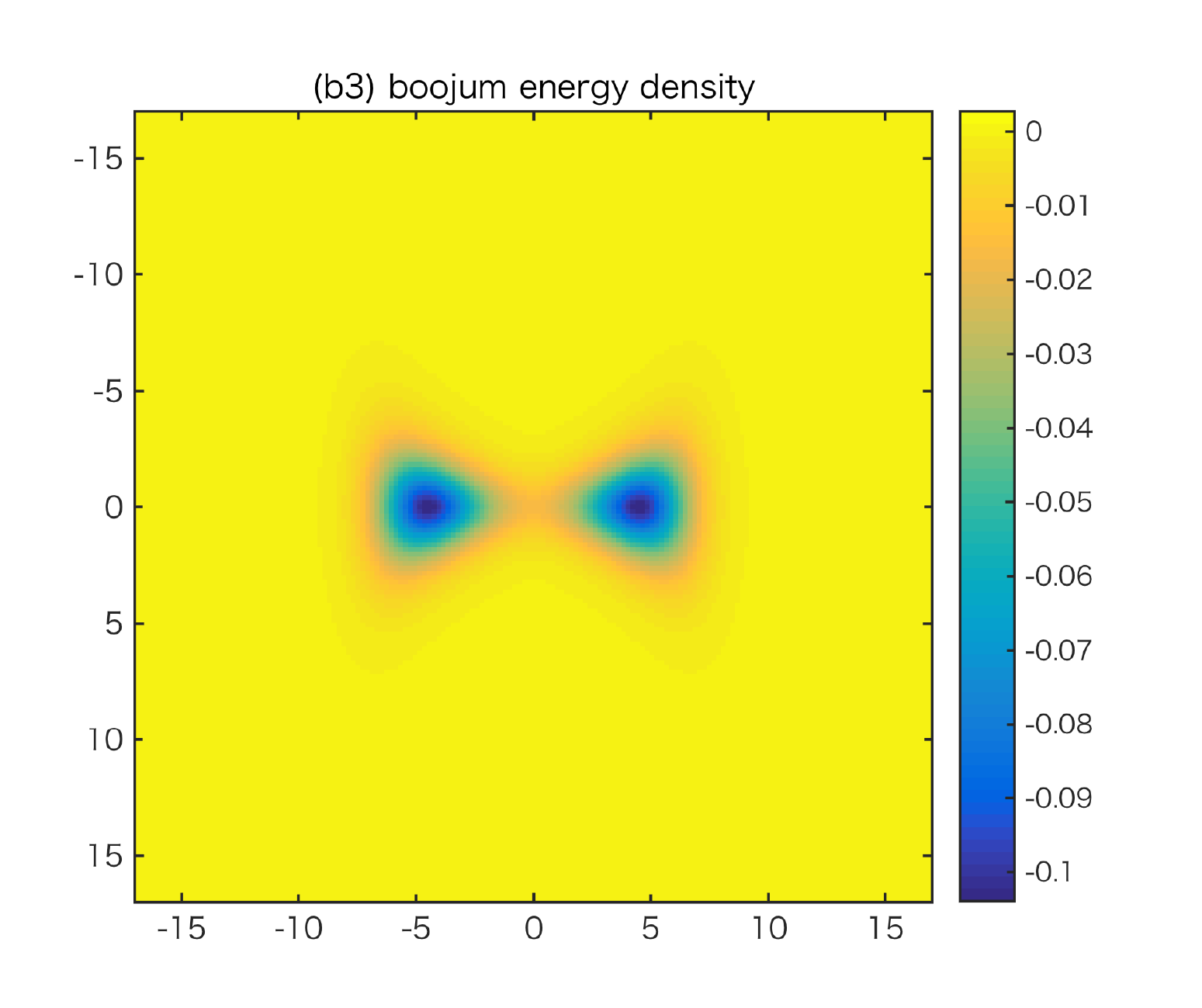}
  \end{center}
 \end{minipage}
 \begin{minipage}{0.5\hsize}
  \begin{center}
   \includegraphics[width=7cm]{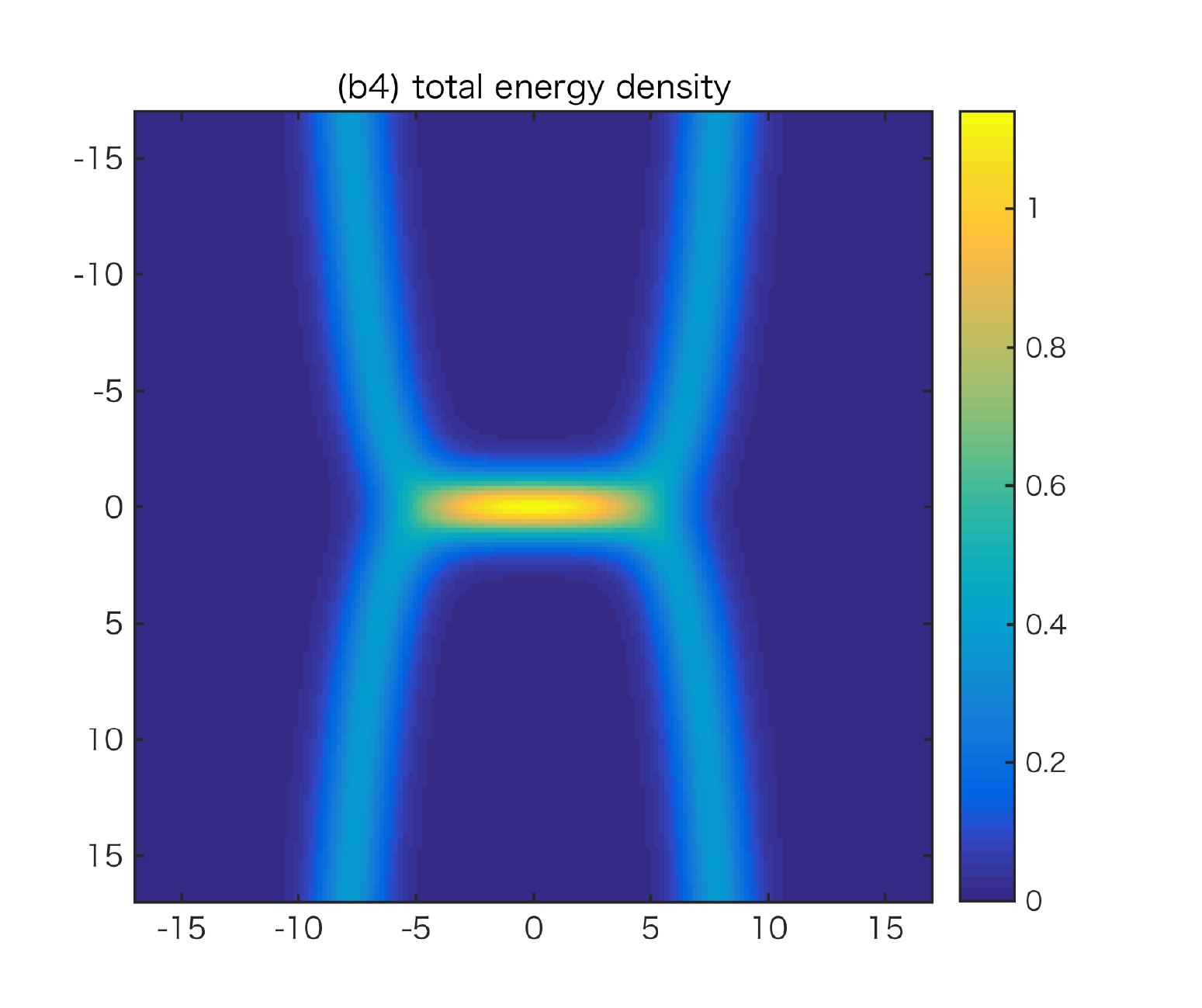}
  \end{center}
 \end{minipage}
  \caption{The plots show the energy density isosurfaces of one vortex stretching between two walls (a1, a2), where the blue and the red curves show magnetic fluxes, the wall energy density (b1), the vortex energy density (b2), the boojum energy density (b3) and the total energy density (b4) with the half distance between two vortices $R=5$.}
\label{fig:2_wall_1vor}
\end{figure}

In Fig.~\ref{fig:2_wall_1vor}, we show the numerical solution for $n_1=n_3=0$ and $n_2=1$ with
$\delta = e^5$ for the model with $\tilde m = 2$. Namely, we put a single vortex string in the middle
vacuum $\left<2\right>$.
Reasons for taking $\tilde m=2$ here is, first, that the mass of each domain wall is $\tilde m/2 = 1$,
and, second, that we have the exact solution $u_W$ to the
domain wall master equation in this case \cite{Sakai3}. Having the exact solution benefits our
numerical works. We will mention more about this in Sec.~\ref{sec:APP}.
The panel (b2) of Fig.~\ref{fig:2_wall_1vor} clearly shows that the vortex string of the finite length
$\ell = \frac{2}{\tilde m}\log \delta^2 \simeq 10$ exists in the middle vacuum. The string length $\ell$ is the 
same as the domain wall distance with no vortex strings.
The magnetic fluxes expressed by the blue curve in the panels (a1) and (a2) in the domain wall
at the negative side of $x^3$ are squeezed and run through the vortex string, and then expand 
inside the domain wall at the positive side of $x^3$. The endpoints of the vortex strings on the
domain walls are accompanied with the boojums as shown in the panel (b3).

\begin{figure}[h]
 \begin{minipage}{0.5\hsize}
  \begin{center}
   \includegraphics[width=8cm]{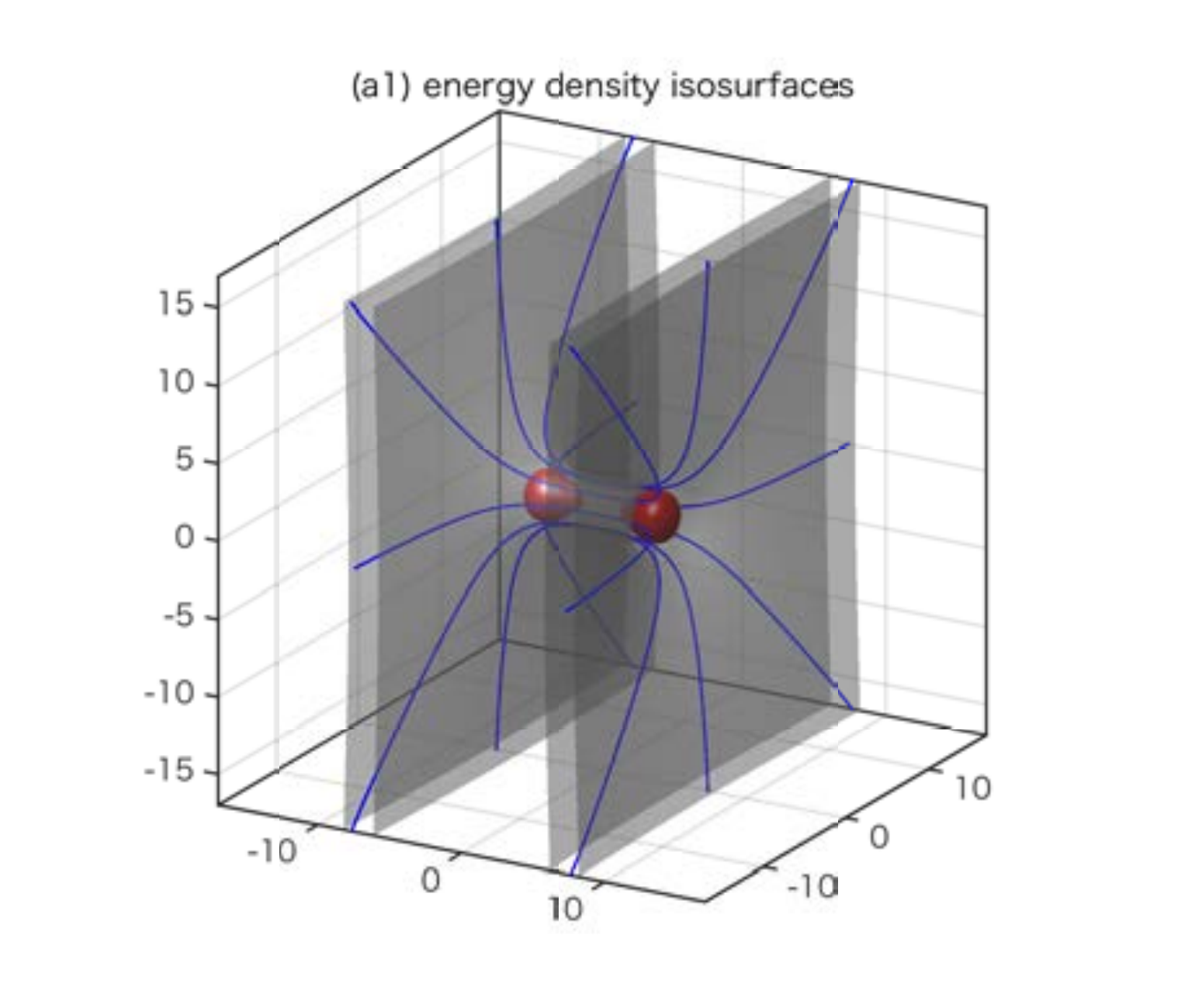}
  \end{center}
 \end{minipage}
  \begin{minipage}{0.5\hsize}
  \begin{center}
   \includegraphics[width=8cm]{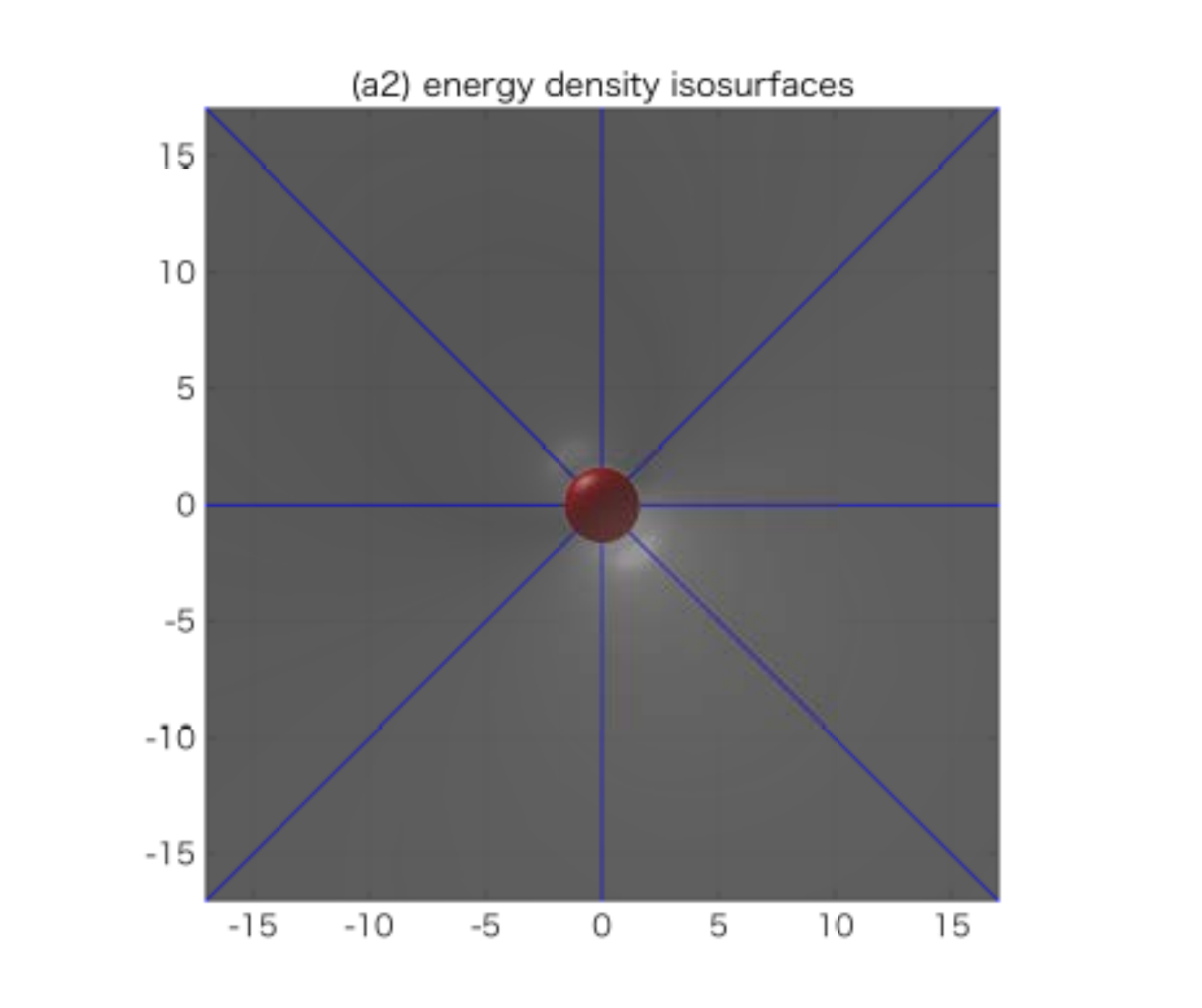}
  \end{center}
 \end{minipage}\\
 \begin{minipage}{0.5\hsize}
  \begin{center}
   \includegraphics[width=8cm]{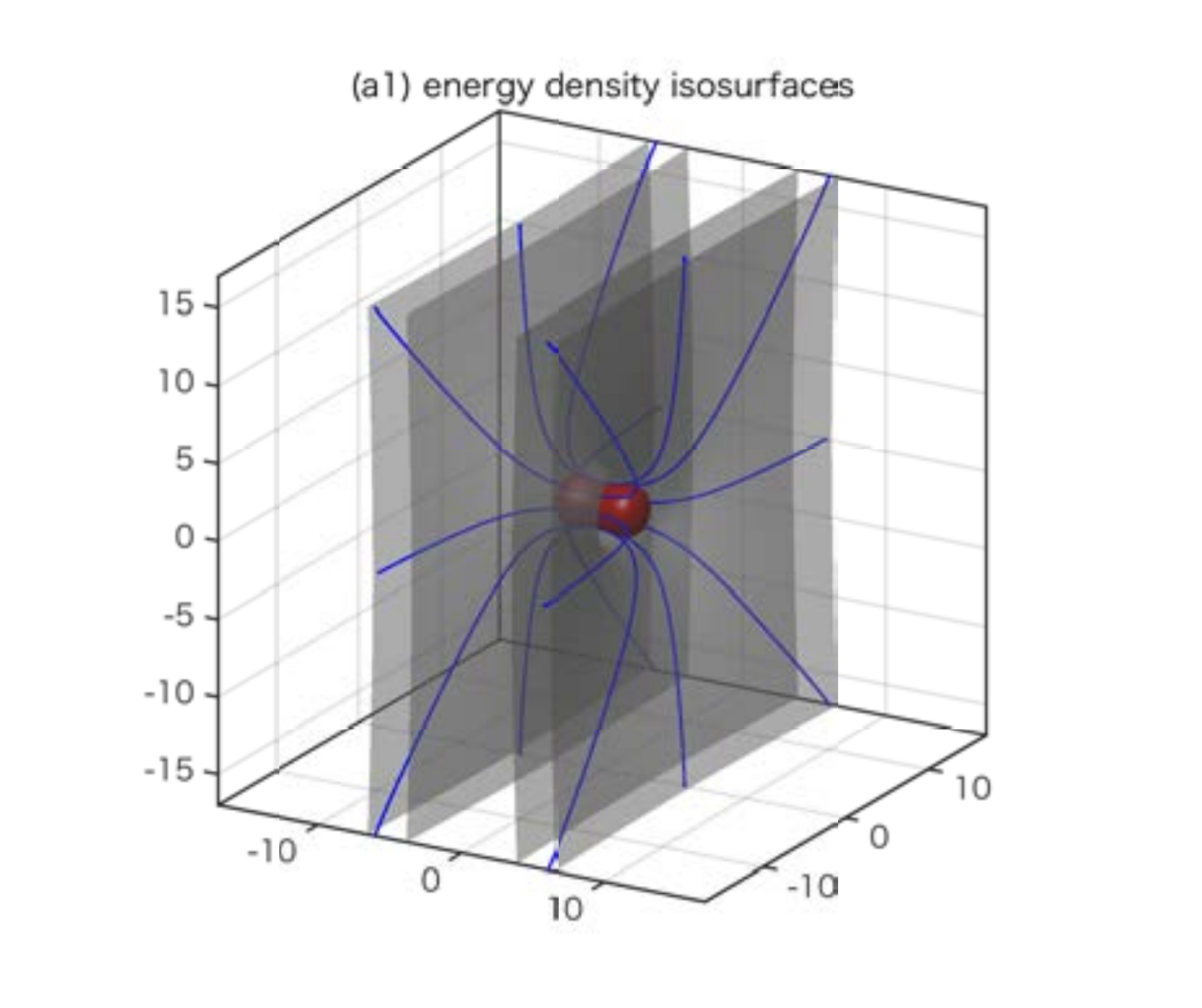}
  \end{center}
 \end{minipage}
 \begin{minipage}{0.5\hsize}
  \begin{center}
   \includegraphics[width=8cm]{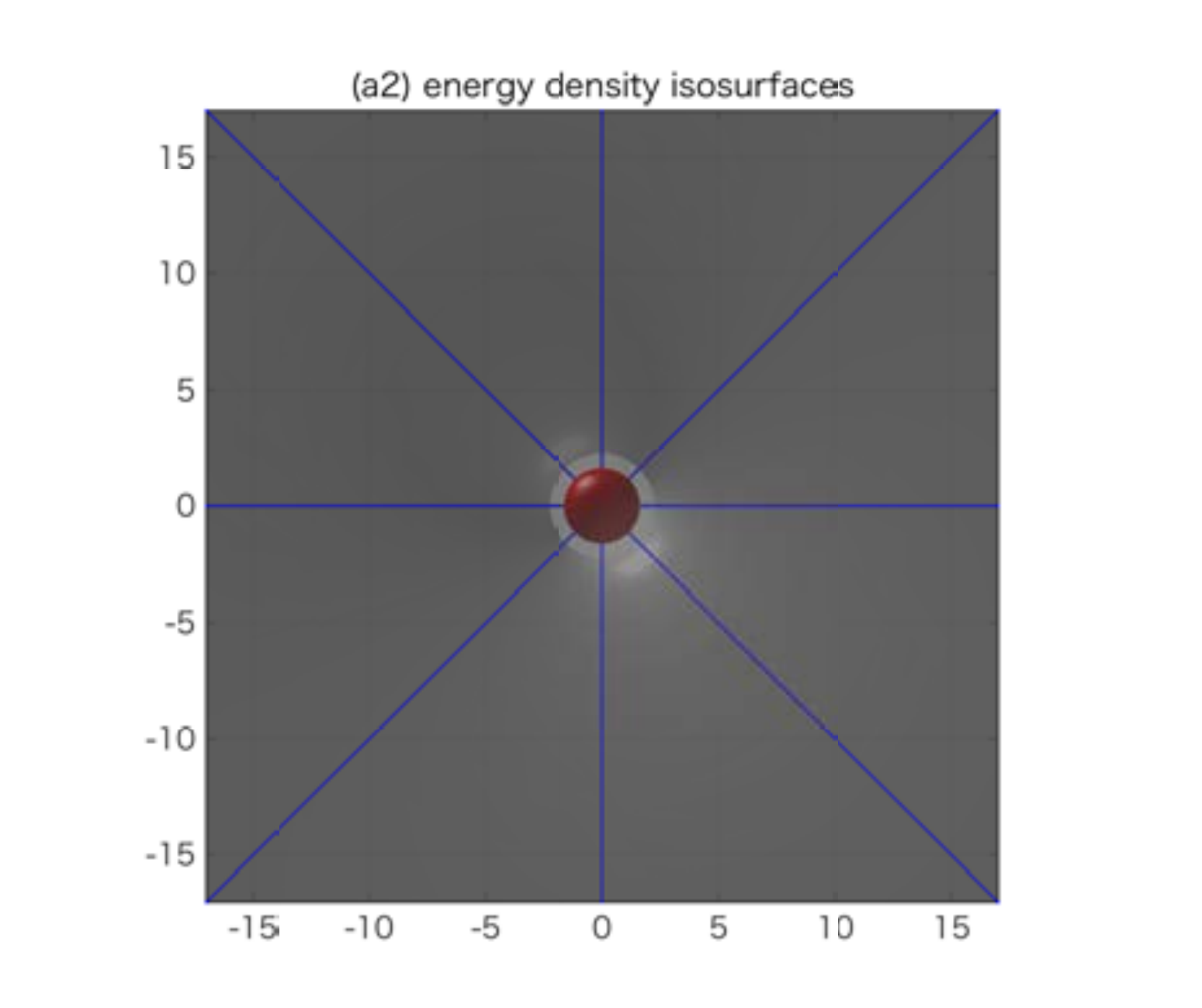}
  \end{center}
 \end{minipage}\\
  \begin{minipage}{0.5\hsize}
  \begin{center}
   \includegraphics[width=8cm]{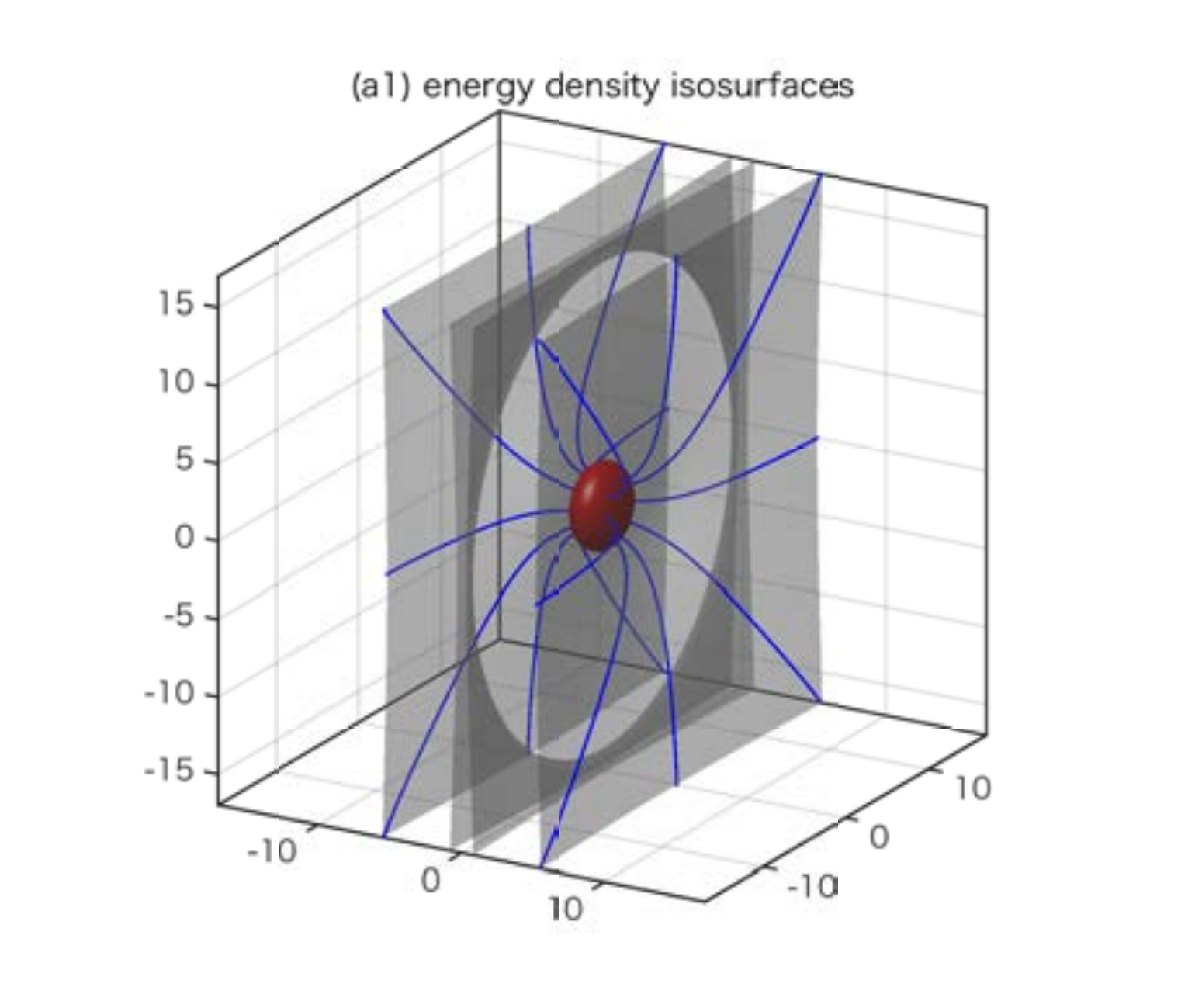}
  \end{center}
 \end{minipage}
 \begin{minipage}{0.5\hsize}
  \begin{center}
   \includegraphics[width=8cm]{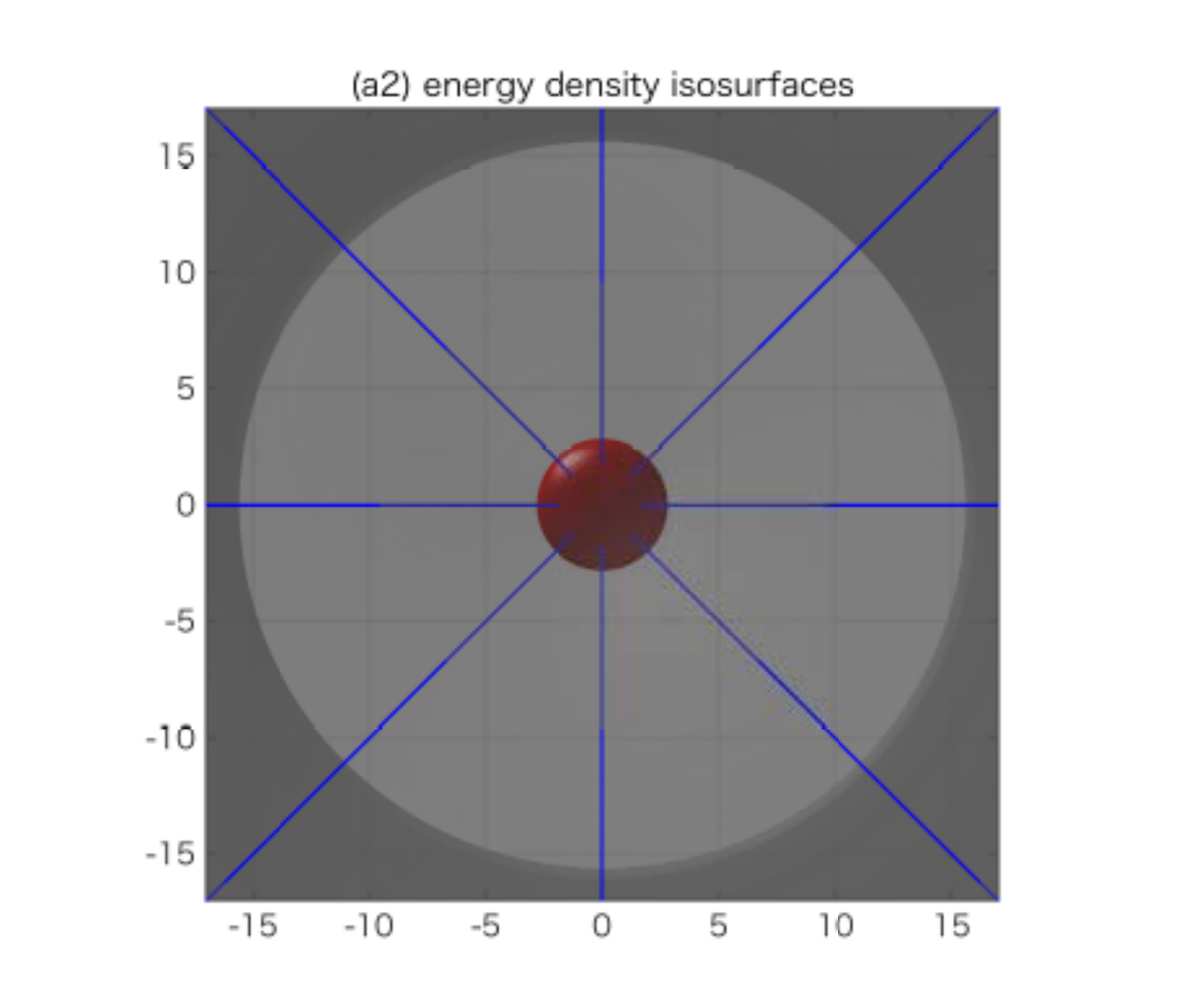}
  \end{center}
 \end{minipage}
\caption{The plots show the energy density isosurfaces of one vortex ending on two walls. 
The half distance between two vortices is taken to be $R=4, 2, 0$ from top to bottom.}
\label{fig:2_wall_1vor_B}
\end{figure}

\begin{figure}[h]
 \begin{minipage}{0.5\hsize}
  \begin{center}
   \includegraphics[width=7cm]{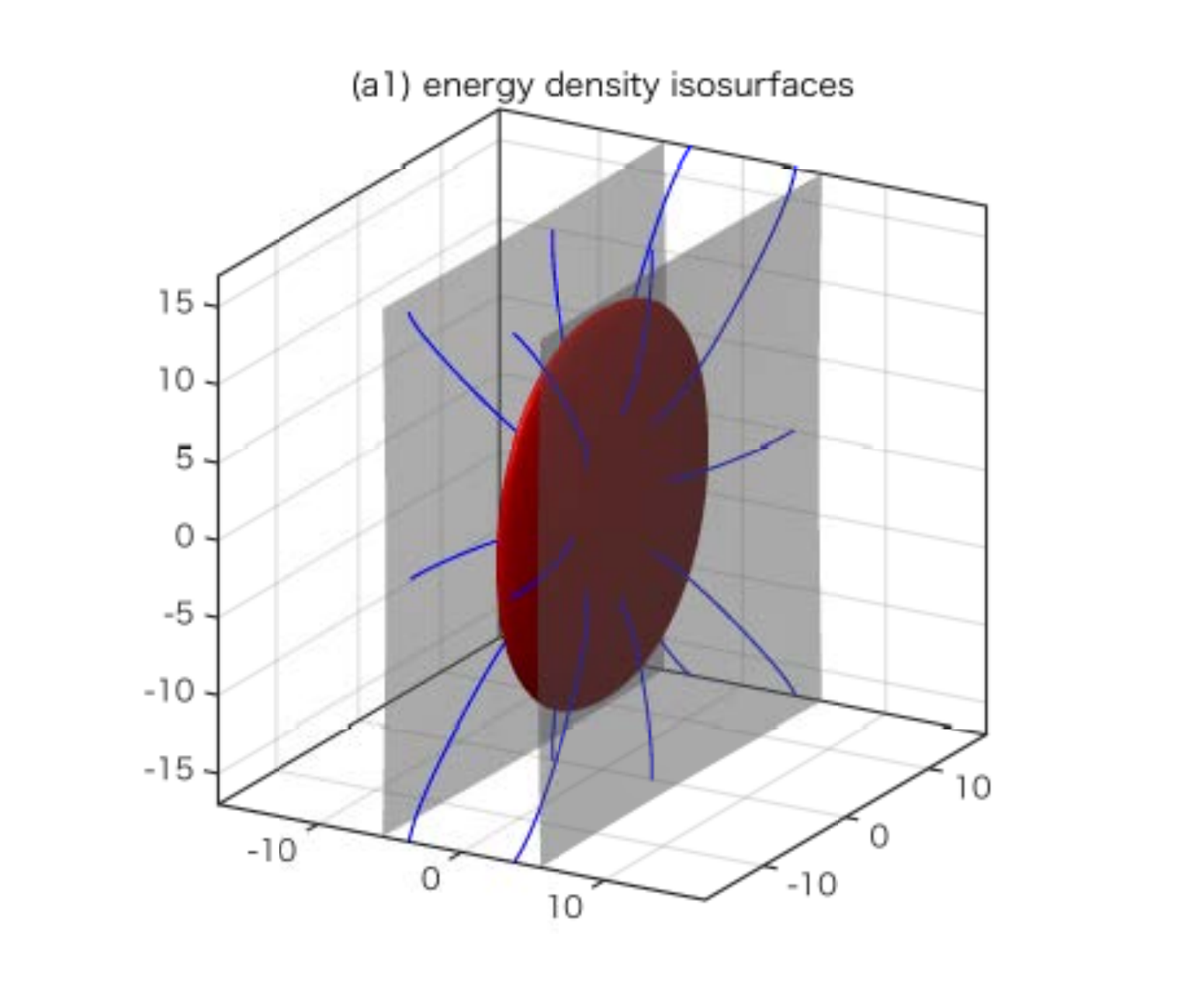}
  \end{center}
 \end{minipage}
  \begin{minipage}{0.5\hsize}
  \begin{center}
   \includegraphics[width=7cm]{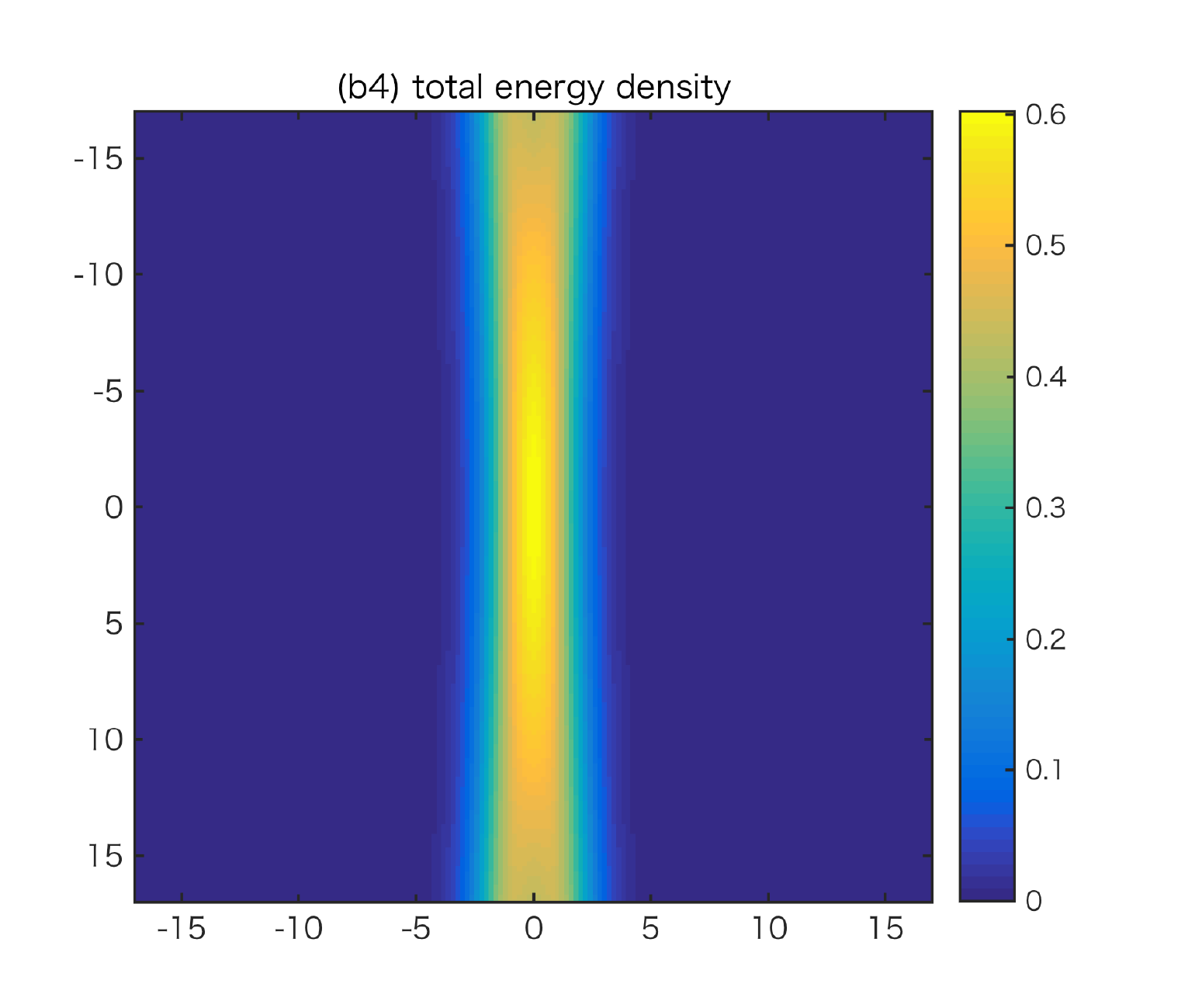}
  \end{center}
 \end{minipage}\\
 \begin{minipage}{0.5\hsize}
  \begin{center}
   \includegraphics[width=7cm]{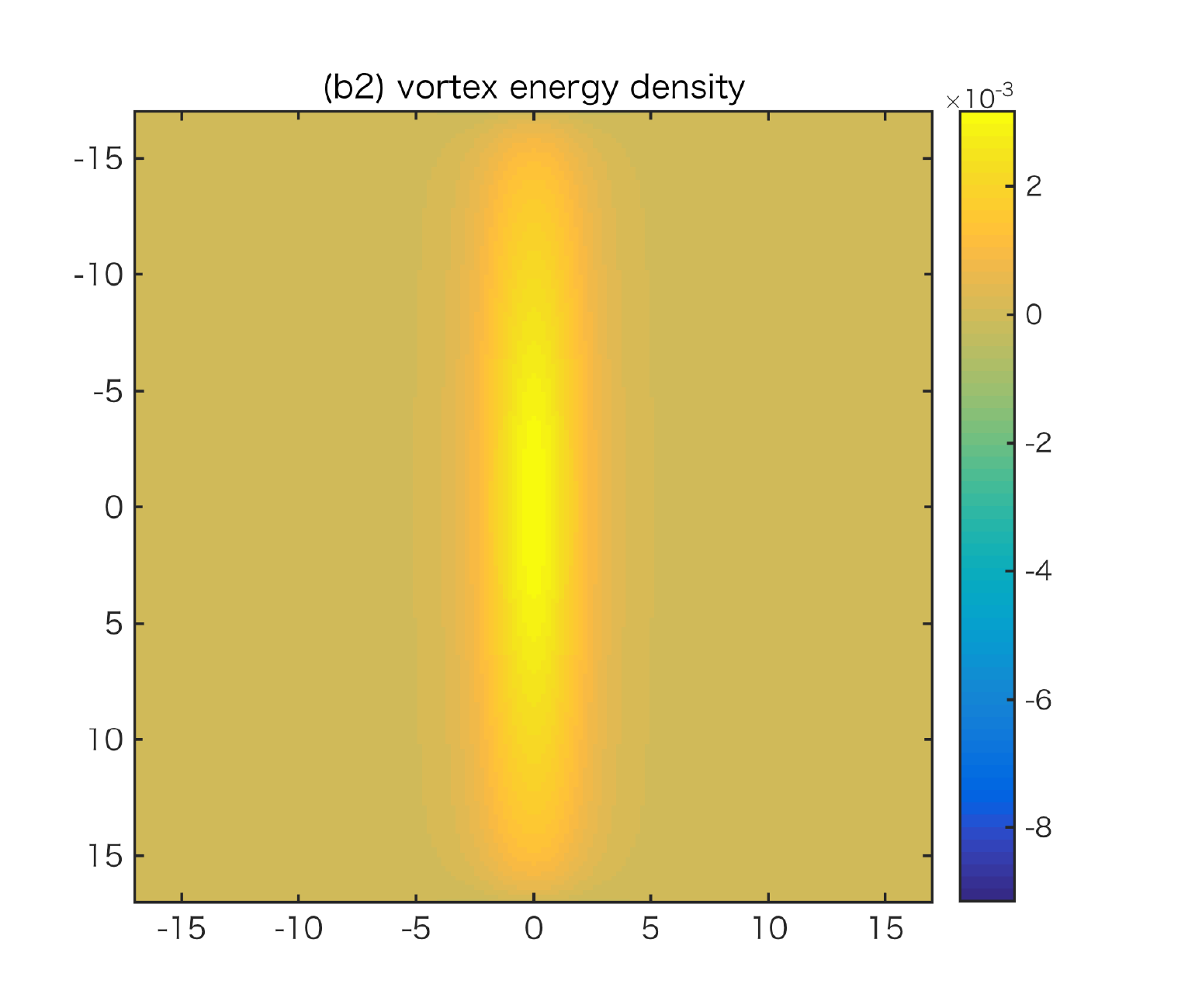}
  \end{center}
 \end{minipage}
 \begin{minipage}{0.5\hsize}
  \begin{center}
   \includegraphics[width=7cm]{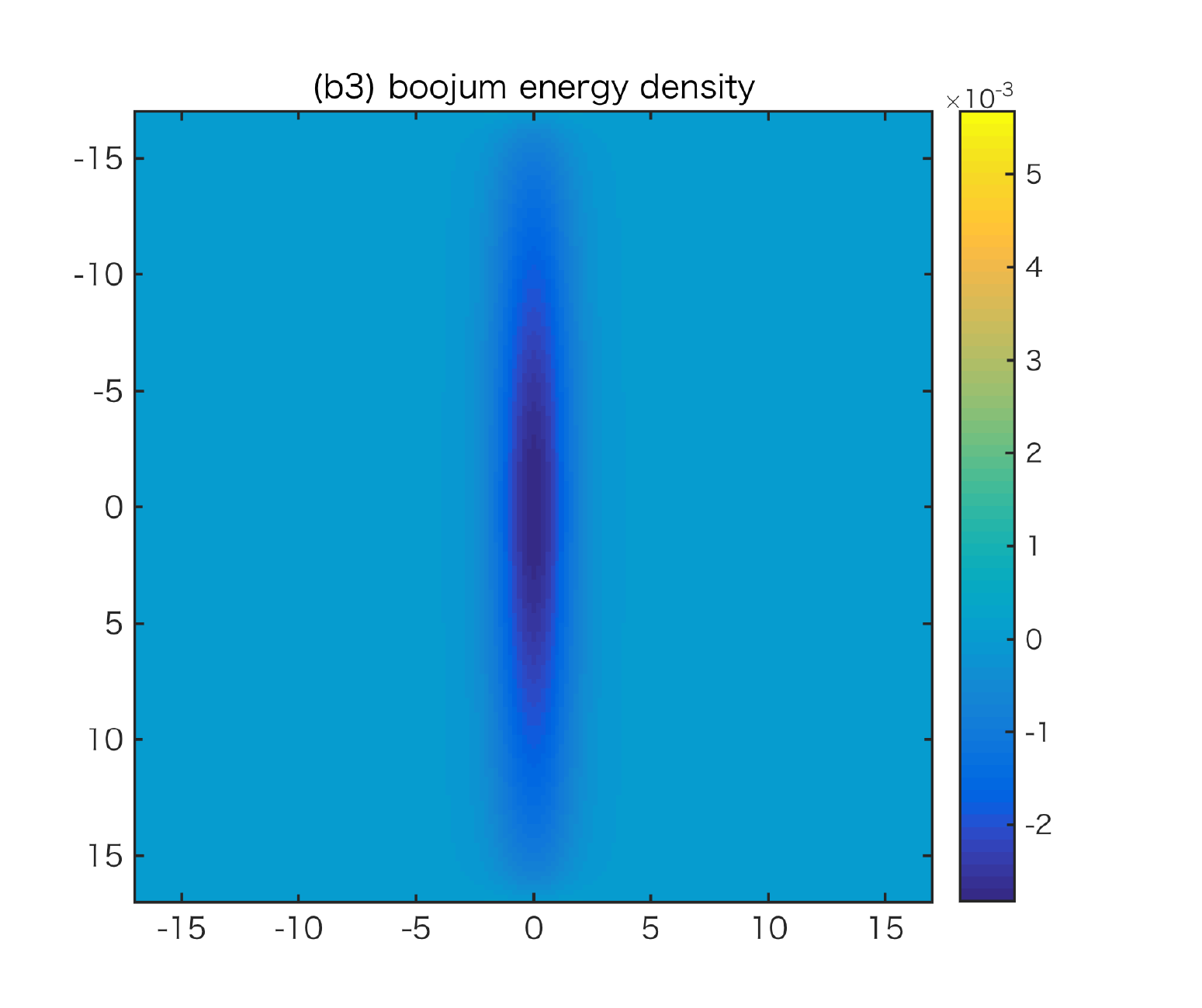}
  \end{center}
 \end{minipage}
\caption{The plots show the energy density isosurfaces of two vortices ending on one wall from two sides (a1), where the blue and the red curves show magnetic fluxes, the vortex energy density (b2), the boojum energy density (b3) and the total energy density (b4) with $\delta =e^{-2}$.}
\label{fig:2_wall_1vor_C}
\ \\\ \\
 \begin{minipage}{0.32\hsize}
  \begin{center}
   \includegraphics[width=5cm]{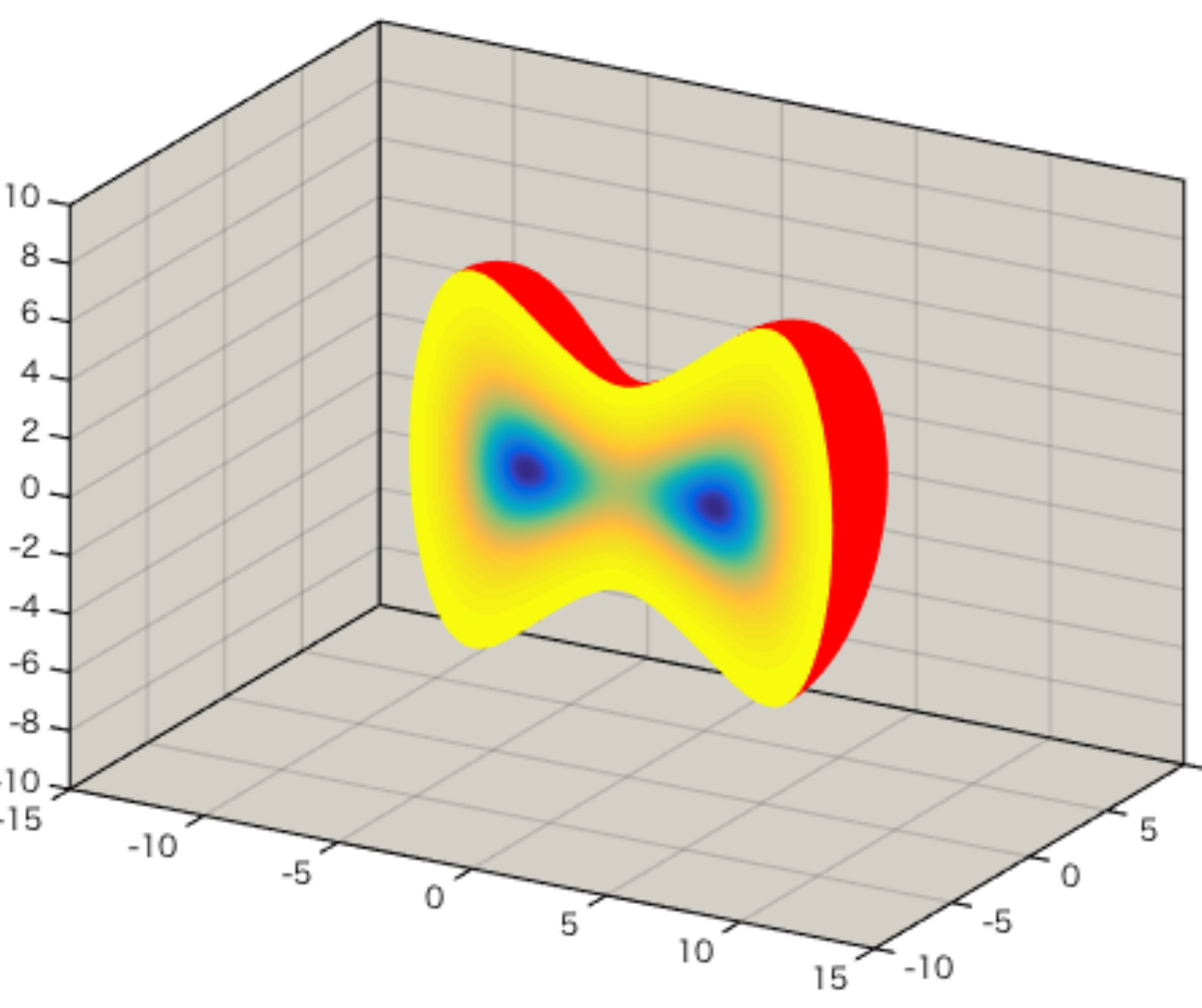}
  \end{center}
 \end{minipage}
  \begin{minipage}{0.32\hsize}
  \begin{center}
   \includegraphics[width=5cm]{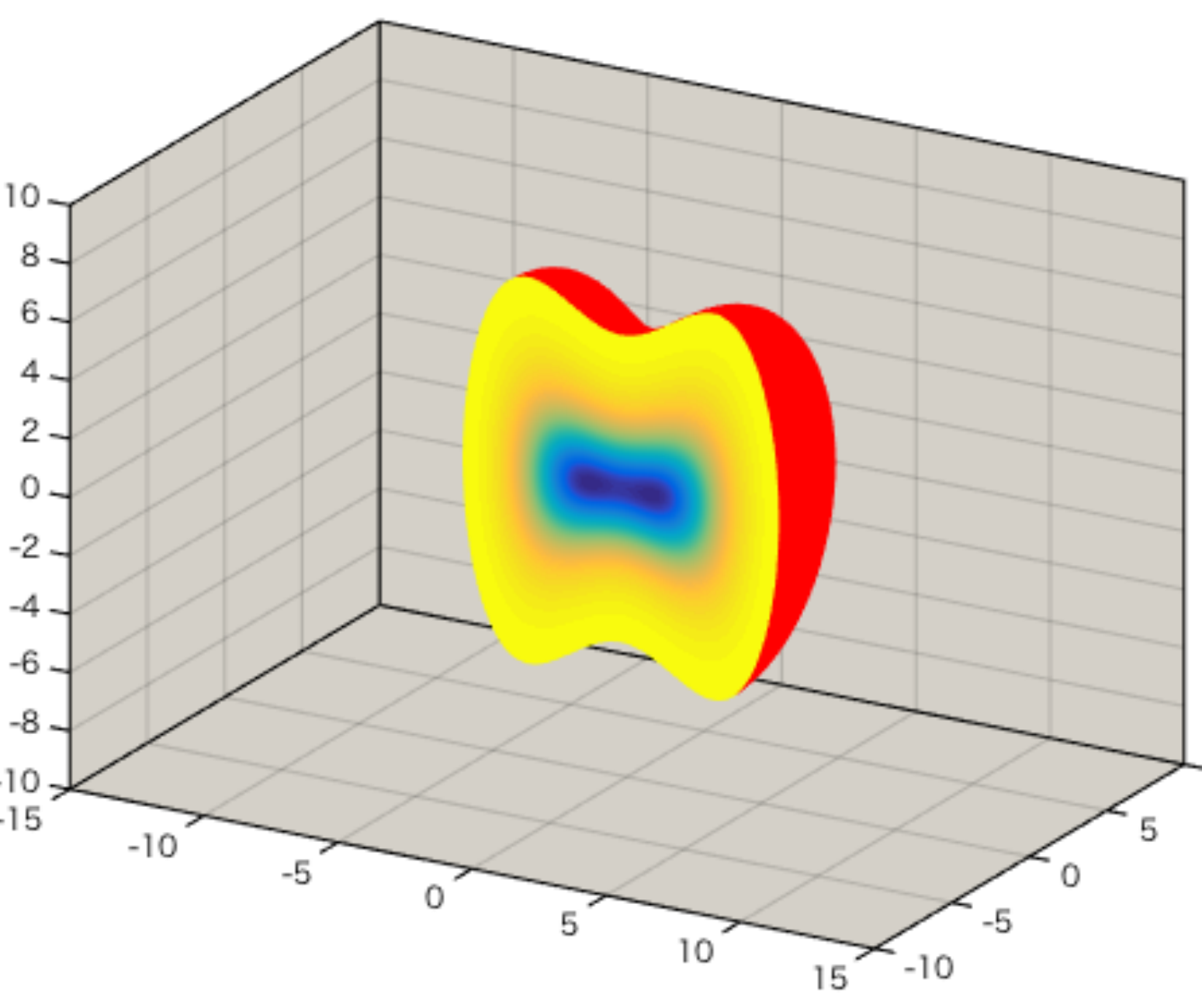}
  \end{center}
 \end{minipage}
 \begin{minipage}{0.32\hsize}
  \begin{center}
   \includegraphics[width=5cm]{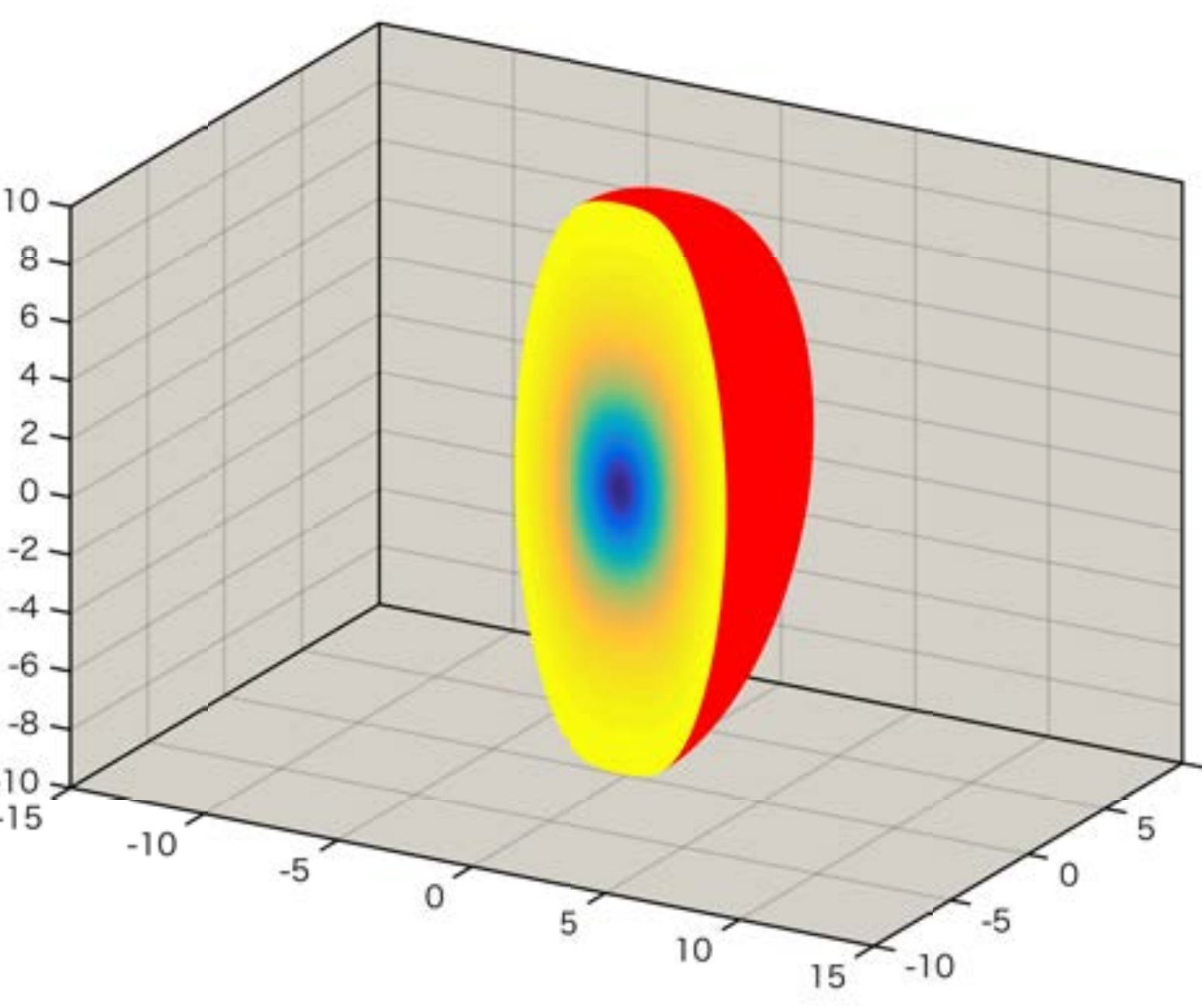}
  \end{center}
 \end{minipage}
\caption{The plots of the boojum energy density for the case that two vortices ending on one wall from one side (top), two vortices ending on one wall from two sides (middle) and one vortex ending on two walls (bottom). The half-distance of two boojums is $R=4, 2, 0$ from left to right.}
\label{fig:boojum_two_vor}
\end{figure}

As $\delta$ gets small, the domain walls get close. At the same time, the
distance between the boojums becomes smaller and the vortex string gets shorter.
We show the solutions for $\delta = e^{4,2,0}$ in Fig.~\ref{fig:2_wall_1vor_B}.
When the domain walls start to merge it is difficult to
find a clear boundary between them. In this situation, it is difficult to speak about the length of the string. 
However, the diameter of the string becomes very wide, while the magnitude of the magnetic flux weakens, see Fig.~\ref{fig:2_wall_1vor_C} for
$\delta = e^{-2}$. The panels (b2) and (b3) in Fig.~\ref{fig:2_wall_1vor_C} show the vortex and boojum charges and we see that 
the vortex is indeed sandwiched by the domain walls. Ultimately, at $\delta=0$, the middle vacuum completely disappears, so that the vortex vanishes as well.
Fig.~\ref{fig:boojum_two_vor} shows changes in the shape of two boojums living in the different domain walls.
As the domain walls close on each other, the boojums merge into a flat disk.

\clearpage 

\clearpage

\clearpage

\clearpage

\begin{figure}[t]
\begin{center}
\includegraphics[width=10cm]{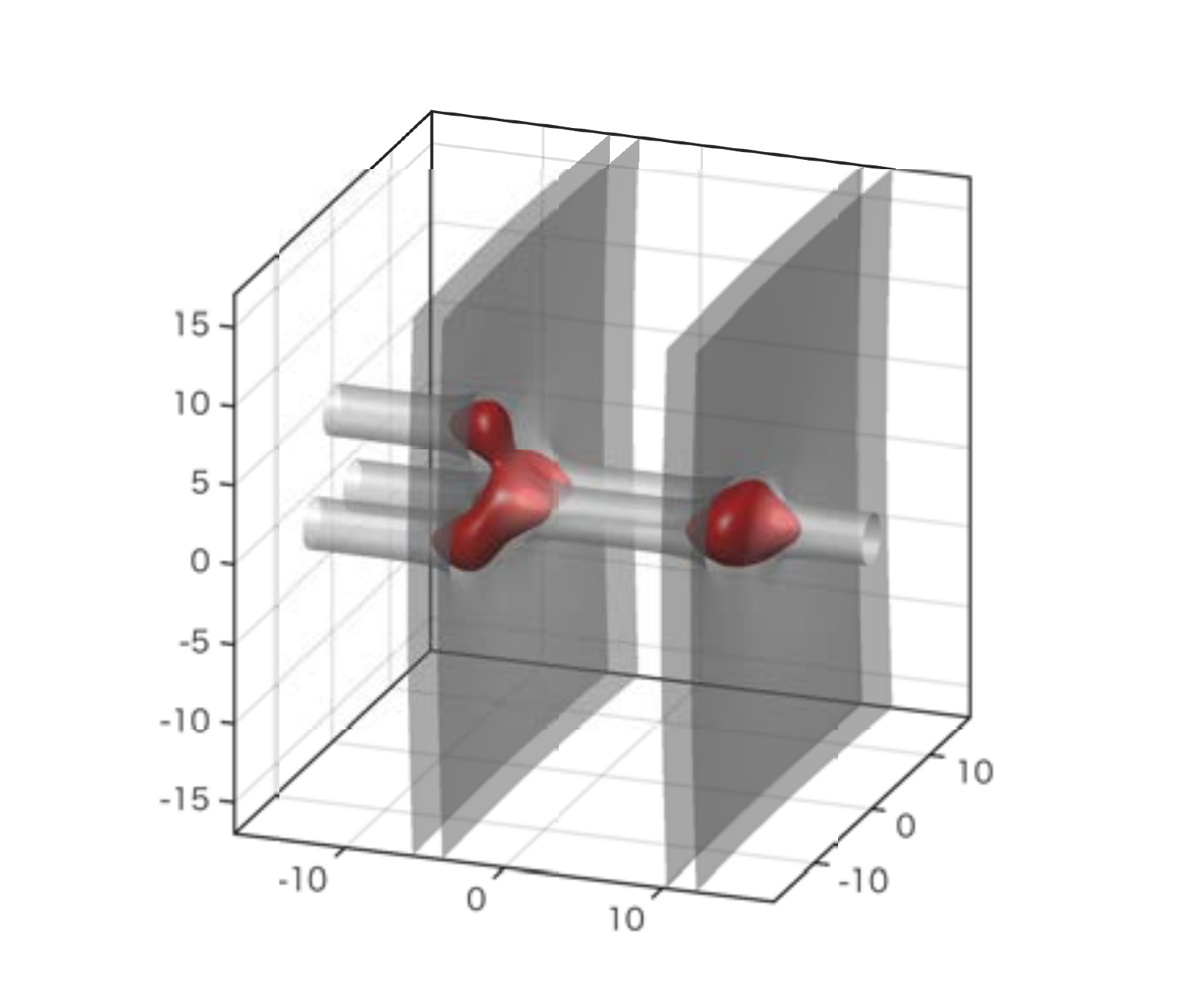}
\caption{The two logarithmically bent but parallel domain walls in $N_F=3$ model.
The number of the vortices are 3, 2 and 1 in the vacua $\left<1\right>$, $\left<2\right>$ and $\left<3\right>$, respectively.}
\label{fig:2w_123vor}
\end{center}
\end{figure}

As an example for more complicated configuration, in Fig.~\ref{fig:2w_123vor}, 
we show the solution which have single, double and triple
vortex strings in the vacuum $\left<1\right>$, $\left<2\right>$ and $\left<3\right>$, respectively.
The number of vortex strings satisfies the relation $2n_2 = n_1+n_3$, so that the domain walls are logarithmically
bending but are asymptotically parallel.

\subsection{Partially degenerate masses }

\begin{figure}[t]
\begin{center}
\includegraphics[height=6cm]{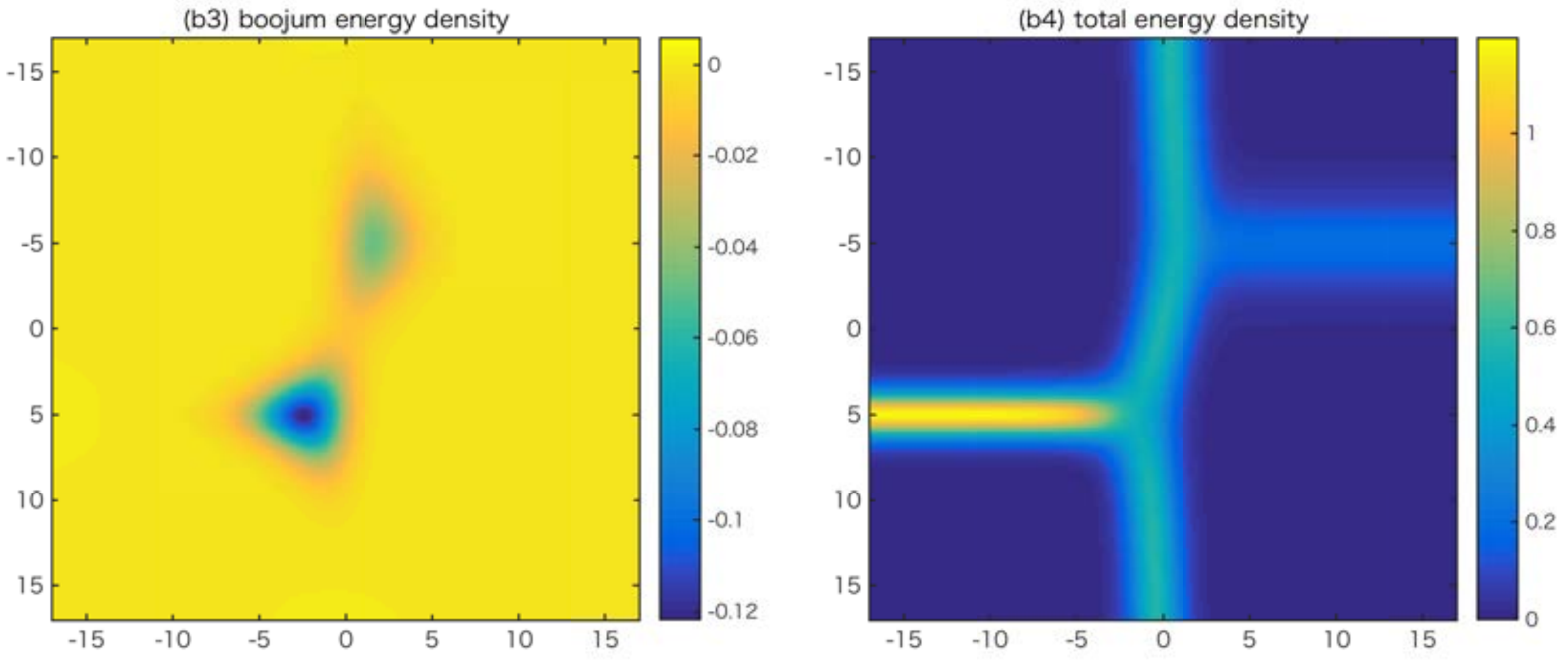}
\caption{The semi-local vortex string with $a=3$
at $z=-5$ ending of the domain wall from the $x^3$ positive side and
the local vortex string at $z=5$ ending on the domain wall from the $x^3$ negative side.
The panel (b3) shows the boojum charge density and (b4) shows the total energy density on the $x^2=0$ plane.}
\label{fig:s_sls_a3_r5}
\end{center}
\end{figure}

When some masses are partially degenerate, the corresponding vacua are degenerate. Then
the vortex strings there become semi-local vortex strings, which we discuss in the second paper \cite{Boojum2}.
As before, we consider
the minimal model $N_F =3$
with $\tilde M = {\rm diag}(\tilde m/2,\tilde m/2,-\tilde m/2)$.
The relevant moduli matrix is given by
\begin{eqnarray}
H_0 = (P_{n_1}(z),\ Q_{n_1-1}(z),\ P_{n_2}(z)),
\end{eqnarray}
where $P_n(z)$ is a monomial of power $n$ and $Q_n(z)$ stands for a polynomial of power $n$.
The $n_1$ semi-local vortex strings in the degenerate vacuum $\left<1\right>$ 
are determined by $P_{n_1}$ and $Q_{n_1-1}$, while the $n_2$ local vortex strings 
in the vacuum $\left<2\right>$ are determined by $P_{n_2}$.

The gradient flow equation to be solved is
\begin{eqnarray}
\p_k^2 U - 1 + \left((|P_{n_1}|^2 + |Q_{n_1-1}|^2) e^{\tilde m x^3}  + |P_{n_2}|^2 e^{-\tilde mx^3}\right) e^{-U} = \p_t U.
\end{eqnarray}
This is identical to Eq.~(\ref{eq:gf_two_vor}) if we replace $|P_{n_1}|^2 \to |P_{n_1}|^2 + |Q_{n_1-1}|^2$. Thus, a suitable
initial configuration is given by
\begin{eqnarray}
{\mathcal U}(x^k) = u_W\left(x^3+\frac{u_{SLS}^{(n_1)}-u_S^{(n_2)}}{2\tilde m}\right)+\frac{u_{SLS}^{(n_1)}+u_S^{(n_2)}}{2},
\end{eqnarray}
where $u_{SLS}^{(n_1)}$ stands for a solution to the master equation for the semi-local vortex string
\begin{eqnarray}
\p_a^2 u_{SLS} - 1 + (|P_{n_1}|^2 + |Q_{n_1-1}|^2) e^{-u_{SLS}} = 0.
\end{eqnarray}
In Fig.~\ref{fig:s_sls_a3_r5}, we show a solution for a generic choice
\begin{eqnarray}
P_{n_1=1} = z + L,\ 
Q_{n_1-1=0} = a,\ 
P_{n_2=1} = z - L,
\label{eq:PQP}
\end{eqnarray}
with $a = 3$ and $L=5$.

\begin{figure}[h]
 \begin{minipage}{0.5\hsize}
  \begin{center}
   \includegraphics[width=6.5cm]{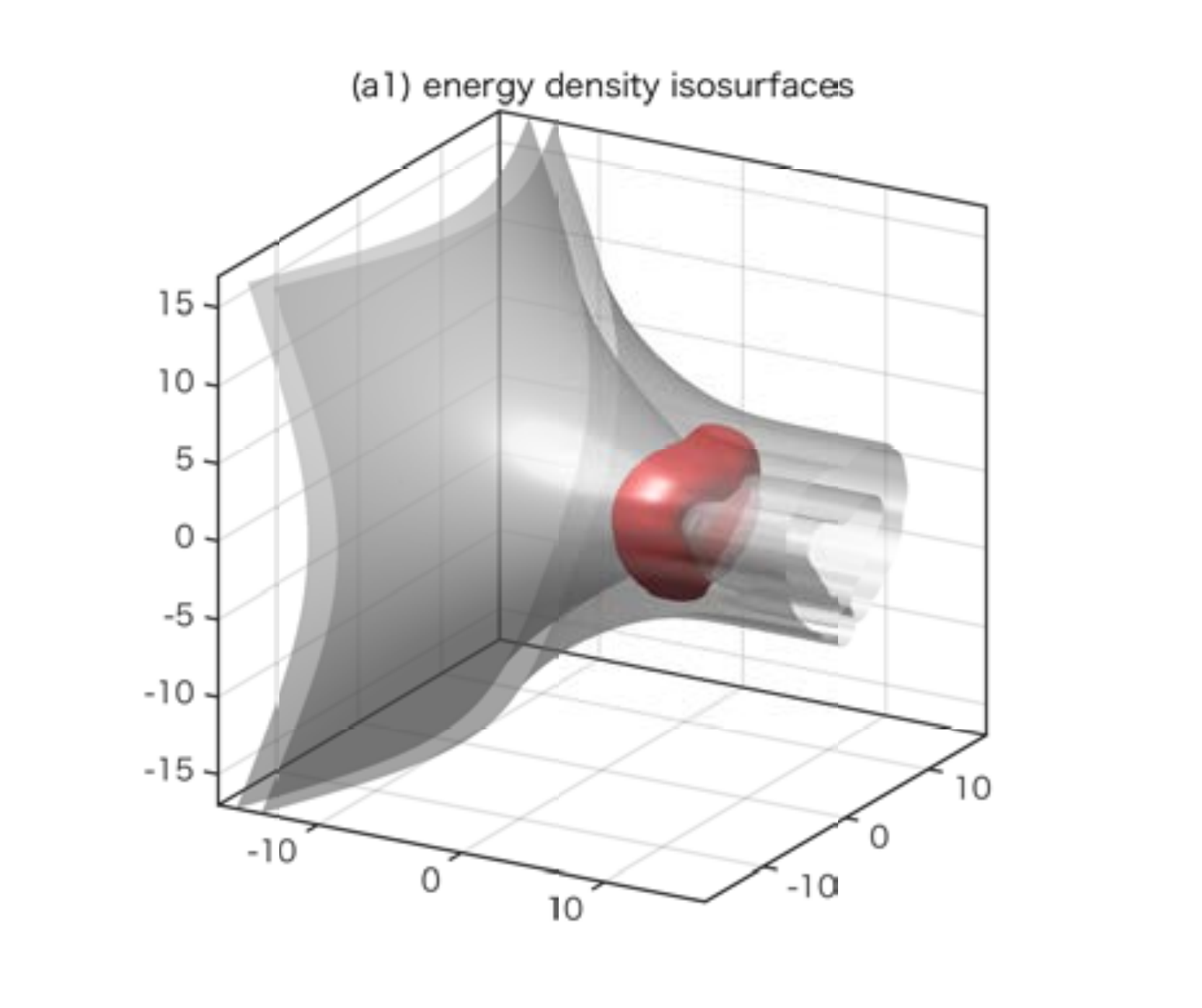}
  \end{center}
 \end{minipage}
  \begin{minipage}{0.5\hsize}
  \begin{center}
   \includegraphics[width=6.5cm]{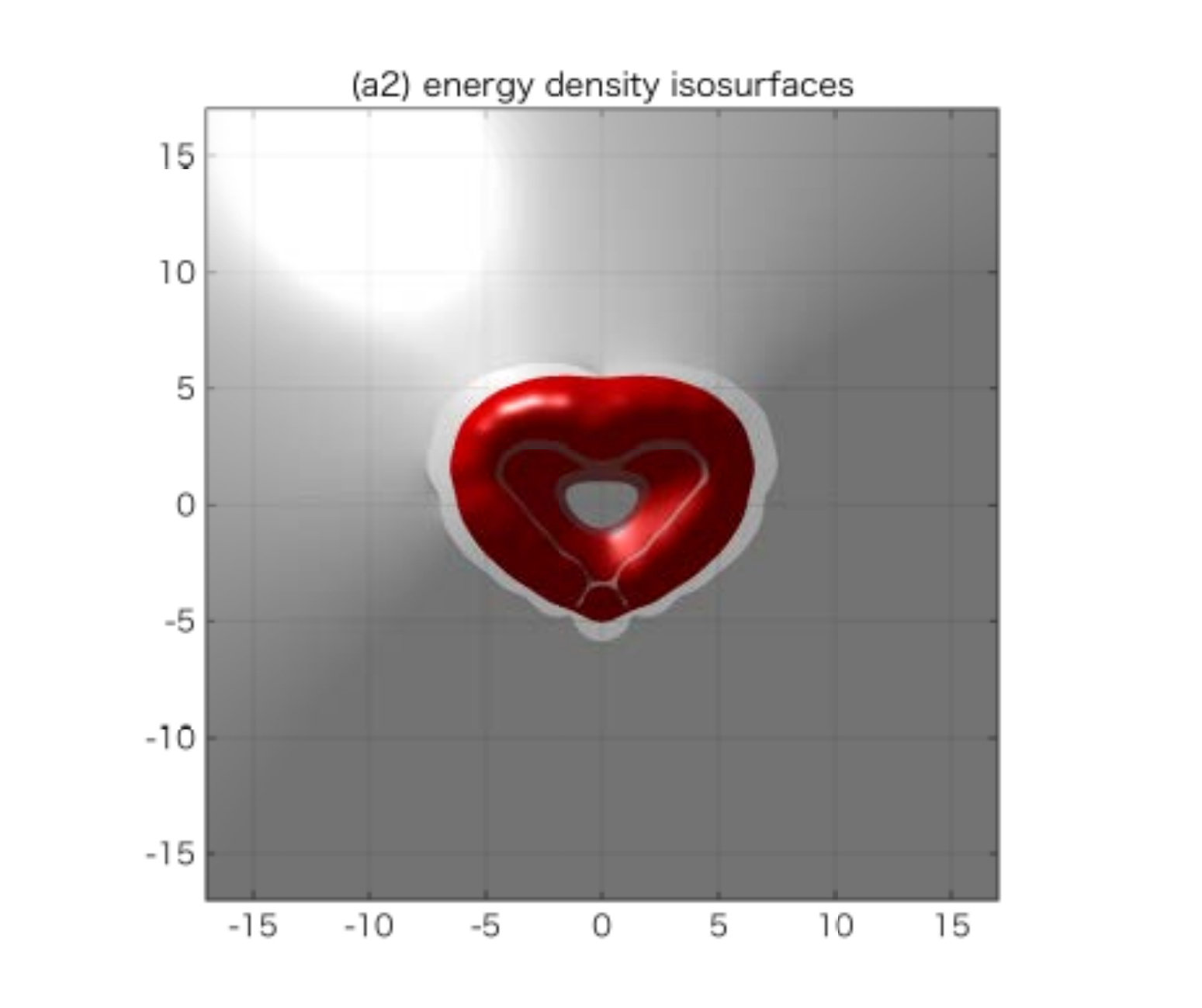}
  \end{center}
 \end{minipage}\\

 \begin{minipage}{0.5\hsize}
  \begin{center}
   \includegraphics[width=6.5cm]{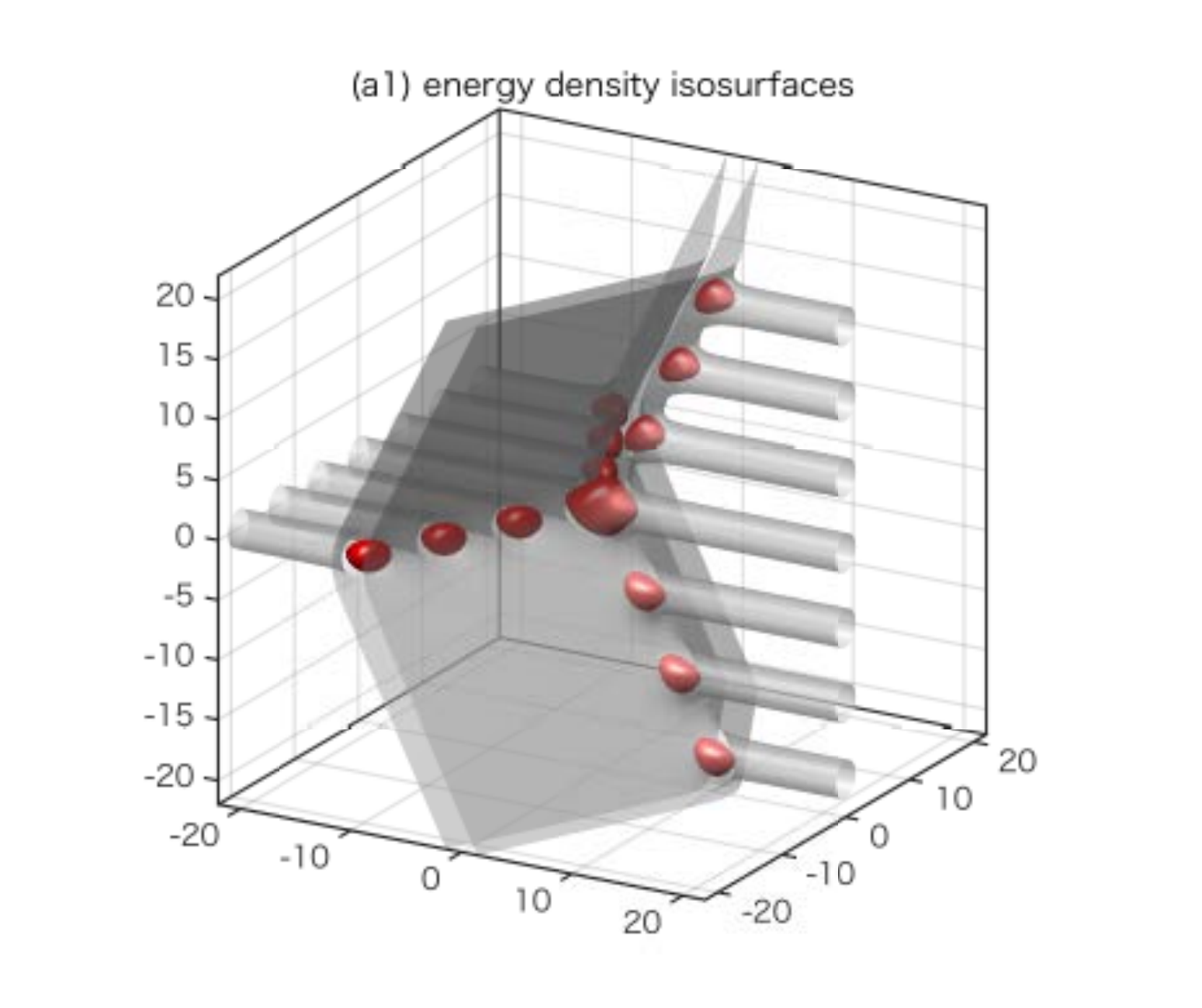}
  \end{center}
 \end{minipage}
  \begin{minipage}{0.5\hsize}
  \begin{center}
   \includegraphics[width=6.5cm]{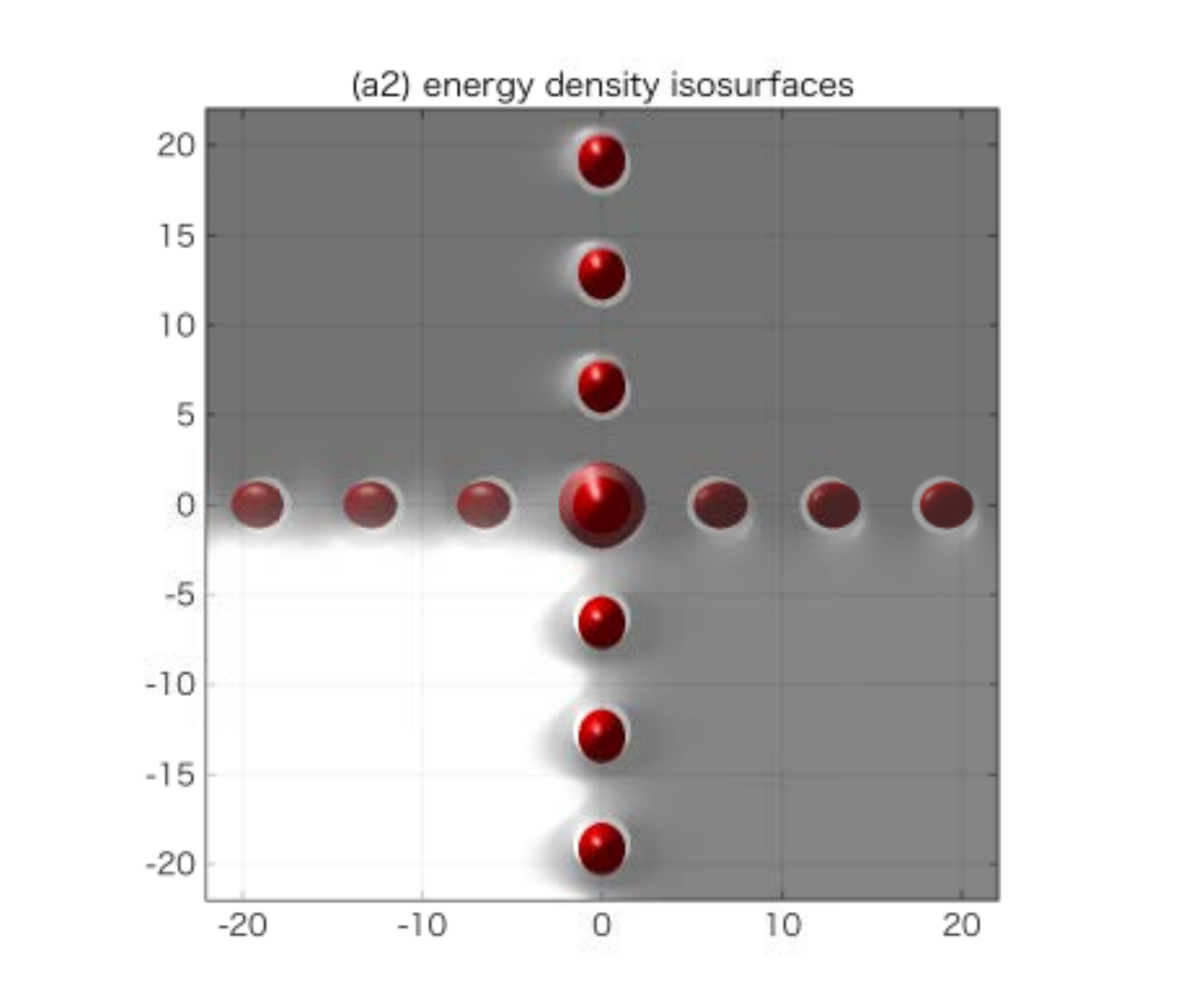}
  \end{center}
 \end{minipage}

 \begin{minipage}{0.5\hsize}
  \begin{center}
   \includegraphics[width=6.5cm]{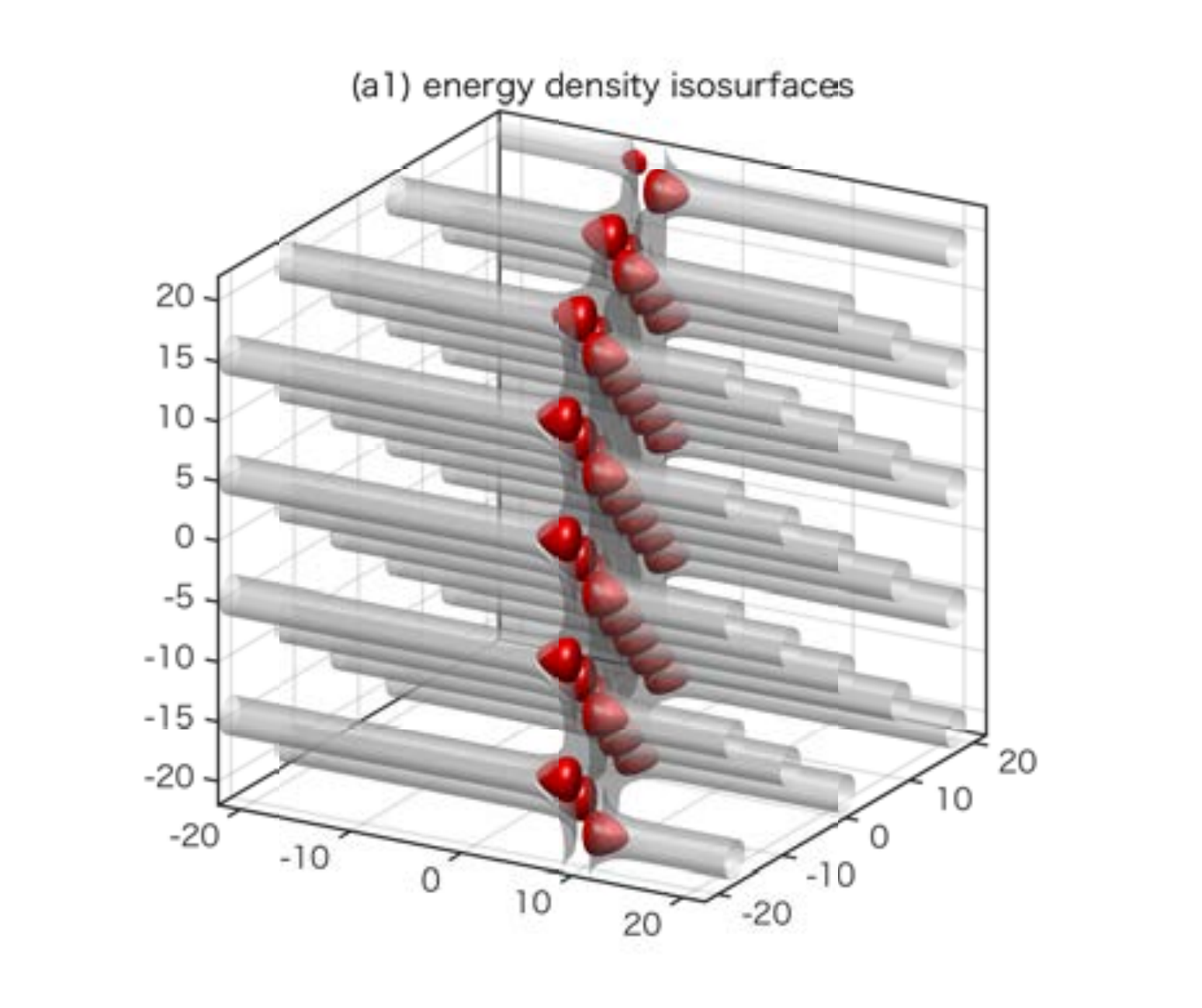}
  \end{center}
 \end{minipage}
  \begin{minipage}{0.5\hsize}
  \begin{center}
   \includegraphics[width=6.5cm]{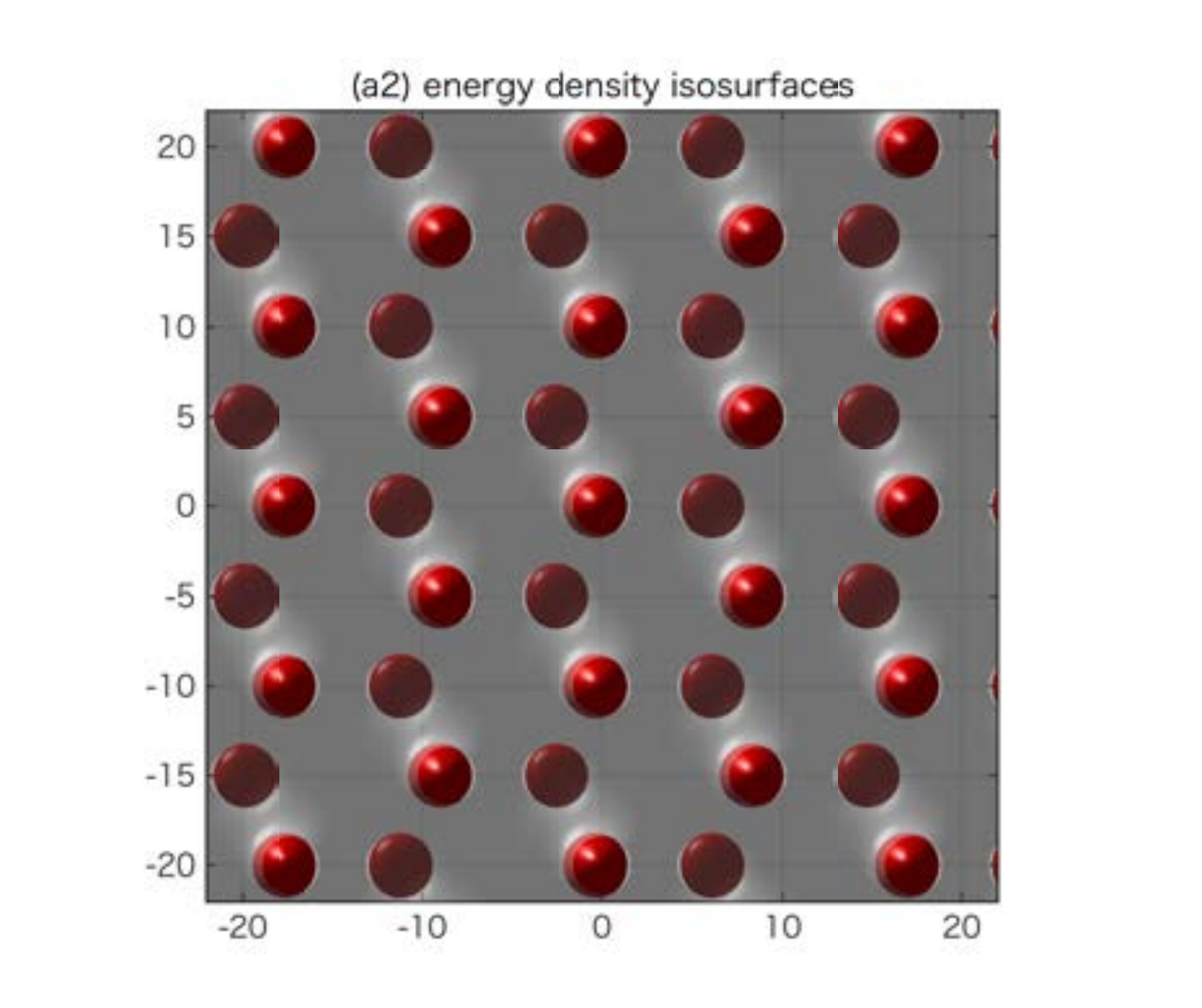}
  \end{center}
 \end{minipage}
 \caption{The plots show the energy density isosurfaces of various configurations of the vortices-wall-boojum system. The top plots show that the energy density isosurface of the boojum is a heart mark shape. The middle plots show that the energy density isosurface of periodic vortices on cross lines ending on one wall from two sides. The bottom plots show that the energy density isosurface of the so-called Abrikosov lattices ending on one wall from two sides.}
\label{fig:funny}
\end{figure}

For closing this section, we show several funny solutions in order to demonstrate that
we can have any kind of numerical configurations, see Fig.~\ref{fig:funny}.

\clearpage

\clearpage


\section{Analytic approximations to 1/2 BPS solitons}\label{sec:APP2}

The 1/2 BPS pieces entering the global approximations of 1/4 BPS solutions summarized in Sec.~\ref{sec:APP} are solutions to single domain wall $u_W(x^3)$, double domain wall $u_{W}(x^3;\delta)$, vortex strings $u_S^{(n)}$ and semi-local vortex strings $u_{SLS}^{(n)}$. These pieces can be supplied as numerical solutions of corresponding 1/2 BPS master equations, which are generally much easier to solve than the full three-dimensional 1/4 BPS master equations.
A complementary way is to obtain \emph{analytic} approximations to 1/2 BPS solitons. This is the goal of this Appendix. In the following subsections, we will develop accurate analytic approximations to both vortex string and domain wall. 

Especially accurate approximations can be found for vortex string master equation Eq.~\refer{eq:appstring}, since there is no intrinsic mass scale as in the case of domain wall master equation Eq.~\refer{eq:appwall}. This allows easy use of the so-called global Pad\'e method, which we describe in the next subsection.

\subsection{Approximations to ANO vortex}

The master equation for single vortex string reads
\begin{equation}\label{eq:vortexapp}
\partial_{\rho}^2 u_S+\frac{1}{\rho}\partial_{\rho}u_S = 1-\rho^2 e^{-u_S}\,.
\end{equation}

No analytic solution of Eq.~\refer{eq:vortexapp} is known.
In order to construct global approximations, we take the advantage from what we can learn about $u_S$ locally. In particular, let us investigate behaviour of $u_S$ close to $\rho= 0$ and for $\rho \gg 1$. Expanding $u_S$ in Taylor series around origin and matching coefficients on both sides of Eq.~\refer{eq:vortexapp} we obtain
\begin{equation}\label{eq:taylors}
u_S(\rho) = u_0+\frac{1}{4}\rho^2-\frac{e^{-u_0}}{16}\rho^4+\frac{e^{-u_0}}{144}\rho^6+O(\rho^8)\,,
\end{equation} 
where $u_0\equiv u_S(0)$. Note that have used the regularity condition $u_S^{\prime} (0) = 0$.
On the other hand, the asymptotic behaviour of $u_S$ is
\begin{equation}\label{eq:asympt}
u_S(\rho) \sim 2\log(\rho)+q K_0(\rho) \sim \log\bigl(\rho^2+q \rho^{3/2}e^{-\rho}\bigr) \hspace{3mm} \mbox{as } \rho \to \infty\,.
\end{equation}
Here, $K_0(\rho)$ is the modified Bessel function and $q>0$ is some constant. 

Both $u_0$ and $q$ are locally undetermined; their values cannot be ascertained neither by Taylor series nor by asymptotic series. In other words, these numbers characterize global properties of $u_S$ and as such, they are very hard to study analytically. Recently, a detailed study of Eq.~\refer{eq:vortexapp} was published \cite{Ohashi}, where based on a perturbative expansion around a small winding number, the numbers $u_0$ and $q$ ($2D_1$ and $2C_1$ in their notation) were obtained with high precision, both analytically and numerically. In the following, we shall adopt the values $u_0 = 1.01072165$ and $q = 3.41572835$ (Eq.~(2.66) in \cite{Ohashi}) in all our approximations.

A (global) Pad\'e approximant  is a widely used tool to model a function based on its series expansion(s). The basic idea is to find a rational function which, when expanded, match all series to given orders. Generally, such approximant tends to be very accurate even outside radii of convergence of individual series and for sufficiently nice functions it remains close to the true solution everywhere. 

Nevertheless, there is only so much a rational function can do. In particular, as we see from Eq.~\refer{eq:asympt}, the asymptotic formula for $u_S$ contains log, exp and square root (suggesting essential singularity at infinity), neither of which can be globally approximated by rational functions. Therefore, simple Pad\'e approximants are insufficient in this case. 

A way to overcome these difficulties, originally discussed in \cite{Winitzki}, is to put the Pad\'e approximant directly into the asymptotic series, in such a way that singular terms are replaced by rational functions, which coefficients are fixed to match Taylor series at the origin. In this way, the resulting function is guaranteed to behave in a desired way both at the origin and at infinity.

Let us, therefore, propose the following ansatz:
\begin{equation}
u_{(p+3,p)} = \log\Bigl(\rho^2+\, e^{-\rho}\sqrt{\frac{P_{p+3}(\rho)}{Q_p(\rho)}}\Bigr)\,,
\end{equation}
where $P_{p+3}(\rho)$ and $Q_{p}(\rho)$ are polynomials of order $p+3$ and $p$ respectively. The coefficients are chosen such that they give correct large $\rho$ behaviour
\begin{equation}
\frac{P_{p+3}(\rho)}{Q_p(\rho)} \xrightarrow[\rho\to \infty]{} q^2\rho^3\,,
\end{equation} 
and the remaining ones are fixed to match the Taylor series \refer{eq:taylors} up to $O(\rho^{2p+3})$. 
After this procedure is done, it needs to be checked that both $P_{p+3}(\rho)$ and $Q_p(\rho)$ are strictly positive for $\rho \geq 0$, as we require that $u_{(p+3,p)}$ is regular on the positive semi-axis. It turns out that this is satisfied only up to $p = 4$ and for higher $p$ singularities appear.  

We have found the following Pad\'e approximants
{\small
\begin{align}
P_3(\rho)  = & 7.54921285+15.0984 \rho +13.37786594 \rho^2 +11.66720016 \rho^3\,, \\
P_4(\rho)  = & 7.54921285 +17.82022676 \rho + 18.82146806 \rho^2 + 11.44776733 \rho ^3 \nonumber \\
 & +4.20650447 \rho^4\,, \\
P_5(\rho) = & 7.54921285 + 14.95444855 \rho +13.29529829 \rho^2 + 6.7801308\rho^3 \nonumber \\
& +2.05572318\rho^4 + 0.31742204 \rho^5\,, \\
P_6(\rho) = & 7.54921285+16.30163555 \rho + 16.01574815\rho^2 + 9.27493053\rho^3 \nonumber \\
& +3.39472576\rho^4+0.76276929 \rho^5+0.08549070\rho^6\,, \\
P_7(\rho) = & 7.54921285+15.37863036\rho +14.48581026 \rho^2 + 8.23429824 \rho^3 \nonumber \\
& + 3.07646819\rho^4 + 0.76371628 \rho^5 +0.11835967 \rho^6 \nonumber \\
& + 0.00902282 \rho^7\,,
\end{align}
}
and
{\small
\begin{align}
Q_0(\rho)  = & 1\,, \\
Q_1(\rho)  = & 1+ 0.36054103 \rho\,, \\
Q_2(\rho) = & 1-0.01907181 \rho +0.02720636 \rho^2\,, \\
Q_3(\rho) = & 1+0.15938215\rho + 0.03066048\rho^2 + 0.00732744\rho^3\,, \\
Q_4(\rho) = & 1+0.03711706 \rho + 0.07252875 \rho^2 +0.00240865 \rho^3 \nonumber \\
& + 0.00077335 \rho^4\,.
\end{align}}

\begin{figure}
\begin{center}
\includegraphics[width = 0.9\textwidth]{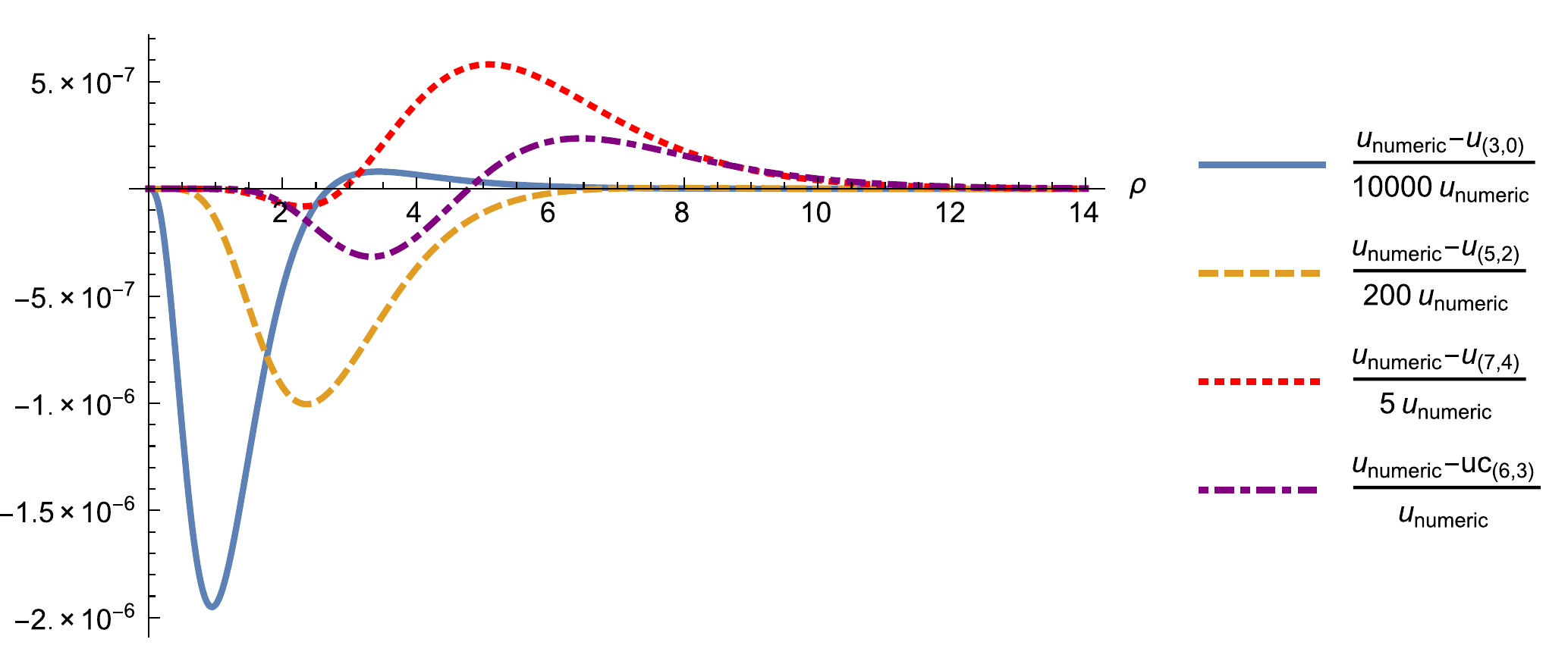}
\caption{\small Relative difference between numerical and approximate solutions $u_{(3,0)}$ (blue solid line), $u_{(5,2)}$ (dashed yellow line), $u_{(7,4)}$ (red dotted line) and $uc_{(6,3)}$ (purple dash-dotted line). Notice that we had to scale down the first graph by a factor 10000, second graph by 200 and the third graph by 5 fit them comfortably into the picture.}
\label{fig:stringpade}
\end{center}
\end{figure}

\begin{figure}
\begin{center}
\includegraphics[width = 0.9\textwidth]{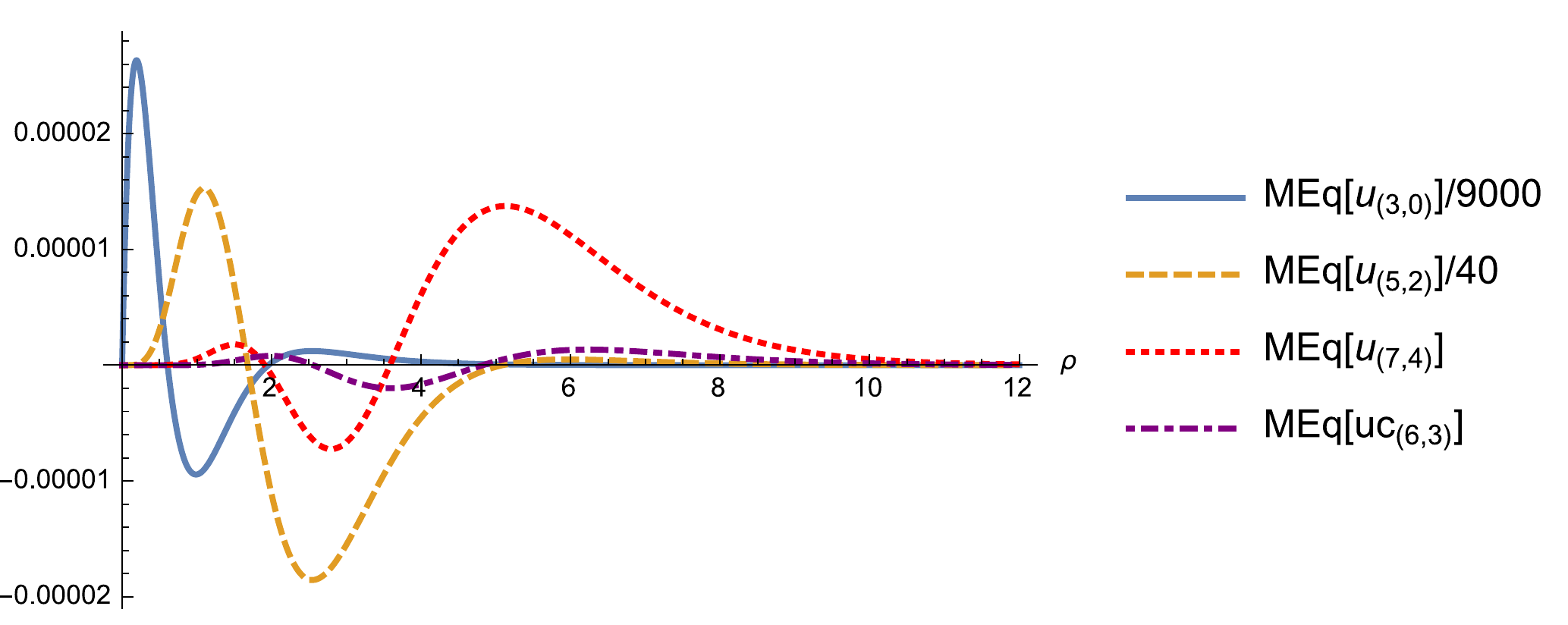}
\caption{\small The master equation for approximate solutions $u_{(3,0)}$ (blue solid line), $u_{(5,2)}$ (dashed yellow line), $u_{(7,4)}$ (red dotted line) and $uc_{(6,3)}$ (purple dash-dotted line). Notice that we had to scale down the first graph by a factor 9000 and the second graph by 20 to fit them comfortably into the picture. 
}
\label{fig:eomstring}
\end{center}
\end{figure}

Let us illustrate the accuracy of our approximations for $p=0, 2, 4$.
In Fig.~\ref{fig:stringpade} we show relative difference between $u_{(p+3,p)}$ and numerical solution and 
in Fig.~\ref{fig:eomstring} we show the master equation evaluated for each approximation. It demonstrates that $u_{(p+3,p)}$ is increasingly more accurate for larger $p$. As we already mentioned $u_{(8,5)}$ is not a regular function. This seems to persist even for higher values of $p$, indicating that capacity of this particular ansatz to approximate true solution has been exhausted. 

Approximations $u_{(p+3,p)}$ have, however, one disadvantage, namely that they depend on odd powers of $\rho$.  In contrast, series \refer{eq:taylors} contains only even powers. The coefficients must be therefore fine-tuned to eliminate all odd powers, especially the first power, since its presence cause a singularity in the master equation \refer{eq:vortexapp} at $\rho=0$. Due to the finite precision of decimal representation of real numbers, this fine tuning is realized only imperfectly, which means that all odd powers, although strongly suppressed, are still present. 

To overcome this technical issue we also considered the following ansatz
\begin{equation}
uc_{(p+3,p)} = \log\biggl(\rho^2+ \frac{1}{\cosh(\rho)}\sqrt[4]{\frac{P_{p+3}(\rho^2)}{Q_p(\rho^2)}}\biggr)\,,
\end{equation}
with the condition
\begin{equation}
\frac{P_{p+3}(\rho^2)}{Q_p(\rho^2)} \xrightarrow[\rho\to \infty]{} q^4\frac{\rho^6}{16}\,.
\end{equation} 
The advantage is that $uc_{(p+3,p)}$ have only even powers of $\rho$ in the expansion around the origin.  Curiously, only for $p=0$ and $p=3$ we get regular functions.
{\small
\begin{align}
P_3(\rho^2) = & 56.99061470+88.00348570 \rho^2 + 49.40766370 \rho^4 + 8.50772247\rho^6\,, \\
P_6(\rho^2) = & 56.99061470+95.87543618 \rho^2+ 61.81840575 \rho^4 + 18.61295952 \rho^6 \nonumber \\
& + 2.46701666\rho^8 + 0.09435195 \rho^{10} + 2.59235211\times 10^{-5} \rho^{12}\,, \\
Q_0(\rho^2) = & 1\,, \\
Q_3(\rho^2) = & 1+ 0.13812713\rho^2 + 0.00447569 \rho^4 + 3.04705768\times 10^{-6} \rho^6\,.
\end{align}} 

While $uc_{(3,0)}$ ranks somewhere between $u_{(3,0)}$ and $u_{(5,2)}$, as far as accuracy is concerned, $uc_{(6,3)}$ is roughly an order of magnitude better than $u_{(7,4)}$ and it is the best approximation to $u_s(\rho)$, which we have found.

\subsection{Approximations to domain wall}

In this subsection we shall for notational convenience relabel the third coordinate as $y\equiv x^3$ and  we also omit the $\tilde{\phantom{m}} $ on the parameter $m \equiv \tilde m$, even though we are still using the dimensionless coordinates.
The master equation for a single domain wall thus reads
\begin{equation}\label{eq:masterwapp}
\partial_{y}^2u_W = 1-\Bigl(e^{my}+e^{-my}\Bigr)e^{-u_W}\,.
\end{equation}
For a generic value of the parameter $m$, no analytic solution is known. 
However, for particular values of $m$, some exact solutions are available \cite{Sakai3}. In our notation, they are given as
\begin{align}
\label{eq:ex1} u_1 & = y+\log\Bigl(1+2e^{-y}+e^{-2y}\Bigr)\,, & m  &=1\,, \\
\label{eq:ex2} u_{3/2} & = \frac{3}{2}y+\log\Bigl(1+3e^{-y}+3e^{-2y}+e^{-3y}\Bigr)\,, & m  &=\frac{3}{2}\,, \\
\label{eq:ex3} u_2 & = 2y+\log\Bigl(1+2\sqrt{6}e^{-y}+8e^{-2y}+2\sqrt{6}e^{-3y}+e^{-4y}\Bigr)\,, & m &= 2\,.
\end{align}

In order to construct analytic approximations, let us see how the generic solution behaves for large positive values of $y$. The dominant term is clearly $u_W \sim m y$. If we set $u_W \sim m y + \delta u$ and neglect higher order terms (i.e. $e^{-2m y}\delta u_W$, $(\delta u_W)^2$) we obtain the equation
\begin{equation}
\partial_{y}^2 \delta u_W = \delta u_W-e^{-2m y}\,,
\end{equation}
which gives  $\delta u_W  = c\, e^{-y}+\tfrac{1}{1-4m^2}e^{-2m y}$, where $c$ is an arbitrary constant (the second independent solution to homogenous equation $e^{y}$ is discarded, since $\delta u_W$ must be negligible compared with $m y$ for large values of $y$). 
This formula does not apply in the case $m=1/2$, where instead we have  $\delta u_W(m=1/2) = c\, e^{-y}+\tfrac{1}{4}e^{-y}(1+2y)$.
As this case is somewhat anomalous, we shall omit it and in the following, we assume that $m>1/2$.

Combining the $\delta u_W$ correction with the dominant term into a single logarithm we obtain the asymptotic form
\begin{equation}
u_W \sim \log\Bigl(e^{my}+c\, e^{(m-1)y}+\frac{1}{1-4m^2}e^{-my}\Bigr)\,, \hspace{5mm} y\gg 1\,.
\end{equation} 
At this point, we do not know the value of 
the constant $c$. 
However, for specific values of $m$, one can exploit reflection symmetry of the solution $u_W(-y) = u_W(y)$ to obtain a rather good estimate. In order to do that, let us look on additional sub-leading terms. 
Repeating the procedure outlined  above, we obtain next order corrections combined into a single logarithm as:
\begin{align}
u_W & \sim \log\Bigl(e^{my}+c\, e^{(m-1)y}+\frac{1}{1-4m^2}e^{-my}+\frac{c^2}{3}e^{(m-2)y}\\
&~~+\frac{c}{(1+m)(1-4m^2)} e^{-(m+1)y} -\frac{4m^2}{(1-4m^2)^2(1-16m^2)}
e^{-3m y}\Bigr)\,. \nonumber
\end{align}
Notice that this asymptotic form contains (upon 
factoring out the 
dominant term $e^{my}$) either powers of $e^{-2m y}$ or powers of $e^{-y}$ or their combinations. 
If we focus on half-integer values of $m$ these factors can be the same, which we use to our advantage, since in that case, we can force the reflection symmetry.
Let us illustrate this on the case $m=1$. We have
\begin{equation}
u_1 \sim \log\Bigl( e^y +c+\frac{c^2-1}{3}e^{-y}-\frac{c}{6}e^{-2y}+\frac{4}{135}e^{-3y}\bigr)\,.
\end{equation}
The strategy of harvesting approximate solutions from the asymptotic forms, which we adopt from now on, is that we discard terms not compatible with the boundary condition $u \to -m y$ as 
$y\rightarrow -\infty$.
In the present case, the terms $e^{-2y}$ and $e^{-3y}$ would grow faster than $e^{-y}$ as  $y\to -\infty$, so we discard them.
Doing so it only remains to ensure the reflection symmetry of $u_1$, which fix the constant to be $c=2$. Thus
\begin{equation}\label{eq:u1}
u_1 \sim \log\Bigl(e^{y}+2+e^{-y}\Bigr)\,.
\end{equation}
Surprisingly, this approximation is, in fact, an exact solution \refer{eq:ex1}! It turns out that terms we have discarded are exactly canceled by sub-leading terms. 

The next case $m=3/2$, including next-to-next leading order corrections to $u$, has the asymptotic form 
with incompatible terms already discarded given as
\begin{equation}
u_{3/2} \sim \log\Bigl(e^{3y/2}+c\, e^{y/2}+\frac{c^2}{3}e^{-y/2}+\frac{c^3-3}{24}e^{-3y/2}\Bigr)\,.
\end{equation}
We see that the choice $c=3$ makes the above approximation an even function of $y$.
Again $u_{3/2}$ with $c=3$ is an exact solution \refer{eq:ex2}. Repeating the same steps for $m=2$, we obtain the form
\begin{equation}\label{eq:u2}
u_2 \sim \log\Bigl(e^{2y}+c\, e^{y}+\frac{c^2}{3}+\frac{c^3}{24}e^{-y}+\frac{c^4-36}{540}e^{-2y}\Bigr)
\end{equation}
and unique value of $c$, which ensures reflection symmetry, is now $c=2\sqrt{6}$. 
Indeed, this choice of $c$ makes $u_2$ an exact solution \refer{eq:ex3}.

All three exact solutions $u_1$, $u_{3/2}$ and $u_2$ were first reported in \cite{Sakai3} and it seems that they are the only ones.
Indeed, if we continue this 
procedure
for $m=5/2$ we obtain the asymptotic form 
\begin{equation}
u_{5/2} \sim \log\Bigl(e^{5y/2}+c\, e^{3 y/2}+\frac{c^2}{3}e^{y/2}+\frac{c^3}{24}e^{-y/2}+\frac{c^4}{540}e^{-3y/2}+\frac{c^5-1080}{25920}e^{-5y/2}\Bigr)
\end{equation}
and upon inspection, we conclude there is no choice of the constant $c$, which would make the above approximation an even function of $y$.

\begin{figure}
\begin{center}
\includegraphics[width = 0.7\textwidth]{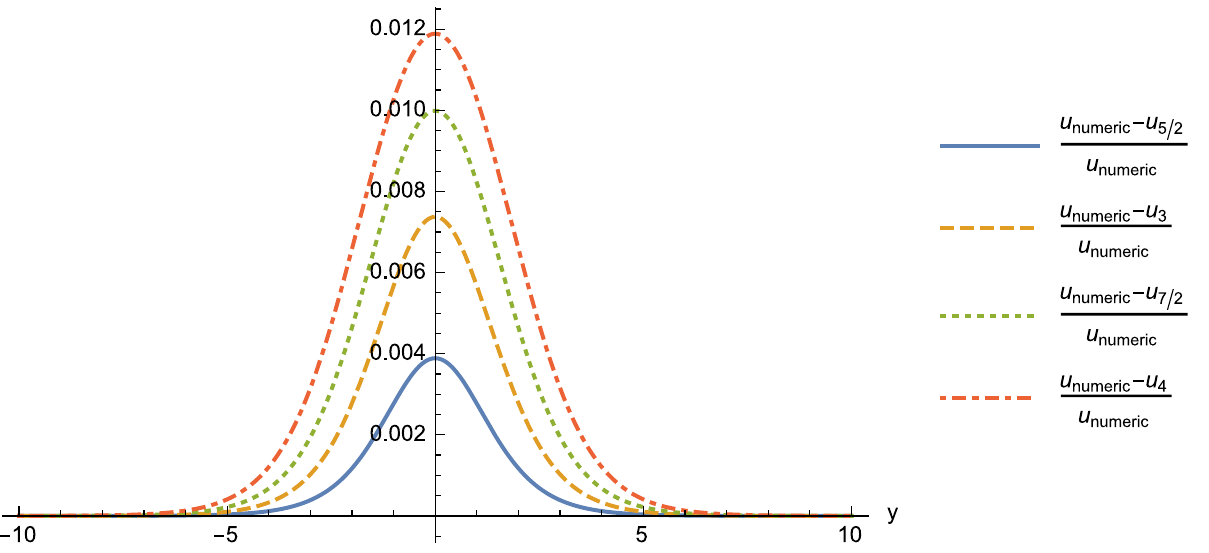}
\caption{\small Relative errors between numerical and approximate solutions $u_{5/2}$, $u_3$, $u_{7/2}$ and $u_{4}$. The area below each curve is 0.013 ($p = 5$, solid blue), 0.023 ($p=6$, dashed yellow), 0.042 ($p=7$, dotted green) and 0.056 ($p=8$, dot-dashed red).}
\label{fig:diff}
\end{center}
\end{figure}

At this point, we adopt the attitude of `patching' the asymptotic form in such a way, that it becomes symmetric under the reflection $y\to -y$. 
The condition for $c$ is chosen in such a way that the patching affects only the innermost terms. Thus, we choose $c^2/3 = c^3/24$, which yields $c=8$. Then we fix the coefficients of $e^{-3y/2}$ and $e^{-5y/2}$ to match their reflected counterparts. In other words, we have
\begin{equation}
u_{5/2} \sim \log\Bigl(e^{5y/2}+8 e^{3y/2}+\frac{64}{3}e^{y/2}+\frac{64}{3}e^{-y/2}+8 e^{-3y/2}+e^{-5/2}\Bigr)\,,
\end{equation} 
which is however only approximative solution, but surprisingly accurate one, as can be seen on Fig.~\ref{fig:diff}.

The process of creating approximations for higher half-integer masses can be now automated. Let us denote $m=p/2$. Based on the previous analysis let us  adopt the ansatz:
\begin{equation}
u_{p/2} \sim \frac{p}{2}y + \log\Bigl(1+\frac{c}{c_1} e^{-y}+\ldots \frac{c^{p-1}}{c_{p-1}}e^{-(p-1)y}+e^{-p y}\Bigr)\,,
\end{equation}
where $c$ and $c_k$, $k=1,\ldots, p-1$ are constants. The knowledge of exact solutions allow us to fix $c_1 = 1$, $c_2 = 3$ and $c_3 = 24$ (see Eq.~\refer{eq:u2}).
The remaining coefficients $c_k$ can be determined by the requirement of reflection symmetry. In particular, we must demand 
\begin{equation}
\frac{c^{p-1}}{c_{p-1}} = c\,, \hspace{5mm} \frac{c^{p-2}}{c_{p-2}} = \frac{c^2}{3}\,.
\end{equation}
These gives a recursion relation
\begin{equation}
c_{p-1} = \Bigl(\frac{c_{p-2}}{3}\Bigr)^{\frac{p-2}{p-4}}\,, \hspace{5mm} (p\geq 5)
\end{equation}
with the solution
\begin{align}
c_p & = \Bigl(\frac{8^{p-2}}{3^{p-4}}\Bigr)^{\frac{p-1}{2}} \hspace{5mm} (p\geq 1)\,, \\
c & = \Bigl(\frac{8^{p-2}}{3^{p-4}}\Bigr)^{\frac{1}{2}} \hspace{8mm} (p\geq 2)\,.
\end{align}
Remarkably, all additional symmetry conditions, such as $c^{p-k}/c_{p-k} =c^{k}/c_k$ for $k=3,\ldots\, p-3$ are satisfied as well.

In summary, we have obtain the following family of approximate solutions
\begin{equation}\label{eq:approx}
u_{p/2}  = \frac{p}{2}y +\log\Bigl(1+\frac{9}{8}\sum\limits_{k=1}^{p-1}\Bigl(\frac{8}{3}\Bigr)^{\frac{k(p-k)}{2}} e^{-k y}+e^{-p y}\Bigr)\,, \hspace{5mm} (p\geq 3) 
\end{equation}
which are entirely determined by the reflection symmetry and few bits of "initial"  data, namely the first three coefficients $c_1 = 1, c_2 = 3$ and $c_3 = 24$. Note that \refer{eq:approx} does not include $m=1(p=2)$ case. It should be treated separately as in  (\ref{eq:u1}). 

Given the spartan amount of information used in the construction of these approximations, it is remarkable how accurate they are. We illustrate this fact on  Fig.~\ref{fig:diff}, where we compare $u_{p/2}$ to numerical solutions.

\bibliographystyle{jhep}
\bibliography{references}
\end{document}